**Aleksandr G. Dodonov, Dmitry V. Lande,**

**Vitaliy V. Tsyganok, Oleh V. Andriichuk,**

**Sergii V. Kadenko, Anastasia N. Graivoronskaya**

# INFORMATION OPERATIONS RECOGNITION:

# FROM NONLINEAR ANALYSIS TO DECISION-MAKING

The book is dedicated to the issues of information operations recognition based on analysis of information space, particularly, web-resources, social networks, and blogs. In this context, open source intelligence technology (OSINT) solves the problem of initial analysis of modern-time information flows. The book provides a detailed description of mathematical principles of information operations recognition, based on mathematical statistics, nonlinear dynamics, complex networks theory, information and mathematical modeling, sociology. A separate chapter covers the applications of approaches from expert estimation theory and decision-making support to information operation recognition.

The book is addressed to a broad circle of specialists from information technology and security domains.

# Table of contents

# Introduction

Any information operation is a component of information war. Today the information operation concept is as relevant as ever. Almost all of us involuntarily become witnesses and participants of various kinds of information combats. In a classical sense, information war is one of information combat forms, a complex of activities intended to produce information impact upon mass conscience in order to change the behavior of the public and impose goals that do not correspond to its interests, as well as (naturally) ensure protection against such impacts.

As we know, information war is a series of activities, initiated in order to achieve informational dominance through damaging information, information-based and information system-based processes of the enemy, and, at the same time, to protect one's own information, information-based and information system-based processes. The key methods of information war are: blocking and distortion of information flows and decision-making processes by the enemy.

We should also note that in contrast to press, in modern-time science and military doctrines wars taking place in the information space are called information operations [Ландэ, 2009]. This definition stresses the fact that they are just components of real multi-aspect struggle. Information operations accompany much more complex processes and represent their components. At the same time, the arena of information operations is information space. On the one hand, it is the place of information battles, while on the other it is the environment to reflect real military action.

Information operations are defined as actions, intended to influence information and information systems of the enemy and to protect one's own information and information systems [Roadmap, 2003]. Implications of information operations can be witnessed in multiple spheres – military, social, economic etc.

The key task of information operations is mass conscience manipulation. Particular goals can vary and include the following:

- Introduction of particular ideas and views into mass conscience and conscience of specific individuals;
- Public disinformation and disorientation;
- Weakening of certain public beliefs and foundations of the society;
- Intimidation of the masses.

There is no need to prove that information aspects of many social phenomena are important for understanding, performing, and counteracting information operations. Indeed, it is difficult to imagine, for instance, voters who make their choices outside information context of the electoral campaign, or promotion of some product without active impact upon its potential consumers.

The basis of information operations includes synergetic principles, emergence concept, and consideration of "system effects". It is assumed that information impacts, launched as a result of special campaigns, must self-develop, expand in avalanche-like fashion, and lead their initiators to desired outcomes. Synergetic approaches are based upon consideration of the society as a complex system, where each component has many degrees of freedom, and that is why they guarantee correctness of modeling results only at the qualitative level.

Information operations differ in terms of their nature, and this nature is extremely complex. That is why they are problematic to model and analyze. Such complexity results from two groups of factors:

- Subjective factors, related to conscious, targeted activity of individuals, involved in information operations;
- Objective factors, related to "system effects" and statistical patterns, applying to a social system, consisting of a large number of components.

Today mathematical modeling is widely used in many areas of science and technology. However, modeling of social processes (which include information operations) remains a problem that is open for researchers to explore.

In social systems, among many other characteristics, one of the most clearly visible ones is integrity, i.e., presence of features that are not inherent to any of the system's components if taken separately. This property, called "emergence", is a result of development of special synergetic connections between the system's components. The term "emergence", first introduced by G. H. Lewes in 1875, means that in physical systems an entity is something more than just a sum of its parts, i.e. on each level of complexity new properties emerge, which were not inherent to separate components. Emergence of a social system does not allow us to confine ourselves to study of its components and connections between them, but calls for integral analysis of the system as a whole. Until late 20th century mostly reductionism-based approach was used for analysis of complex systems (including social systems). The approach explains the properties of complex systems with properties of their components – atoms and molecules. The situation changed drastically after development of system analysis, emergence of the science of complexity, and breakthrough in the area of computational capacities. Today we are witnessing the development of chaos theory, complex networks, non-linear and self-organizing systems. Computer modeling of complex systems indicated that many of their properties could not be deduced from some pre- defined set of dynamic equations. On the opposite, the equations could be obtained only as a result of numeric modeling.

At the same time, obviously, it is impossible to develop and utilize some universal methodology for modeling of information operations, because many concepts and factors are problematic to formalize. In each particular case we have to trust the intuition and competence of the analysts, who handle the issues of organization and prevention of information operations at professional level.

Sometimes they manage to predict certain trends in processes that are clearly visible within social practice.

The situation with objective factors is completely different. They can be statistically analyzed, and they allow us to build numerical estimates that, in their turn, can be used for building of credible forecasts. Contemporary applied statistics includes rich arsenal of thoroughly developed and approbated methods. However, statistics allows us to describe only formal aspects of the matters under consideration, leaving content-related aspects "overboard". That is why it is necessary to expand the set of tools for analysis and modeling of information operations. Mathematical modeling is, undoubtedly, one of the most interesting and promising directions in this context. Its key advantage is the opportunity to consider both formal and content-related aspects that determine the dynamics of the processes being studied. Indeed, the structure of a well-substantiated model always reflects the essential points that the model's authors are capable of understanding. Beside that, adequate models are built using approbated methods, ensuring their formal rigidity.

When it comes to information operations, the most promising approach is modeling, based on certain realistic behavioral rules applying to separate components of the system, and refined by certain parameters, changed in the process of modeling. In this case the inverse problem becomes relevant: estimation of the model's parameters based on real behavior of some dependency.

Information operations modeling attempts started long ago. However, they were restrained by computational difficulties, especially when it was necessary to describe dynamics of systems with feedback. Presently, we have quite a lot of opportunities for efficient computer processing of data. This allows us to prepare sets of input parameters based on analysis of statistical research results, and, at the same time – to solve formalized problems with sufficient degree of precision within acceptable time frames. Based on above-mentioned considerations we can assume that in the nearest time mathematical modeling will become the key means for planning and prevention of information operations.

The research addresses some models of information flows in the Internet that both influence information operations and more and more often provide the environment for their launching. At the same time, much attention is given to such specific issues as technologies and modeling of information operations, information impacts, information flows, and (as a result) information operation recognition.

While studying information operations we should take objective criteria into account. Among these criteria we can consider the dynamics of dissemination of information scenarios within the respective fragment of information space.

At the same time, selection of information flow topics, allowing us to detect information operations, represents a complex labor-intensive and ambiguous task.

*This book contains the results of research conducted under grant support of The State Fund for Fundamental Research of Ukraine as part of competitive project # F73/23558 "Development of Decision-making Support Methods and Means for Detection of Informational Operations".*

# 1. Information operations

In recent years, thanks to numerous documents and publications of US Department of Defense, the term "information operation" has gained popularity, first and foremost, because information technologies are playing an increasingly important role in military operations. At the same time, information operations are defined as actions, targeted at enemy's information and information systems, as well as protection of one's own information and systems [Roadmap, 2003]. Information operations are considered a combination of key capacities of radio-electronic warfare, computer network operations, psychological operations, military action, and operations for ensuring security, intended to influence, destroy, and distort the information, that the enemy needs for decision-making, as well as to protect one's own information.

Information operations cover a whole complex of processes in the most diverse spheres. At the same time it should be noted that they are an important and traditional component of military operations. While the formal definition in the documents of US Department of Defense is focused on military aspects, it is quite applicable to almost every sphere of life.

Below we are going to consider information operations, implemented in the form of information impacts on human conscience.

Information reflects the meanings it contains. That is why today information turned from an abstract term into the object, the purpose, and the means of information operations, became a critical concept within the problematic scope of security. On March 18, 1999, William Cohen, US Secretary of Defense at the time, stated that the ability of an army to use information to dominate in future combats would give the US a new key to victory for many years (if not for several generations) to come [Hill, 2000].

While modeling and performing information operations, we should consider the value of information for decision makers (DM). The value of information includes its relevance, accuracy and "analytic property". In practical terms, the value of information can also be defined as its significance or usability (availability for usage). By usability of information we understand ensuring the access of a DM to information that is ready for usage. ISO 9241 standard defines usability as the effectiveness, efficiency and satisfaction with which specified users achieve specified goals in particular environments. In practice, DM gets most of useful information from information and analysis systems, providing guidance for a given situation (situational awareness), as well as decision-making support. According to the field manual (FM 100-6) of the US Department of the Army "Information Operations", situational awareness is a combination of clear understanding of the disposition of one's own and enemy forces, and assessment of the situation and intentions of the commanders.

Information operations are performed in a specific social environment. In order to ensure their success, it is necessary to adapt to this environment, overcome a certain barrier of insufficient attention to information impact. This barrier emerges as a result of the so-called, environmental immunity system that can block information impacts if it is powerful enough and/or if it has already learned to protect itself from such impacts. Preparatory actions before launching an information operation can include creation of "immune deficiency" in the social environment by influencing the information space, for example, using materials in the media. Very often information impacts use "virus marketing" mechanisms, for example, in the form of rumors, when disinformation presented as sensation spreads with tremendous speed. The immunity system fights back information operations just like this one. Quite often immunity system of the society is associated with the government that has to ensure security of this society, i.e. under strong government apparatus, the chances of success of anti-social information operations significantly decrease. The reader knows well, how counteraction to such informational processes was organized in totalitarian states. In democratic societies totalitarian methods are, naturally, inapplicable. In this case immunity is achieved through "learning", i.e. a democratic society must go through many information attacks, impacts, influences of stereotypes, in order to work out the necessary immunity.

Today, the level of readiness for performing information operations is considered a key success factor of any social procedure or campaign.

Targets of information operations include information and analysis systems of the impact subject. By influencing such systems, one can force decision-makers from enemy camp draw inadequate conclusions, so that the targeted social process changes its trajectory and follows the direction, required by the impacting party [Горбулін, 2009] (fig. 1.1).

In this case immediate information impacts can include placing documents, that compromising the opponents, in the information space, advertisement (including hidden ads) of one's advantages, distortion of data about external environment, distortion of information concerning one's intentions etc.

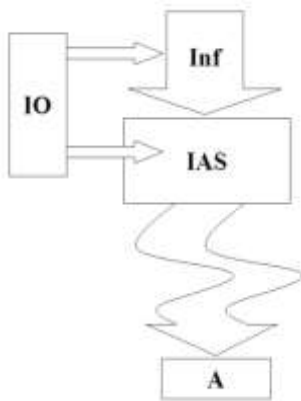

Fig. 1 – Impacting the enemy's information and analysis system:  Inf – information space;  IAS – information and analysis system;  A — system user (DM);  IO — information impacts

As a rule, social procedures and processes are difficult to evaluate and model, because their results are psychological and sociological, rather than physical. This fact also makes it problematic to predict the results of information operation modeling. Beside that, experimentation with information impacts in the framework of information operations is more complex and dangerous, than modeling of physical processes.

In order to ensure efficient influence upon enemy's decision-making processes it is, sometimes, necessary to keep taking certain actions for a long time before they start producing effect.

## 1.1. Information influence

One of the key information operation components is the social impact, including the whole spectrum of impact processes. Considerable changes in public opinions or attitudes towards a certain problem or event are expected to entail changes in behavior, related to this problem.

In 1948 Harold D. Laswell [Lasswell, 1948] developed a model of communication transmission, including five components:

- source – a person that influences or convinces other persons;
- message – the means, with which the source tries to convince the target;
- target – a person that the source tries to influence;
- channel – a method of message communication;
- impact – reaction of the target to the message.

Although Laswell was interested, primarily, in mass communication, his model of information transmission can be applied to interpersonal communication, belonging to the type of circular models of Shannon–Weaver and Osgood–Schramm, who include feedback loops in communication process, stating that communication is a circular rather than linear process [Schramm, 1974], [Osgood, 1954].

Modeling of objective factors of social impacts calls for interdisciplinary approaches, related to informatics, marketing, political science, social psychology. The most renown models of public opinion formation and social impact are based on Latane's dynamic social impact theory [Latane, 1981], [Latane, 1997], elaborated by many other authors, particularly by [Nowak,1990], [Lewenstein, 1993], [Kacperski,2000], [Sobkowicz,2003].

Trying to validate the mechanism of social impact of messages, [Latane, 1981] stressed the importance of the three indicators of the source's relation to the target:

- strength – social force, probability or level of impact upon individuals;
- immediacy – physical or psychological distance between individuals;
- number of sources – number of sources, aimed at the target.

Present-day situation with information operations modeling is characterized by a set of open problems, primarily, those concerning understanding of information impact and influence concepts.

Universal characteristics of an object are its state and ability to influence other objects. Implementation of influence potential requires certain conditions that are called impact. An object capable of imposing its will is called a subject, while control is denoted as influencing the impact object with some specific purpose.

When an individual is the impact object of one or several sources, dynamic social impact theory states that the level of social impact upon the individual can be expressed by an equation, providing the basis for the so called individual-oriented model:

$$I_i = -S_i\beta - \sum_{j=1, j\neq i}^{N} \frac{S_j O_j O_i}{d_{i,j}^{\alpha}},$$

where $I_i$ is the intensity (amount) of the social pressure upon the individual $i$ $(-\infty < I_i < \infty)$. $O_i$ is the individual's opinion (±1) about the topical question, where +1 and -1 represent support of or objection to the given question, respectively. $S_i$ is the strength of individual $i$ or impact $(S_i > 0)$, $\beta$ is the individual's resistance to change $(\beta > 0)$, $d_{i,j}^{\alpha}$ is the distance between individuals $i$ and $j$ ($d_{i,j}^{\alpha} \geq 1$), $\alpha$ is the indicator of distance reduction $(\alpha \geq 2)$, $N$ is the total number of agents (individuals that constitute the population). The value of $\beta$, the tendency towards maintaining one's opinion or resistance to change, defines, whether higher or lower social pressure is required to change the opinions of individuals within the model. Larger values of $\alpha$ represent the effect of increase of distance between the source and the target; as a result, the level of social pressure on the target also changes.

Based on the suggested terms, the concept of "information field of the object" is introduced by [Кононов, 2003], and its characteristics are described. This provides an opportunity to define information impact as impact upon the object's information field. By researching information fields of objects, we can define information impacts and control actions. In this process information can be considered the object and the means of influence. In order to use information as impact means, in the process of control we should prepare the data, produce the respective information, and only then – implement it in the form of influence (impact).

One of the key information operation implementation methods is information influence, exerted for the purpose of information control. In this case by information control we mean the control mechanism, where control action bears implicit character, and control object is presented a certain information picture. Based on this picture, the object forms its behavior pattern. Thus, information control is the means of impact, creating an incentive for the public to behave in an organized manner, i.e. to perform required actions.

According to [Кононов, 2003], [Кульба, 2004], it is appropriate to decompose the process of information influence of one object upon others into the following phases:

- the source generates influence of data, information elements, and information sets;
- the influence source transmits information;
- the recipient gets the information;
- data set, information components, and new sets of the influence object are generated;
- the influence object takes action.

Information impacts upon system components can be classified according to such criteria as sources of origin, impact duration, nature of origin, etc.

In order to select specific ways of information control implementation, we should specify the problems, solved through information impact, analyze the process of information operation formation, and come up with criteria for their evaluation. Information control is considered as a process, embracing three interconnected directions:

- control over data exchange between the real world and virtual world of the influence subject;
- control over virtual world of influence subjects and decision-making mechanisms;
- control over the process of turning decisions into real-world actions by the influence subject.

Information impact may belong to one of the two basic types:

1. Desirable change of data, used by the information and analysis system of the impact object during decision-making.

2. Immediate influence upon decision-making process of the impact object, for example, upon decision-making procedures or specific decision-makers.

Factors of particular importance for implementation of information operations include environment, condition of information impact objects, their mutual influence. Particularly, if the information operation object is some electoral field, then it is important to consider all electoral populations, belonging to this field, and representing supporters (or opponents) of these or that political forces. While further on we will consider some models, where the homogeneity of the environment is explicitly postulated, in the general case, the environment can consist of several spheres, related to information operations:

- dominant perception;
- increased sensitivity;
- indifference to respective information impacts.

If mass media system is used, external impact is implemented through a set of information impacts, i.e. manipulations with information (dissemination, withholding/concealment, change).

Characteristic features of modern-time information operations:

- provide an opportunity to conceal the fact of their implementation, while still produce the desired effect;
- signify targeted activity of a broader character;
- ensure considerably lower cost of goal achievement in comparison to traditional means.

Usage of online mass media within management cycle, among other things, provides the opportunity to communicate the decisions of the management to broad audience. Beside that, these media reflect public reaction to the decisions being made. According to a study by E. Noelle-Newman [Noelle, 1973], many people try to hide their views and opinions if they contradict opinions of the majority. According to this theory, mass media are an effective instrument for influencing public opinion. Many researchers of information combat in open systems stick to this view as well. For instance, S. Ball-Rokeach and M. DeFleur [Ball-Rokeach, 1984] underline the strong connection between social system, media, and audience. Dependence of the audience on the media is defined by individual differences of the recipients, the scale of covered events, the amount and degree of centralization of information functions, fulfilled by the media. News messages often include not only information about some event, but also about authors' attitude to this information [Лозовский, 2000], [Лозовский, 2006], [Лозовский, 2003], [Лозовский, 2000], [Урсу, 2011], [Урсу, 2012]. For example, in the coverage of the events in eastern Ukraine, the editors' attitude can be defined by how the combating sides are called when mentioned: "militia/volunteers" – "guerillas", "punitive forces" – "antiterrorist operation forces" etc.

In order to implement efficient information impact upon conscience of the public, it is necessary to ensure that the message is read by as many people as possible [Brosius, 1992] [Wanta, 1994]. As, nowadays, more than 100, 000 news reports appear in online media just within Ukrainian Internet segment, even the latest news of a significant event can remain unnoticed.

A research by S. Ball-Rokeach, M. Rokeach, and J. Grube [Ball, 1976] indicate that people tend to revise their opinions and behavior models if they feature certain contradictions. Results of this research are used to perform permanent, gradual change of public attitude through a sequence of information impacts.

From information operations viewpoint, online media have a set of advantages in comparison to traditional ones [Манойло, 2007], such as:

- efficiency, accessibility, economy, information dissemination;
- potentially unlimited audience;
- complexity of information representation and perception;
- opportunity to create media, available to any organization or individual.

Presently, information operations are widely used [Почепцов, 2001]; in some countries special information operation units are created, and guidelines for their implementation in time of peace and war are developed, as well as instructions for cooperation of the respective units etc. Beside that, special non-governmental structures are created, that have an opportunity to effectively implement information operations.

## 1.2. Information operation phases

Let us now address the phases of information operations. Naturally, there is no unified "standard" plan for implementation of information operations, both offensive and defensive. We can only consider the approximate sequence of actions taken in the process of information operation implementation. These actions can be reconstructed based on generalized experience of some of the already implemented information operations.

In practice, information operation as a process of information impact upon public consciousness is, usually, implemented as follows. As a result of preliminary intelligence phase, the plan of the next phase – operational control – is developed, and respective operational intelligence measures are planned, which provide an approximate decision model. After that operational control is applied to the enemy. During operation intelligence phase, the level of deviation of the initial model from reality is defined. If the deviation is insignificant, then the initial plan is implemented. Otherwise, a new plan is developed for operational control and control of the enemy. Then the cycle is repeated until operational intelligence confirms the model. At the same time, the final decision is made with a certain degree of operational risk.

Information impact process includes the following key phases [Чхартишвили, 2004]:

- Preliminary intelligence;
- Detection of current situation, condition of the enemy;
- Control of the enemy (information impact upon the enemy, aimed to communicate the information, corresponding to the control subject's intention);

- Operational intelligence (verification of reflexive control results);
- Operational control – actions of the control subject, aimed at achievement of the required goal.

While planning and modeling social processes, particularly, information operations, we should always keep in mind, that general behavior of social systems cannot be defined based exclusively on refined mathematical models. The main reason for this phenomenon is that such behavioral processes are mostly guided by socio-psychological factors.

The two basic information operation types are offensive and defensive operations. However, in practice most information operations belong to a mixed type. Moreover, the majority of information operations are both offensive and defensive at the same time. Each of information operation types includes the above-mentioned phases, but also envisions certain specific features and clarifications.

A peculiar feature of offensive information operations (information attacks) is that impact objects of such operations are clearly defined, and planning process is based on rather specific information about these objects. An information attack, usually, requires the attacker to find or create and informational cause (for defensive information operations the cause may be the enemy information attack itself), promote this cause, i.e. organize its propaganda (as opposed to counter-propaganda used in defensive information operations), and take measures to suppress information counteraction.

Operational control of information operations using information and analysis systems can be illustrated by the diagram, displayed on fig. 1.2.

According to this diagram, from the real world (R) information gets into information space, particularly, to the media (I), or directly to the experts (E) (also through the media). From the experts or directly from the information space, information gets into the information and analysis system (IAS), for example, with the help of content-monitoring tools. The information and analysis system communicates the data that defines the information impact measures, to decision-makers (P). Information impact measures are information operations (IO), targeted at information space and directly at real-world objects (people, environment, computer systems, etc).

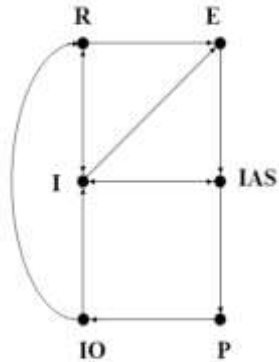

Fig. 1.2 Diagram of operational control using information and analysis systems

An approximate algorithm of information operation assessment was presented in the information operations center of US Department of Defense [Roadmap, 2003] in the form of eight consecutive steps including specific actions:

1 – characterization of information space;

2 – inclusion of information operation assessment into military operation plan;

3 – study of the needs for information, required for assessment of the information operation, and planning of data collection;

4 – compiling initial data for information operation assessment;

5 – implementation of activities for information operation and intelligence support;

6 – observation and collection of data for information operation assessment;

7 – analysis of obtained data; assessment of the information operation;

8 – reporting the obtained results to and providing recommendations for the commander.

While most information operations represent activities, in which only non-lethal weapons are used, they allow their initiators to achieve the results that are the same as in the case of traditional weapon implementation, and, at the same time, minimize or totally eliminate potential negative consequences.

According to US DD analyst Lara M. Dadkhah, who studied the war in Afghanistan, capacities of "Taliban" guerillas for launching information operations were, sometimes, "the most effective air defense means". After all, it doesn't really matter, how the enemy aircraft is neutralized, i.e. whether it is hit by a missile or landed as a result of enemy propaganda.

Developing this topic further, US Armed Forces command stated that the purpose of information operations force would be to establish more accurately the balance in implementation of lethal and non-lethal impacts in order to enhance the efficiency of information capacities of the troops, and, thus, to achieve the purpose of a military operation in shorter time. At the same time, the command recommended to consider the above-mentioned components in the next edition of the doctrine JP 3-13 "Information operations", as well as in other documents, featured in the series JP 3-13.XXX and immediately concerning information operations. From their side, US military specialists stressed the extreme need for preparation and release of a new document on information operation assessment.

## 1.3. Information operations modeling

Modeling can be considered one of the ways of solving real-world problems, particularly, those, emerging while planning and implementing information operations. Most often, modeling is used when experimentation with real objects is either impossible or too costly. Modeling includes bringing of the real problem into the world of abstraction, analysis, and optimization of the model, as well as bringing of the optimal solution back into the real world.

There are two alternative approaches to modeling – analytical and imitation (simulation) modeling. Ideal analytic models allow us to get strict analytical solutions or, at least, problem statement, for instance, in the form of differential equation systems. However, analytic solutions are not always possible to find. That is why, especially nowadays, and particularly while solving problems from social dynamics domain, imitational modeling (*Simulation Modeling*) methods are used more and more often. Simulation modeling is a powerful and virtually irreplaceable means of social procedure analysis. A simulation model can be considered as a set of rules, defining the future system state based on its present state. Modeling process is observation of the system's evolution (according to the given rules), and, respectively, (if possible) assessment of the model's adequacy.

The most promising direction of information operations modeling is mathematical description of information dissemination and perception environment self-organization, based on current situation. Self-organizing domains, in which there is no centralized control mechanism, but evolution goes on as a result of multiple local interactions, are studied by the theory of complex systems. This theory covers such knowledge fields as nonlinear physics, thermodynamics of non-equilibrium processes, dynamic systems theory. Mutual impacts of separate elements of complex systems define the emergence of complex behavior under absence of centralized control. Research of such behavior calls for usage of the most contemporary methods, covered by the interdisciplinary basis of present-day methodology – the concept of complexity. At present, theoretical and technological basics of this concept include the theories of dynamic chaos and complex networks, synergetic studies, fractal and wavelet analysis, multi-agent modeling, self-organized criticality theory (studying dynamic development up to critical state, characterized by strong temporal and space fluctuations, without external control [Bak, 1996]), percolation theory etc.

Modeling of social procedures (undoubtedly, including information operations) calls for numeric experiments, as, most often, substantial limitations are witnessed, making it problematic to conduct "natural" experiments in the field.

When modeling information operations, a numeric experiment allows us to reduce the number of operations on limitation clarification, initial data selection, as well as selection of model component functioning rules etc. In such a case we get an opportunity to consider cases, which are problematic to implement in practice, using real data only to identify parameters of the mathematical model. At the same time, mathematical modeling has its limitations, and real world turns out to be too complex to be modeled with a sufficient degree of detail, precision, etc., i.e. more or less credible mathematical models are so complex and multi-parametric that they are problematic to analyze and evaluate using exact methods.

In order to detect information influence in social networks, we can use different models. Let us consider several of these models [Wasserman, 1994].

- Models with thresholds, including linear thresholds, where an agent can be in active and passive states, while transition is possible only from passive to active state (inverse transition is not allowed);
- Models of independent cascades, belonging to the so-called "interacting particle systems";
- Models of infiltration and contamination;
- Ising models;
- Models based on cell automata;
- Markov chain-based models.

Particularly, according to dominant nature of information impacts, information operations in the web can be nominally classified as:

- "propaganda", when information impacts bear mostly promotional character, for instance, when

elections are held and prepared, or in the process of political struggle;

- "des-information", when the primary goal is to disorient the enemy; these information operations are used in both political and economic spheres, as well as during military conflicts;

-  "manipulative", when the key task is to put the competitor's behavior under control using various information impact techniques (persuasion, manipulation, attitude modification);

- defensive ("counter-operations"), when the purpose is to neutralize the enemy's information impact (say, through counter-propaganda), to protect oneself from such influence, to respond to information impact of the competitor [Манойло, 2003].

If information impact depends only on awareness and interconnections between the agents, then a classical model from game theory can be used. The award, received by an agent depends on the activities of its "friends". Game theory-based information impact models include the following ones [Губанов, 2009]:

- mutual awareness;
- coordinated collective actions;
- communications and minimum sufficient network search problem;
- network stability;
- information influence and control;
- information combat.

Most models focus, primarily, on the agent interaction rules, but when it comes to the influence network itself and its properties, analysis of respective models yields very modest results. That is why in [Əliquliyev, 2010] it is suggested to define the so-called "information archetype" of social network users and agents' states. Just like in many other models, an agent's state can be active and passive, while an active agent's ability to spread information through the network depends on multiple factors.

Different approaches to classification of information impact models, mentioned above, indicate that currently available models adequately reflect many properties and effects, witnessed in social networks. At the same time, all models of information influence and struggle should be formulated according to the specificity of an applied problem being solved, the possibility of modeled problem identification, parameters of agent environment, potential actions of agents, their advantages and awareness.

On the whole, we can conclude that modeling of influence in a social network is, currently, becoming an independent study subject, and respective models will soon form a separate branch of research.

During planning of information operations, mathematical models can be refined only in the process of modeling of specific procedures, when they are constantly compared to reality.

The purpose of the information operation assessment methodology is to ensure timely and accurate analysis of potential discrepancies between the planned operation and the actual impact. When substantial differences (that influence the probability of the operation success) are detected, analysis system must inform decision-makers about them, in order to correct current plans and decisions accordingly. At the same time, while planning information operations we cannot rely on trial-and-error method. That is why, it is necessary to develop methods that would allow us to aggregate retrospective data and verify model adequacy on their basis.

Synergic approaches provide the basis for viable information operation models. Indeed, a society is a complex system, where each component is characterized by a multitude of features, and has many degrees of freedom. An important property of this system is self-organization, resulting from interaction of such components as randomness, recurrence, positive and negative feedbacks.

Relative simplicity of interpretation of obtained results can be considered the specific feature of information operations modeling. Such concepts as "electorate number", "political weight" etc are perceived intuitively, even without knowledge of exact definitions (if applicable). This allows us to make detailed analysis the subject of broad discussion.

As some decisions are unstable in respect to their parameters, values of these parameters have to be defined with high precision. This requires a complex of methodologies based on processing of large statistical data as well as on versatile sociological research.

Presently, the most realistic problem statement is based on mathematical model usage for predicting the dynamics of social processes at qualitative level. In this formulation dynamics modeling occupies a kind of intermediary level between what we are going to set forth here and accurate forecasting. However, we will need to select parameter values that (to some reasonable approximation extent) correspond to the situation under consideration. Moreover, usage of relative values seems to be the most productive approach in most cases. Although it is impossible to get credible data on future turns of events based on this approach, most probably, it might allow us to reconstruct a more or less clear picture of what can happen and how. And that is already something.

In order to succeed, we need to consider separate information impacts as parts of a unified information operation (similarly, artillery shelling or aviation strikes can be considered coordinated parts of a military operation).

Information operations have the following basic features:

- information operations represent an inter-disciplinary set of methods and technologies from such spheres as informatics, sociology, psychology, international relations, communications, defense;
- even now there are still no standards for information operation implementation;

- not just defense departments, but also numerous governmental and commercial structures are interested in development of information operation technologies;
- the problem of developing a scientific approach to information operations is topical and relevant.

While implementing information operations, it is important to detect the content (knowledge), input into information, concerning various aspects – social, political, religious, historical, economic, psychological, mentality-related, cultural – specific to different social strata. That is why, now it makes sense to consider information operations in a broader sense, as knowledge-based operations (Knowledge Operations) [Burke, 2001].

Nowadays, an ordinary information operation in web environment is performed as follows: as a rule, a special web-site is created (let us call it the "source of origin"), that functions during certain time, and publishes rather specific information. At the moment of time X a special document appears on the web-page. Usually, it contains some compromising evidence (credible or falsified) against the attack object. Then the so-called "information laundering" is performed. The document is re-published online by sources of two types: either those interested in the attack or those in need of any information to fill their information field with. In case of complaints, these re-publishers just refer claimants to the "source of origin", and, in the worst case, remove the material from their web-sites on request from the attack object. The source of origin, if necessary, also removes the information, or even gets eliminated (afterwards it often turns out that the source is registered under the name of some nonexistent person). However, the information has already spread, and the original source's task is completed: the attack is launched.

Contemporary information space presents a unique opportunity to get any information on some given inquiry (providing the respective tools are available), allowing us to analyze interconnection between possible events or events already happening and information activity of a certain group of information sources. On the other hand, during retrospective analysis of any process or phenomenon, some characteristics of its development represent particular interest:

- quantitative dynamics, inherent to the process or phenomenon, such as the number of events within a unit of time, or the number of events, related to it;
- definition of critical, threshold points, reflecting quantitative dynamics of the phenomenon;
- definition of occurrences in critical points, such as detection of key subjects in media publications, concerning the chosen process or event;
- once the key manifestations of the phenomenon in critical points are detected, these manifestations are ranked, and dynamics of development of certain manifestations before and after critical points is studied;
- statistical, correlative, and fractal analysis of the general dynamics and dynamics of separate manifestations is performed. Based on this analysis, attempts to predict future development of the phenomenon or its specific manifestations are made.

In order to study the interconnection between actual events and publications about them in the Internet, the authors have used InfoStream system, ensuring integration and monitoring of online information resources.

The daily number of web-publications on some topic, and, especially, change (dynamics) of this value sometimes allow even average-level specialists in the subject domain to draw more or less accurate conclusions.

We can obtain the data on such dynamics, for example, through daily visits to web-sites of news integrators (news.yandex.ru, webground.su, uaport.net). Of course, users of professional monitoring systems, such as Integrum or InfoStream, are in better position. It is InfoStream that allowed us to obtain amazing statistics on the number of web-publications on influenza epidemics in different time periods.

During modeling and implementation of information operations, we should keep in mind the importance of the value of information for decision-makers. The value of information includes its relevance, accuracy, and "analytic property". From practical viewpoint, the value of information can also be defined as its significance or usability (availability for usage). By usability of information we understand ensuring the access of a DM to information that is ready for usage.

The first step in creation of a network multi-agent information dissemination model is formation of a realistic virtual information space, populated by virtual agents, with which separate information messages in social networks are associated. Beside other information, these messages encapsulate links to information resources from the Internet.

Below we suggest a model of thematic information flow (TIF) formation based on network multi-agent model [Додонов, 2015а], within which separate documents, presenting information subject, are associated with agents, while agent lifecycle is associated with document lifecycle in the information space. It is assumed that the population of agents evolves with time.

In this book we consider a multi-agent information dissemination model, where the key element is a message. Every message can provoke different reaction types, such as positive/negative comments, expressions of approval or disapproval (like/dislike), while the message text can be copied and reposted; beside that, one message can contain links to other messages. That is why we are going to consider a message as an agent in the model, while an agent's evolution will be associated with events, happening to it. Let us denote the basic characteristic feature of an agent as "energy" ($E$) that reflects the message relevance and the degree of interest it provokes. Naturally, obsolescence of information or negative response decreases the energy of a message, while positive reaction or new references to the message increase its energy.

## *Rules of evolution of an agent in the model*

An agent emerges with initial energy $E_0$ and with every discrete time unit its energy decreases by 1. We will consider events, typical for social networks: like, dislike, repost, link (when one agent makes a reference to another agent). These events influence the agent's energy level as follows: a "like" increases the energy by 1, a "dislike" decreases it by 1, a repost increases it by 2, a link increases it by 1. On the other hand, the probability of occurrence of any of these events depends on the relevance of the message and on the level of interest to the information it contains; these factors are also expressed in energy terms within the model. So, let us denote the probabilities of events that can happen to a message with energy $E$.

$$p_{like}^{(E)} = p_{l_0}\varphi(E); \quad p_{dislike}^{(E)} = p_{d_0}\varphi(E); \quad p_{repost}^{(E)} = p_{r_0}\varphi(E),$$

where $p_{l_0}$, $p_{d_0}$, $p_{r_0}$ are model parameters while $\varphi$ is some monotonously non-decreasing function of current energy of the agent, assuming values within [0, 1] range. Once the energy reaches 0, the agent "dies" and is no longer taken into consideration.

Fig. 1.3 displays an example of potential dynamics of a multi-agent system: processes of production of new agents by existing ones are marked by continuous lines, while processes of making references to agents are marked by punctured lines; living agents are black circles, while "dead" agents (as of the moment $t = 5$) are empty circles.

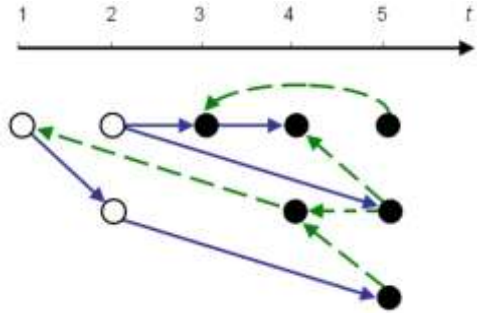

Fig. 1.3 – Multi-agent space fragment

## *Modeling of information flow dynamics*

Fig. 1.4 displays the results of numeric modeling of agent number (ordinate axis on the graph) in the multi-agent system under consideration as function of model cycles (abscissa axis on the graph).

The considered model of agent space evolution under different values of controlling parameters is consistent with dynamics of real topical information flows, defined using InfoStream system.

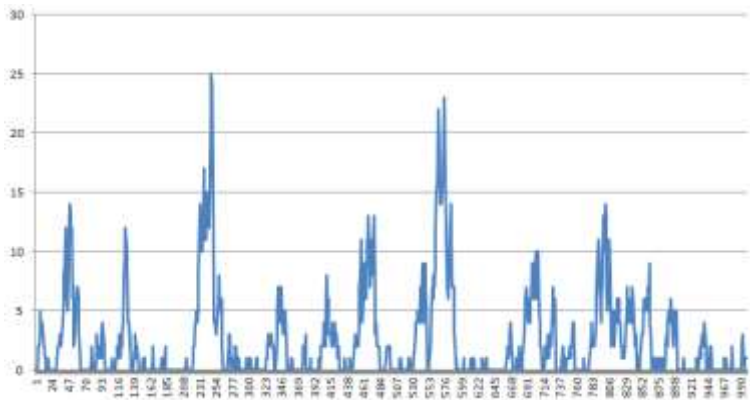

Fig. 1.4 – Dynamics of change of agent number in the model

Modeling of dynamics of the whole information flow starts with a single agent. A new agent can emerge in one of the two ways. First, an existing agent can be copied using repost operation. Beside that, "self-generation" of an agent is also possible (it represents publication of a new message). Thus, with each new moment of time, any of the listed events can occur to every agent with a certain probability. Also, at any moment of time a new agent can emerge as a result of self-generation with a probability $p_s$.

Let us now consider the life cycle of a single agent [Ландэ, 2016a]. An agent emerges with initial energy level $E_0$, then its energy changes, depending on events that happen to it. Let us assume that two events are possible: a "like" and a "repost". Within one time unit three options are possible: both events can happen, one of the two events, or none of them.

Let us record the agent's energy level $\varepsilon_t$ at the moment $t$. Then, the energy of the agent at the next moment can be expressed as follows.

$$\varepsilon_{t+1} = \varepsilon_t + \delta_t,$$

where $\delta_t$ is a random value from the set $\{-1, 0, 1, 2\}$. According to energy change rules, introduced above, energy increase by 2 represents the situation when both "like" and "repost" take place; increase by 1 – to repost-only; the energy doesn't change if a "like" takes place, and decreases by 1, if none of the events occurred. Consequently, we can denote the conditional distribution under known energy level:

$$P\left(\delta_t = 2 \mid \varepsilon_t = E\right) = p_{like}^{(E)} p_{repost}^{(E)};$$

$$P\left(\delta_t = 1 \mid \varepsilon_t = E\right) = \left(1 - p_{like}^{(E)}\right) p_{repost}^{(E)};$$

$$P\left(\delta_t = 0 \mid \varepsilon_t = E\right) = p_{like}^{(E)} \left(1 - p_{repost}^{(E)}\right);$$

$$P\left(\delta_t = -1 \mid \varepsilon_t = E\right) = \left(1 - p_{like}^{(E)}\right)\left(1 - p_{repost}^{(E)}\right).$$

These formulas apply when *E>0*. Further on we are going to use the denotation: $P_\Delta^{(E)} = P\left(\delta = \Delta \mid \varepsilon = E\right)$.

The process of agent's energy change can be considered as whole-number random walk with transition probabilities

$$p_{ij} = \begin{cases} P_{j-i}^{(i)}, & (j-i) \in \{-1,0,0,1,2\}, \quad i > 0; \\ 1, & i = j = 0; \\ 0, & else. \end{cases}$$

As energy value at the next moment of time depends only on energy level at current moment of time, the stochastic sequence $\left(\varepsilon_0, \varepsilon_1, ..., \varepsilon_t, ...\right)$ is a Markov chain with transition probabilities $p_{ij}$. The graph, representing such Markov chain, is shown on fig. 1.5.

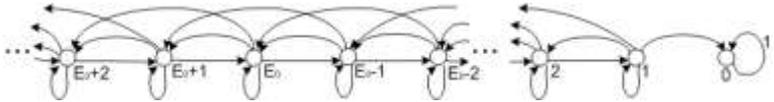

Fig. 1.5 – A graph, representing a random walk of agent's energy

Probability distribution of the sequence can be explicitly shown using transition probabilities:

$$P\left(\left(\varepsilon_0, \varepsilon_1, ..., \varepsilon_t, ...\right) = \left(E_0, E_1, ..., E_t, ...\right)\right) = \prod_{i=1}^{\infty} p_{E_{i-1}E_i}.$$

Or a similar distribution for $\left(\delta_0, \delta_1, ..., \delta_t, ...\right)$:

$$P\left(\left(\delta_1, \delta_2, ..., \delta_t, ...\right) = \left(\Delta_1, \Delta_2, ..., \Delta_t, ...\right)\right) = \prod_{i=1}^{\infty} P_{\Delta_i}^{\left(E_0 + \sum_{j=1}^{i-1} \Delta_j\right)},$$

providing $\Delta_i \in \{-1, 0, 1, 2\}$ and for some $k: \sum_{i=1}^{k} \Delta_i = -E_0$ (the agent's energy fell to 0 on step *k*), then

$\Delta_l = 0, \quad \forall l > k$. For series $(\Delta_0, \Delta_1, ..., \Delta_t, ...)$, that do not satisfy these conditions, the probability equals 0. Further on we are going to consider only series that satisfy the listed conditions.

Theoretically, there are series $(\Delta_0, \Delta_1, ..., \Delta_t, ...)$, that occur with positive probability, while $k : \sum_{i=1}^{k} \Delta_i = -E_0$ can rise to any values without limitation. We should note that $k$ corresponds to agent lifespan duration and denotes the time period, during which the information message remains relevant. Consequently, agent's lifespan should be finite with large probability.

Let us denote the agent lifespan duration with initial energy amount $E_0$ as $\tau_{E_0}$; thus, we denote the time during which from $E_0$ we get to 0. In the realistic model it would be good to have an estimate $P\left(\tau_{E_0} > T_{\max}\right) < \varepsilon$ for not very large values of $\varepsilon$; this would allow us to consider finite series $(\Delta_0, \Delta_1, ..., \Delta_t, ...)$ instead of infinite ones.

Let us consider function $\rho_T(E) = P(\tau_E > T)$. The following recurrent relation is relevant.

$$\rho_T(E) = P_2^{(E)} \rho_{T-1}(E+2) + P_1^{(E)} \rho_{T-1}(E+1) + P_0^{(E)} \rho_{T-1}(E) + P_{-1}^{(E)} \rho_{T-1}(E-1).$$

Such a system of recurrent relations can be solved using initial conditions:

$$\rho_0(E) = \begin{cases} 0, & E = 0, \\ 1, & E \neq 0. \end{cases}$$

Under initial parameters $p_{l_0} = 0.4, \quad p_{r_0} = 0.1$, the following estimate can be obtained from the recurrent equation solution: $P\left(\tau_{E_0} > 1.5E_0\right) < 10^{-3}$. That is, agent's lifespan duration is limited by $1.5E_0$ and, consequently, in order to get $(\delta_0, \delta_1, ..., \delta_t, ...)$ distribution estimates, that are accurate enough, we can consider vectors with finite length $T_{max} = 1.5E_0$.

The number of events, which occur to agents, is also an issue, deserving attention. Let us consider the distribution of "like" numbers. We should note, that if at the moment $t$ a "like" occurred, then $\delta_t \in \{0, 2\}$; otherwise $\delta_t \in \{-1, 1\}$. Let $(\Delta'_1, ..., \Delta'_{T_{max}})$ be the vector, that satisfies the condition that $\Delta'_t \in \{0, 2\}$, if $t = t_1, ..., t_n$ and $\Delta'_t \in \{-1, 1\}$ otherwise, where $0 < t_1 < \cdots < t_n < T_{max}$. So, for an agent that received a "like" the following formula applies:

$$P\{like\} = \sum_{t_1 < ... < t_n} \sum_{\{\Delta_1, ..., \Delta_{T_{max}}\}} \prod_{i=1}^{T_{max}} P_{\Delta_i}^{\left(E_0 + \sum_{j=1}^{i-1} \Delta_j\right)}.$$

## Numeric results

Fig. 1.6 illustrates the obtained distribution density under initial parameters $p_{l_0} = 0.4$, $p_{r_0} = 0.1$. The points, connected by straight lines, denote the obtained values $P$ {agent received $n$ "likes"}.

The smooth curve corresponds to Weibull distribution density:

$$f(x) = \begin{cases} \dfrac{k}{\lambda} \left(\dfrac{x}{\lambda}\right)^{k-1} e^{-\left(\frac{x}{\lambda}\right)^k}, & x \geq 0 \\ 0, & x < 0 \end{cases}$$

Weibull distribution parameters $k$ and $\lambda$ were obtained with maximum likelihood estimation method. Under the aforementioned initial parameters, the obtained values are $k = 1.9$, $\lambda = 3.8$.

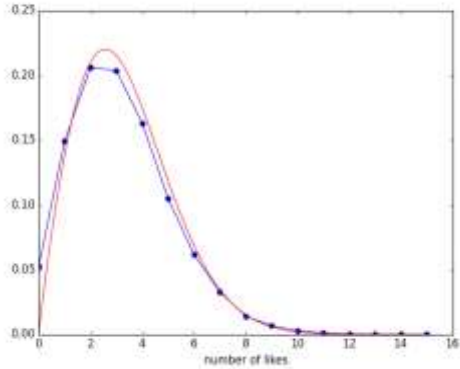

Fig. 1.6 – Distribution density of the number of "likes",
received by the agent under initial parameters
$$p_{l_0} = 0.4, \quad p_{r_0} = 0.1$$

Fig. 1.7 displays a similar result under initial parameters $p_{l_0} = 0.3, \quad p_{r_0} = 0.1$. In this case the obtained distribution parameters are $k = 1.9, \quad \lambda = 3.0$.

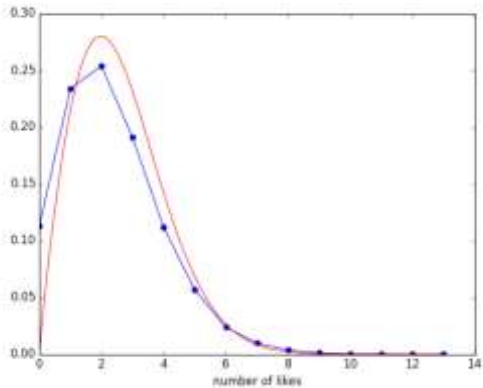

Fig. 1.7 – Distribution of the number of "likes" received by
the agent under initial parameters $p_{l_0} = 0.3, \quad p_{r_0} = 0.1$

## *Study of actual information flows*

Obtained modeling results were compared to the results of the authors' research of news message lifecycles in the microblog network Twitter, where the characteristics of reposts (retweets) of selected messages were analyzed [Li, 2012]. Distribution of "likes" and "retweets" in this case (just like in the model) matches the standard Weibull distribution, while parameter $k$ coincides with the model value with high accuracy (fig. 1.8).

Software tools, developed using $R$ programming language, included three components:

- Tools for scanning and collection of data on growth of retweet numbers for messages of a specific network user (for messages from New-York Times paper, the cycle of 15 sec was selected).

- Processing of collected information through approximation of data on growth of retweet number with Weibull function; calculation of respective scale and form coefficients (nonlinear approximation using least squares method), as well as calculation of the derivative for estimation of retweet number growth speed and building of the necessary graphs.

- Accumulation of obtained results for further analysis. For this purpose, data, obtained using software tools, developed on R, are imported into external databases.

Thus, the records were gathered, which included the text of each message, the time it was published online, values of scale and form coefficients, graphs of growth of the number of retweets and "likes", graphs of approximation of data on retweet number growth by Weibull function, graph of retweet number growth speed etc.

As a result of described research, a model of news lifecycle in information networks has been built.

As a result of modeling, statistical patterns have been detected, relating to the number of "likes" and "reposts" of specific messages. Modeling results indicate that their distribution matches Weibull distribution function.

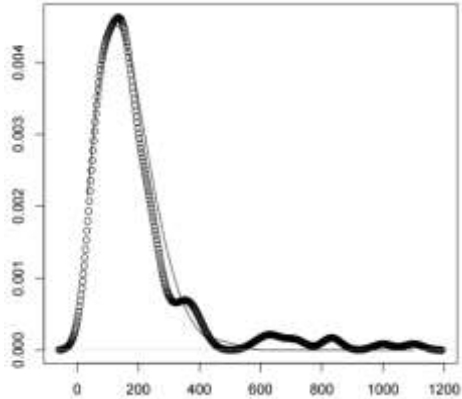

Fig. 1.8 – Distribution density of retweet numbers, obtained from the real network (approximation with Weibull distribution under $k$=1.9, $\lambda$=180)

Modeling data was verified by research of the real microblog network Twitter. Matching between modeling results and real network parameters allows us to speak of a pattern, specific to real networks, and of high model accuracy.

Naturally, orientation towards a single type of sources and mathematical models may lead to deficit of information, necessary for decision-making, inaccuracies, and, sometimes – to disinformation. Only usage of complex systems, based on utilization of multiple sources, data bases, mathematical models, along with above-mentioned capabilities of content-monitoring systems, may guarantee effective information support during information operation counteraction.

## 1.4. Information operation recognition problem

For operational analysis of information situation in the process of information operation detection specialized information space monitoring (content-monitoring) systems are used. First, such systems ensure efficiency that cannot be provided by traditional search systems (network content indexing time in even the best of them ranges from several days to several weeks). Second, they ensure completeness (in terms of both sources and representation of materials of these sources), that is not always provided by ordinary news aggregators. Third, they provide the necessary analytical tools, allowing the user to compile analytical reports, based on publications on given topics for the required time period.

In the context of information operation prevention we should carefully monitor the dynamics of publications on target campaign, and, if possible, take tonality of these publications into account, as well as use available analytical tools, such as wavelet analysis. At the same time we should use existing information attack models as basis. For instance, if a model includes such phases as "background publications" – "calm" – "preparatory shelling" – "calm" – "attack" (Fig. 1.9), then even the first three components allow us to predict further events with high probability.

The abovementioned plan is, obviously, an optimal one, based exclusively on web-resource content-monitoring data.

Of course, users of professional content-monitoring systems are in better position. Many contemporary information and analysis systems incorporate tools for displaying statistics of inclusions of concepts, corresponding to users' inquiries, into databases. Particularly, the authors used the statistics sub-system of web-space content-monitoring system InfoStream, in which the respective functionality was implemented.

When information operation trends are analyzed as time series, it is the series of thematic publication numbers during fixed time period (usually, during one day), associated with a particular operation, that are considered. That is why, for elicitation of trends, we are studying information streams, associated with information operation topics – thematic information streams.

Numerous academic studies are dedicated to research of information operations; in [Corso, 2005], [Kleinberg, 2006], [Ландэ, 2006], [Rakesh, 2014] it is shown that in typical situations the dynamics of news (information precedent) dissemination bears the wavelike character with distinctive periods of impact increase and gradual decline.

As a result of analysis of numerous behavior diagrams of thematic information streams (TIF), the most typical basic behavior profiles were detected [Ландэ, 2012]. Some scenarios develop as follows: after a quick preparatory information surge there is a gradual recession (for instance, in case of reports on natural catastrophes), while others, on the opposite, start with gradual informational preparation, followed by sharp downfall (for example, in case of publications on events planned in advance). Some thematic flows are characterized by a symmetrical dynamics curve that can be narrow (short-time) or broad (long-time).

In case of information streams, associated with specific thematic streams, we should describe dynamics of each of these streams separately, keeping in mind that increase of one of them might automatically lead to decrease of others and vice versa. That is why limitations on information streams across all topics apply to the whole set of information subjects. In case of research of the general information stream we are witnessing the "re-streaming" of publication volumes from subjects whose relevance is decreasing, into other subjects.

Message trends, set forth in [Горбулін, 2009], and representing information operation stages, are shown on Fig. 1.9. In the listed cases analysts should focus on such models, for example, when monitoring allows them to detect the key phases ("background" – "calm" – "preparatory bombardment" – "calm" – "attack"), so that even the first three components allow them to predict future events with high probability.

We should note that such dynamics of topical message numbers during information operation implementation is well described by a famous equation of electromagnetic wave propagation:

$$y = A + Bx\sin(x),$$

where $x$ is time, $A$ and $B$ are empirically defined constants.

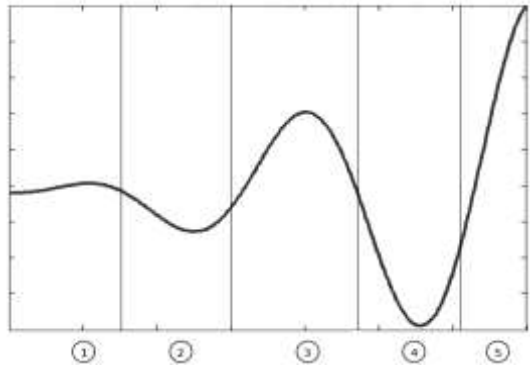

Fig. 1.9 – Dynamics of thematic message numbers during information operation implementation: 1 – background; 2 – calm; 3 – "preparatory shelling"; 4 – calm; 5 – attack/increase trigger

As we know, present-day innovative activity is also implicitly measured by the number of innovation-related publications; there are a few innovation process models, among which we can mention the innovation diffusion model [Bhargava, 1993]. At the same time, innovation implementation can also be considered an information operation. So, let us look at the respective research results. Fig. 1.10 displays a diagram, substantiated in [Хорошевский, 2012], and showing the number of publications, related to innovative activity.

By uniting the graphs, illustrating information operation start (Fig. 1.9) and information activity trend (Fig. 1.10), we can get a complete graph, displaying the representation of information operations in the information space (Fig. 1.11).

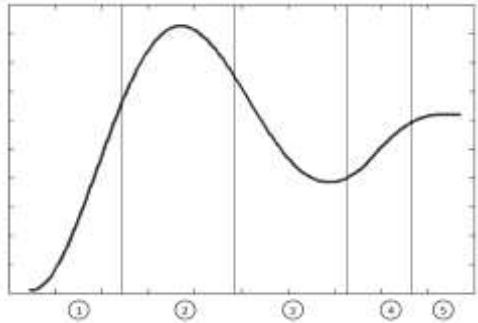

Fig. 1.10 – Diagram of the number of publications, corresponding to innovative activity trend: 1 – attack/growth trigger; 2 – overestimated expectation peak; 3 – loss of illusions; 4 – public realization; 5 – productivity/background

It is assumed that systematic violation of typical dynamics of certain information subjects in the open information space might indicate both information operations [Горбулін, 2009] and existence of an information reservation. During research of information operations much attention is given to information subject analysis [Додонов, 2013].

Suggested models fully correspond to real data, extracted by content-monitoring systems [Додонов, 2009], [Ландэ, 2007]. That is why the listed patterns can be used as templates for information operation detection, both through analysis of retrospective fund of online publications and through online real-time monitoring of some of their indicators. As we know, for information operation detection we should carefully monitor the dynamics of publications on the target topic, and, if possible, use the available analytical tools, as well as digital data processing and image recognition tools, such as wavelet analysis or Kuntchenko's polynomials [Чертов, 2009].

We should note that the suggested model allows us to differentiate information streams, whose behavior is determined by natural patterns of the information space, from streams, induced by influence of external factors.

Particularly, in case of information reservations, the following indicators can be used: deviation of dynamic trends of certain information subjects from characteristic distribution patterns, emergence of periodic value instability zones, or (the opposite) untypical local stability of these values.

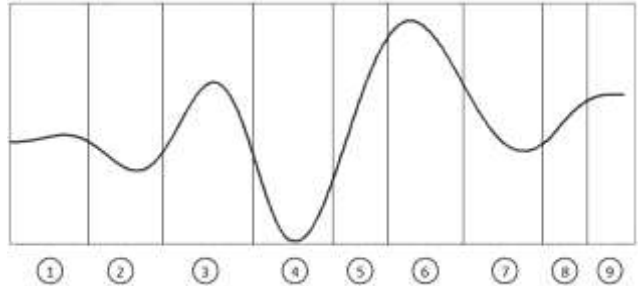

Fig. 1.11 – Generalized diagram, illustrating all the stages of information operation lifecycle: 1 – background; 2 – calm; 3 – "preparatory shelling"; 4 – calm; 5 – attack/growth trigger; 6 – overestimated expectations peak; 7 – loss of illusions; 8 – public realization; 9 – productivity/background

As an example, Fig. 1.12 shows the dynamics of publications in RUNet (thematic information streams) on such inquiries as "Cyprus banks", "Offshore", "Virgin Islands" for March and April 2013, the well-known crisis period. The data was obtained using InfoStream system. As we can see from Fig 1.12, publications concerning Cyprus bank crisis peaked on March 17-18, 2013, while most publications on Virgin Islands appeared on April 4-5, when events, similar to Cyprus situation, only in much smaller scale, started there. We should note the weak correlation of dynamics of information streams related to Cyprus and Virgin Islands. Mutual correlation coefficient of the respective time series amounted only to 0.3. However, we can stress the high mutual correlation level of time series, associated with such topics as "Offshore" and "Cyprus banks" (0.73), as well as "Offshore" and "Virgin islands" (0.77).

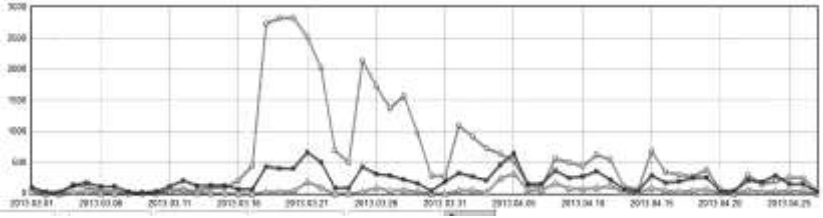

Fig. 1.12 – Diagram of topical information streams related to queries: o – "Cyprus banks"; Δ – "Virgin Islands"; x – "Offshore"

Obviously, manifestations of information operations related to offshore banks, in this case, can be better witnessed during analysis of the more general topic – "Offshore". On the diagram of the respective time series two local extreme zones are clearly visible; they represent crisis situations on Cyprus and Virgin Islands; beside that we can see phases, representing "calms" and "preparatory bombardments".

We can presume that if at a certain point dynamics of a specific information stream starts to deviate significantly from dynamics of the stream, representing the more general subject (in the listed case these are "Cyprus banks" and "offshore"), then it is, possibly, an indication of a newly-launched information operation, related to the narrower topical scope.

During wavelet-analysis [Астафьева , 1996], [Buckheit, 1995] (Fig. 1.13) it was decided to use a "Mexican Sombrero" type wavelet, as its shape was very close to the diagram, shown on Fig. 1.11.

The processes under consideration are clearly visible on both wavelet-spectrograms and their respective skeletons (graphs of extreme lines).

The listed models and methods are applicable for description of general dynamic trends of information processes; however, the forecasting problem remains open. Evidently, more realistic models can be built if an additional set of factors (most of which are not reproduced in time) is taken into account. All the same, the structure of rules, providing the functioning basis for most of available models, allows us to introduce respective corrections, for instance – to artificially model random deviations.

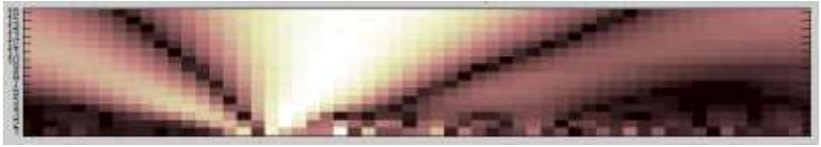

a)

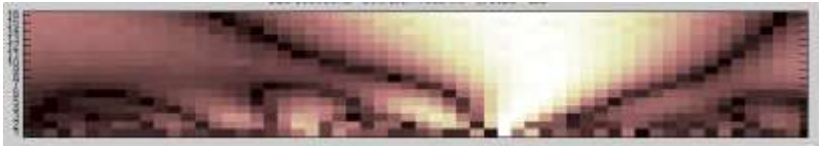

b)

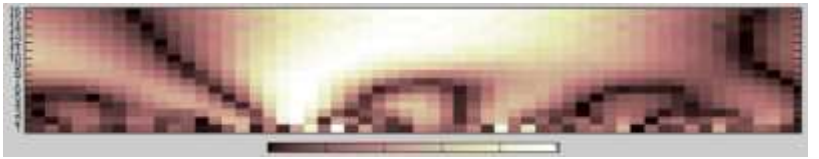

c)

Fig. 1.13 – Wavelet-scalograms, corresponding to dynamics of information streams on queries: a – "Cyprus banks"; b – "Virgin Islands"; c – "Offshore"

Let us now consider another example: the results of express analysis of the thematic information stream, associated with such object as the National academy of sciences of Ukraine, considering the risks, faced by the Academy in the second half of 2015. As a result of analysis, content-monitoring system InfoStream allowed us to formulate a query for the period between 01.07.2015 and 31.12.2015. As a result of the query processing a topical information stream of 1932 documents from Ukrainian web segment was compiled. For detection of information operations, publication dynamics on target subject was analyzed using the available analytical tools (Fig. 1.14).

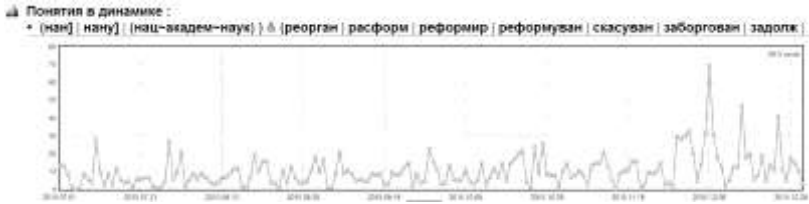

Fig. 1.14 – Topical information stream dynamics

To define the degree of "proximity" of the analyzed time series to information operation diagram in different scales it is suggested to use "wavelet-analysis" that is, nowadays, widely used in both natural sciences and sociology.

Wavelet-coefficients indicate how close the actual process behavior is to the wavelet in the given scale. On wavelet-scalogram (Fig. 1.15) we can see all characteristic features of the initial series: scale and intensity of periodic changes, trend directions and values, location and duration of local phenomena.

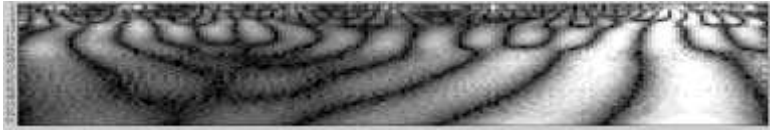

Fig. 1.15 – Wavelet-scalogram (Morlet wavelet) of the analyzed information stream

Once critical points are defined, the content-monitoring system builds basic scenario chains of messages related to the query for selected dates. Thus, the key events for selected dates are detected: *They want to let the National academy of sciences go with the wind… (Vector News) 2015.12.03*

- *Young scientists rally under Rada (Left bank) 2015.12.08*
- *Ukrainian government destroys Ukrainian science (2000.ua) 2015.12.08*
- *"The hands of academia are tied": employees of the NAS of Ukraine rally under Presidential Administration (UNIAN) 2015.12.16*

Conducted analysis, particularly, indicates that a targeted information operation is launched against the NAS of Ukraine, and counteraction to this operation is widely covered in the national information space.

As part of conducted research, a set of information operation templates of different scales was developed for information operations modeling. It corresponds to phases 1-6 of the shape, presented on Fig. 1.11. Examples of templates of different length are shown on Fig. 1.16.

Comparison of the developed template system with the time series associated with the actual topical information stream (Fig. 1.17) in the newly-built model is performed using correlation analysis.

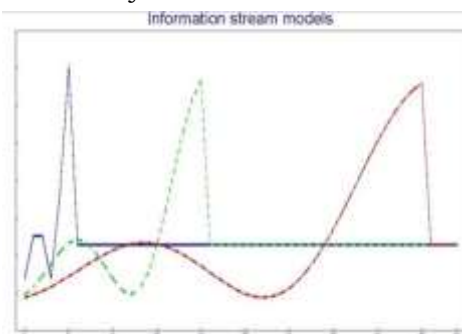

Fig. 1.16 – Template examples with length 5, 20, and 45

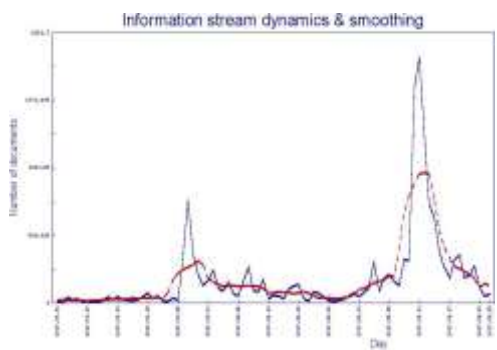

Fig. 1.17 – Time series, associated with the real thematic information stream, and the smoothing curve

The respective correlation diagram, where the horizontal axes represents the time and vertical axes – the scale (template length), while coloring represents correlation coefficients, is shown on Fig. 1.18. The diagram allows us to clearly visualize the phases of information operation implementation, corresponding to certain templates under different observation scales.

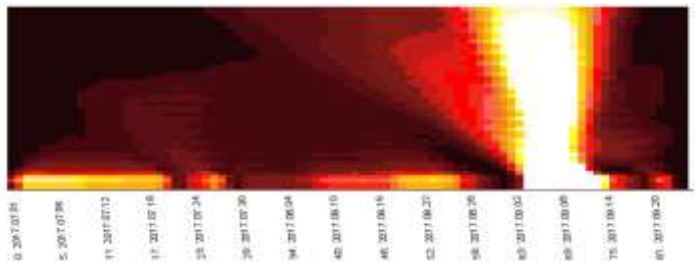

Fig. 1.18 – Correlation diagram, representing the correlation between thematic information stream and the abovementioned template system

We should note that reproduction of the results in time presents a serious problem during information process modeling, and provides the basis for a scientific methodology. Presently, retrospective analysis of the already implemented information operations remains the only relatively credible way of their verification.

A linear combination of linearly independent transforms $f_1(e), f_2(e),..., f_n(e)$ of the respective generating element $e$ can be built as approximation polynomial $P_n$ of order $n$ for the part of output signal $f_s(e)$:

$$P_n = \sum_{\substack{k=0, \\ k \neq s}}^{n} c_k f_k(e),$$

where coefficients $c_k$ are defined based on condition of minimum distance between the polynomial being built and the signal. Element $c_0$ is defined from the expression:

$$c_0 = \frac{\left\langle f_s(e), f_0(e) \right\rangle - \sum\limits_{k=1,\ k \neq s}^{n} c_k \left\langle f_k(e), f_0(e) \right\rangle}{\left\langle f_0(e), f_0(e) \right\rangle},$$

while other coefficients $c_k$ – are calculated as a solution of the linear equation system:

$$\sum_{k=1,\ k \neq s}^{n} c_k F_{i,k} = F_{i,s}, \ \ i = 1,...,n, \ \ i \neq s,$$

where centered correlants $F_{i,k}$ are also calculated through respective transforms:

$$F_{i,k} = \left\langle f_i(e), f_k(e) \right\rangle - \frac{\left\langle f_i(e), f_0(e) \right\rangle \cdot \left\langle f_k(e), f_0(e) \right\rangle}{\left\langle f_0(e), f_0(e) \right\rangle}.$$

Efficiency indicator $d_n$ can be used as a numeric characteristic of a quality criterion during comparison of the signal to a selected template (i.e. as approximation measure for Kuntchenko's polynomial $P_n$ approximating the signal $f_s(e)$):

$$d_n = \frac{\sum\limits_{k=1,\ k \neq s}^{n} c_k \left\langle f_k(e), f_s(e) \right\rangle}{\left\langle f_s(e), f_s(e) \right\rangle}.$$

The described method of recognition of certain patterns through construction of a space with a generating element and search for coefficients of the respective Kuntchenko's polynomial can be used in any problem domain, where characteristic templates can be distinguished within a time series a priori.

Thus, having built typical models for topical publication intensity series behavior during information operation implementation, and having compared templates obtained on their basis, we can use Kuntchenko polynomial method for detection (and prevention) of potential information attacks.

Thematic information stream dynamics is determined by a complex of both internal and external nonlinear mechanisms that should be considered (possibly, implicitly) during modeling. Often, a satisfactory option is to stick to a simplified understanding of a thematic information stream as some time-dependant value, behavior of which is analytically described by nonlinear equations. Today, mostly analytic nonlinear models are used for information stream modeling. The most commonly utilized approaches include nonlinear dynamics, cell automata theory, percolation, self-organized criticality etc [Ландэ, 2009], [Додонов, 2011].

According to [Расторгуев, 2014], it is appropriate to start detection of information threats (and any other threats as well) with analysis of approaches and methods of targeted reprogramming of information systems, such as technical systems, social structures, or individuals. Outlining of a series of information operation-specific events and their further detection within the general "life noise" is a classical way of solving such problems. The approach includes the following steps.

1. A series of events, presenting a threat, is always associated with some goal. Achievement of this goal calls for reprogramming of information combat subjects. In order to understand the extent of goal achievement, it is desirable to be able to evaluate the results of reprogramming and your own capabilities, i.e. – the degree of potential damage that information weapon can bring to this or that system. Destruction and reprogramming of some objects is easy, while in case of other objects the task may be impossible to complete within required resource and time limitations.

2. Estimation and substantiation of event occurrence frequency, under which we can talk about launching of a reprogramming information operation.

3. Providing the user with efficient and easy access to information in the case when dangerous trends are detected.

## 1.5. Information operation counteraction problem

The analyzed practical examples allowed us to work out a certain general methodology for defensive information operation implementation using a web-resource content-monitoring system. Let us assume, a company "ABC" is an object of an aggressive information operation. The 12 steps for counteraction are as follows:

1) Collection of information with publications in "alien" media (unrelated to "ABC", non-affiliated) about the company.

2) Building a diagram of publication of messages on "ABC" company in online media.

3) Analyzing the dynamics with a retrospective period of 6-12 months using time series analysis methods. Publication content in threshold points is analyzed; points, duration and regularity of impacts are defined; impact points are "tied" to other events from the object's area of interest.

4) Defining sources that publish most publications with negative tonality concerning "ABC".

5) Detecting the "sources of origin" of publications in the media, i.e., sources that were the first to publish the negative information.

6) Defining potential "customers" – owners or stakeholders, who influence the publication policy of certain media.

7) Defining the spheres of common interest of "ABC" and potential "customers" that ordered the operation (through detection of common information characteristics – intersections of "information portraits" built by InfoStream system for the object and the "customer"), ranking of potential "customers" according to their interests.

8) Defining information impact criteria based on top-rating interests.

9) Modeling of information impacts. For this end the "customer's" connections are detected (i.e. persons and

organizations that are most closely related to the "customer"), dynamics of "customer's" impact is analyzed and forecast of this dynamics is made; publication content in threshold points of dynamics curve is analyzed – critical impact points are defined.

10) Further impact steps are predicted through analysis of similar publication dynamics for other companies from retrospective database of InfoStream system.

11) With realities and publications from the retrospective database taken into account, potential operation consequences are evaluated.

12) Informational (and other) counteraction is organized. Examples of publications related to counteraction context are stored in the retrospective database.

## 1.6. Document analysis

US armed forces are currently among the leaders in the sphere of information combat theory and practice.

Information operation ideas were considered based on provisions of such documents as US National Security Strategy 2010, The National Military Strategy of The United States of America 2011, "Sustaining US Global Leadership: Priorities for 21st Century Defense".

The essence of global information operations is set forth in the concept of building and use of the US armed forces "Joint Force 2020". In its provisions the following issues are addressed: use of armed forces, disseminated across the globe, creation of unified formations, "tailored" to solve particular problems and capable of acting as one. According to the US military doctrine, presently there are 14 fields of activity in the sphere of information operations (Table 1). During the last few years the following documents were adopted by US Armed Forces:

– the new edition of JP 3-13 doctrine "Information operations" (January 2006);

- Doctrine for Joint Psychological Operations JP 3-53 (September 2003);
- Field Manual 3-05.30 "Psychological operations" (April 2005);
- Field Manual 3-05.301 "Psychological Operations Process Tactics, Techniques, and Procedures" (December 2003, already revised);
- Field Manual 3-05.302 "Tactical Psychological Operations Tactics, Techniques, and Procedures" (October 2005);
- Pocket guide to (handbook of) technical PsyOp means: types, tactical, and technical characteristics and capacities (April 2005);
- "Psychological Operations Leaders Planning Guide" (August 2005), containing excerpts from manuals FM 3-05.301 and FM 3-05.302. Both guides are intended for PsyOp units;
- A whole series of military training programs (ARTEP), including:
  - ARTEP 33-712-MTP, "MTP for Headquarters and Headquarters Company of the PSYOP Group and Headquarters and Support Company of the PSYOP Battalion", (April 2006);
  - ARTEP 33-715-MTP "MTP for the Psychological Operations Dissemination Battalion" (September 2006);
  - ARTEP 33-737-30-MTP "MTP for the Tactical PSYOP Company" (2007);
  - ARTEP 33-727-MTP " MTP for the Regional PSYOP Company" (2007);
- A series of documents, regulating the issues of military training of PSYOP personnel, including:
  - STP 33-37II-OFS "Officer Foundation Standards II Psychological Operations (37A) Officer's Manual" (2006), intended for meddle-rank officers (captain to lieutenant-colonel) and including the key tasks, military training issues and requirements to knowledge levels that should be met by PSYOP officers of the listed categories;

- Soldier's Manual and Trainer's Guide MOS 37F STP 33-37F14-SM-TG Psychological Operations Specialist Skill Levels 1 Through 4 (August 2008).

Active scientific research work in the interests of PSYOP is supervised by military training and academic research department of US armed forces.

Table 1 Fields of activity of US Armed Forces in the sphere of information operations. Source: Joint Publication 3-08. Interorganizational Coordination During Joint Operations. 24 June 2011. p. 324 (D-10, JP 3-08)

| (1) Strategic Communication (SC) |
| --- |
| (2) Joint Interagency Coordination Group |
| (3) Public Affairs |
| (4) Civil-Military Operations |
| (5) Cyberspace Operations |
| (6) Information Assurance |
| (7) Space Operations |
| (8) Military Information Support Operations |
| (9) Intelligence (information operations intelligence integration) |
| (10) Military Deception |
| (11) Operations Security |
| (12) Special Technical Operations (STO). |
| (13) Joint Electromagnetic Spectrum Operations |
| (14) Key Leader Engagement (KLE) |

Serious materials, dedicated to different information and psychological operation related topics, are regularly published in leading military-scientific journals of US

Armed forces: Military Review, Parameters, Special Warfare, and others.

In 2010 it was decided to replace the previously used term "psychological operations" with the term "information operations of military support". The primary reason for the change is that the new terminology reflects, mostly, the activities of Defense Department in the area of informing and influencing enemy, neutral, and friendly public opinion, aimed at achievement of strategic goals of US military management. The published memorandum states that the change of terminology will not significantly influence the currently implemented operations. Military units, supporting psychological operations will continue to build their capacity to "persuade-change-influence" during operations of any kind, at any location and time [Memorandum, 2017].

The term "information operations" is introduced by the specialists of US defense department in field manuals FM 100-23 "Peace operations" and FM 33-5 on psychological warfare.

In the listed instructive documents one of the conceptual approaches to definition of "information operation" concept is the opportunity to use methods and means of this information combat form in times of both military action and peace.

In the Air Force Doctrine Document 2-5 "Information operations" of 2005 information operations are defined as measures, taken to influence the enemy information or information systems and, simultaneously, protect one's own information and information systems.

According to the Joint publication 3-13 "Information operations" of 2006, information operations are defined as integrated usage of the key capabilities of electromagnetic tools, computer networks, psychological operations, military skills, and security operations, accompanied by special support and respective capabilities, intended for influencing, destruction, and damage, seizing of decision-making process (by either humans or technical means), and, at the same time, for protection of one's own facilities.

The same guidance from 2012 includes a more general definition. Information operations are defined as integrated usage (during military operations) of capabilities, related to information, together with other operation means, aimed at destruction, damaging, and seizing of enemy decision-making process, with simultaneous protection one's own. NATO Strategic Communications Policy (2009) defines "information operations" as military recommendations and coordination of NATO information military operations, meant to produce the desired impact upon intensions, understanding (perception), and capabilities of the enemy and other subjects (potential enemy, decision-makers, cultural leaders, international community representatives etc) and support operations, missions, and goals of the Alliance.

In 2010 the Joint Chiefs of Staff of the US Department of Defense published the doctrine of Military Information Support Operations, stating that these operations were a critical element of the US foreign policy. In military conflicts such operations serve as force multiplier, providing an opportunity to decrease the efficiency of enemy's armed forces, limit the intrusion of civilians, reduce the related damage, and increase support of the operation currently in progress by local population [Military, 2011].

On January 14, 2014 the Defense Minister of the Russian Federation signed an order on formation of cyber-headquarters, whose primary task would be to prevent unsanctioned meddling with Russian digital management systems within the General headquarters of the Russian armed forces. On February 22, 2017, during his speech in Gosduma (the State council), he confirmed that information operation units were formed within Russian armed forces.

In Ukraine information operations are still not defined by official doctrines and concepts, however, their potential implementation is based on a series of fundamental documents:

The Law of Ukraine "On Intelligence Agencies of Ukraine" of March 22, 2001 no.2331-III lists the following key tasks of intelligence agencies: "implementation of special activities, intended for support of national interests and state policy of Ukraine in economical, political, military, military-technical, environmental, and *information spheres*…".

In addition, Article 3 of the Law of Ukraine "On Foreign Intelligence Service of Ukraine" of 01.12.2005 no. 3160-IV specifically provides the tasks of the Foreign Intelligence Service of Ukraine, including those concerning "implementation of special impact measures for support of national interests and state policy of Ukraine in economical, political, military-technical, environmental, and *information spheres*…".

Laws of Ukraine "On Security Service of Ukraine" of 25.03.1992 no. 2229-XII and "On the Fight Against Terrorism" of 20.03.2003 no. 638-IV place a whole set of tasks upon the Security Service of Ukraine, concerning *information security of the state*, particularly, during antiterrorist operation implementation.

In the Law of Ukraine "About Armed Forces of Ukraine" (as a result of adoption of the Law of Ukraine "On amendments to some laws of Ukraine concerning the Special Operations Forces of the Armed Forces of Ukraine" of 07.06.2016 no 1437-VIII) parts four and five of Article 1 are edited as follows: "troops, units, and divisions of Armed Forces of Ukraine, can be involved in implementation of measures of legal regimes of martial law and emergency, organization and support of resistance movement, implementation of *information-psychological operations*, as provided by Law…".

# 2. OSINT – open source intelligence

## 2.1. OSINT as intelligence domain

Open source intelligence (OSINT) represents one of the key instruments of information operation implementation. OSINT is one of the intelligence domains, including search, selection, and collection of intelligence information, available from publicly available sources, as well as analysis of this information.

OSINT concept is based on two key terms:

- Open source is an information source that provides it without requirement of confidentiality, i.e. provides information that is not protected from public disclosure. Open sources are associated with publicly available information environment, and access of physical persons to them is not restricted in any way;
- Publicly available information is information that is published or posted for wide usage, i.e. available to public.

According to the CIA analyst Sherman Kent (1947), politicians get up to 80 per cent of information, needed for decision-making in times of peace, from open sources. Later lieutenant-general Samuel Wilson, head of Intelligence department of US Department of defense in 1976—1977, noted that 90 per cent of intelligence data came from open sources, and only 10 – from agents.

OSINT is, usually, performed through monitoring, analysis, and research of information coming from the Internet. Materials, compiled based on information from open sources, support all intelligence methods and activities through accumulation of intelligence knowledge, its analysis and dissemination.

According to [ATP, 2012], OSINT is also one of the ways of intelligence that significantly contributes to planning of military actions, and provides all the necessary information for these actions. It is also noted that:

1) OSINT is one of the methods of intelligence through gathering of information from open sources, its analysis, preparation, and timely submission of the final product to higher management for solution of certain intelligence problems.

2) OSINT is an intelligence method, developed based on collection and analysis of publicly available information, and not subject to direct supervision by US government. OSINT is a result of systematized collection, processing, and analysis of the necessary publicly available information.

American security researcher Marc M. Lowenthal defines open information as any information that can be obtained from open collections: all media types, governmental reports and other documents, academic research and reports, commercial information providers, Internet, etc. The key characteristic of open information is that its procurement does not require illegal collection methods, and that it can be obtained by means, which are fully compliant with copyright laws and commercial conditions of the providers.

International community is using more and more information from open sources to solve a wide spectrum of problems. Particularly, the role of OSINT during implementation of information operations is defined by a set of aspects, including efficiency of information flow, volume, clarity, ease of subsequent usage, cost of obtaining, etc. The following factors influence the process of planning and preparation of OSINT measures:

- Effective information support. Most of the necessary reference materials on information operation objects are gathered from open sources. This base is built through collection of information from the media. Accumulation of data from open sources is the key function of OSINT.
- Relevance. Availability, depth, and scale of publicly available information allow us to find the necessary information without engaging specialized human and technical intelligence means.

- Simplification of data collection processes. OSINT provides the necessary information eliminating the need for engaging redundant technical and human intelligence methods.
- Depth of data analysis. Being a part of intelligence process, OSINT allows managers to perform in-depth analysis of publicly available information in order to make respective decisions.
- Efficiency. Sharp reduction of time of access to information in the Internet. Reduction of the number of man-hours, spent on search for information, people, and their interrelations based on open sources. Quick obtaining of valuable relevant information. Abrupt situation changes during crises are most thoroughly reflected by current news, so (as we know for sure), the downfall of the Berlin wall was witnessed in both Washington and the CIA headquarters in Langley, not through intelligence service reports, but through TV screens, broadcasting CNN reports right from the scene.
- Volume. Opportunity for mass monitoring of certain information sources, intended for search of the needed content, people, and events. Experience shows that proficiently collected information fragments from open sources, when taken as a whole, can prove equivalent to or even more significant than professional intelligence reports.
- Quality. In comparison with reports of special agents, information from open sources turns out to be more preferable, at least because it is unbiased and not mixed with lies.
- Clarity. So, while in the cases when OSINT is used, trustworthiness of open sources can be both clear and unclear, in the case of secretly obtained data, their credibility is always doubtful.
- Usability. Any secrets are supposed to be protected by barriers of "classifications", clearances, restricted access etc. As for OSINT data, it can be easily communicated to any interested organizational bodies.

There is an opportunity to conduct complex research based on data from the Internet.

- Cost. Cost of obtaining data through OSINT is minimum; it is defined only by the price of the service used.

Particularly, software and technological solutions offered for OSINT today, allow for:

- Data collection from social networks, such as Facebook, Twitter or Youtube, analysis of collected data;
- Getting to the core of events based on gathered content;
- Aggregation of information, obtained from the Internet;
- Information influence in the Web;
- Information credibility assessment;
- Monitoring and recognition of identity in the Web, particularly, based on geo-location;
- Working with information, obtained from the unseen web-space (dark web, hidden web, deep web).

## 2.1.1 OSINT application spheres

There are plenty of OSINT applications, including the following ones.

### *Intelligence*

Open sources contain vast amounts of information, needed by intelligence bodies and compliant with their requirements, providing understanding of objective and subjective factors, related, for example, to information operation implementation. At the same time, there is no doubt, that in order to increase the efficiency of intelligence activities, the respective bodies use open information in combination with resources, received from agent networks.

The initiative of the United States Intelligence Community, known as the National Open Source Enterprise, is expressed in the Directive of the Intelligence Community no. 301, published by the Director of National Intelligence [DNI, 2006]. The Directive provides authority and responsibilities of the Deputy Director of National Intelligence for open sources (ADDNI / OS), the Open

Sources Center of DNI and the National Open Sources Enterprise.

## *OSINT in armed forces*

Units of US armed forces that participate in OSINT activities are listed below:

- Unified Combatant Command
- Defense Intelligence Agency
- National Geospatial-Intelligence Agency
- US Army Foreign Military Studies Office
- EUCOM JAC Molesworth
- Foreign Media Monitoring in Support of Information Operations, U.S. Strategic Command

## *National security*

Department of Homeland Security includes an active intelligence unit for working with open sources. On February 14, 2007, «Domestic Open Source Enterprise» was established for support of OSINT department and working with state, local, and tribe partners.

## *Justice*

Community of OSINT authorities uses open source intelligence to forecast, prevent, and investigate crimes, as well as to persecute suspects, including terrorists. Besides, fusion centers around the US are using OSINT more and more often to support their intelligence efforts and investigations.

Examples of successful OSINT authorities include Scotland Yard OSINT; Royal Canadian Mounted Police (RCMP) OSINT.

New-York Police Department (NYPD) includes the OSINT unit, as does Los-Angeles County Sheriff's Department: it is located within Emergency operations Bureau and connected to regional intelligence center of LA.

Within law enforcement activities OSINT can be used to prevent such violations as:

- Organized crime and gangs
- Pedophilia
- Personal data theft and extortion

- Money laundering
- Crimes related to violation of intellectual property rights
- Creation and development of extremist organizations

At the same time, OSINT is used to detect involvement and rising impact in the Web:

- Identification of key figures and activists
- Monitoring of enemy network in real time
- Restriction of information dissemination
- Public opinion formation
- Identification and monitoring of extremist organizations
- Risks for public transport
- Sanctions and legal requirements
- Analysis of enemy databases (HME, IED, TTPs)
- Target geo-location
- Support of military operations

## *Cybernetic security*

As part of OSINT, cyber-security processes are supported. Particularly, based on OSINT information, answers to the following questions from telecommunication network protection domain can be obtained:

- Who is attacking your organization?
- What are their motives?
- How are they organized?
- Which instruments are they using?

## *Business*

OSINT in business sphere includes commercial (business) intelligence, intellectual intelligence, and business analytics, and often represents the key activity domain of private intelligence agencies.

Enterprises can use the services of information brokers and private detectives for collection and analysis of the respective information for business purposes; these may include mass media, deep web, web-2.0, and business content.

## 2.1.2. International experience

Open source intelligence enhances the efficiency of the whole US intelligence community, from national to tactical levels. Below we list some organizations in the US, that collect, purchase, utilize, analyze, and disseminate information from open sources.

- Defense Open Source Council (DOSC);
- Intelligence and Security Command (INSCOM);
- Department of the Army Intelligence Information Service (DA IIS).
- Director of National Intelligence of Open Source Center (DNI OSC).
- Open Source Academy;
- Advanced Systems Department (ASD).
- FBI.
- Federal Research Department (FRD), Congress Library.

In addition to broad OSINT application in the USA, let us provide examples of the technology usage in other countries.

German external intelligence service, Federal intelligence service, also utilizes the advantages of Open Source Intelligence in the units of Abteilung Gesamtlage/FIZ and Unterstützende Fachdienste (GU).

In Australia the primary expert body for open sources is the Office of National Assessments that is one of intelligence structures. In the UK there is the BBC Monitoring information service, focusing on collection of open-access information with the efforts of journalists. The task of analysis of data, collected by the BBC, is placed upon the subscribers, including employees of British secret services.

## 2.2. Information space monitoring

Contemporary content-monitoring methods represent the adaption of Text Mining analysis concept and classical content-analysis methods to conditions of formation and development of dynamic information sets, such as information streams in the Internet. A typical content-monitoring task is to build diagrams of dynamics of references to some terms (reflection of events) in real time. Let us take a look at how content-monitoring system InfoStream [Григорьев, 2007] keeps track of publications, related to spreading of computer virus Petya in the middle of 2017. A special query was formulated as *«Virus&Petya»*, and input into the system's web-interface (Fig. 2.1).

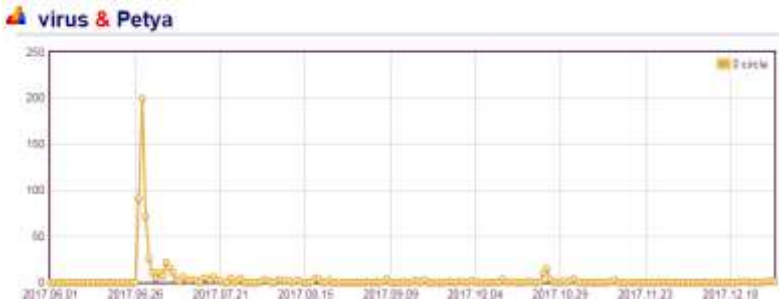

Fig. 2.1 – Diagram of concept dynamics in time

The diagram indicates that the crisis peak fell on June of 2017, while another surge, presumably, caused by the virus clone was witnessed in late October.

Let us consider the example of cyber-security issues to show, how to detect documents featuring maximum amounts of materials on computer viruses, within text information arrays from the Internet.

In order to get the list of key subjects, related to oil product market, we input the query **«Virus & Cyber»**, that was marked by a special indicator **lang.EN & country.US**, denoting search for Russian-language documents in the Ukrainian Web segment by InfoStream system (Fig. 2.2). After that, it was enough to switch to "lookup" mode and analyze the documents, references to which were returned by the system (Fig. 2.3).

Besides, we can switch to "Subjects" mode that envisions clustering of search results with weight coefficients taken into account, which allows the system to show only the most relevant document chains to the user. That is why the system ensures sufficiently high level of correspondence between found documents and the search task, expressed by the query.

Fig. 2.2 – Basic subject chain according to the query

Fig. 2.3 – A fragment of basic subject chain

## 2.3. Information sources

Open information sources include such traditional media as newspapers, journals, radio, TV, public governmental reports, professional and academic reports, official data on budgets, demographic data, press-conference materials, various public statements, observation results –radio-monitoring, usage of publicly available data of remote Earth sensing and aero-photo-shootings (such as, Google Earth), conferences, papers, articles. Besides, open sources also include modern Internet-resources, particularly, web-communities and content, created by users (end user generated content) — social networks, video-hostings, wiki-guides, blogs, web-forums.

During OSINT-based research access to enormous information volumes is provided. Correct target planning and understanding of where to look for the necessary information is an important efficiency factor of experts' (analysts') work.

Two key types of information sources are defined: primary and complementary.

A primary source is a document or a physical object, containing information that was written or created as a result of research and analysis. These sources are immediate "witnesses" of one event or another; they contain "insider information". Usually, this information is fragmented, ambiguous, and difficult to analyze. Primary sources are mentioned as references during subsequent analysis. These sources include:

- Original documents, extracts from them, translations, academic journals, speeches, letters, interviews, news and video footage, official reports etc;
- Creative works: poetry, drama, novels, music, etc;
- Artifacts: ceramics, furniture, clothes, historical knowledge etc;
- Stories and memories of people.

Complimentary or secondary sources include governmental press-services, commercial news agencies, information from press-secretaries and non-governmental organizations.

In the process of intelligence in open sources we should keep in mind the presence of disinformation.

Employees of units, performing intelligence in open sources, usually, do not get information through direct observation. They mostly rely on secondary sources that can intentionally or unintentionally add, delete, change, or filter in some other way the information that is available to broad public. It is important to know the history of open sources and define the purpose of information publication, detect bias, attempts to influence people.

Internet represents a dynamic information environment, constituted by static and moving virtual websites, containing mixture of old and new information. There are many virtual databases that grow according to the geometric progression law.

## 2.3.1. The web-space

Web-resources are indexed by standard information search systems that are, usually, unable to get access to or index all the information, needed for effective open source intelligence. It is assumed that during open source intelligence we can see only a part of the total information volumes stored in the Internet. Three quarters of information are stored in the so-called Deep Web.

Search systems are the key tools, used by OSINT personnel during intelligence and collection of publicly available information from open sources. OSINT personnel actively uses search engines and different search criteria to find text, images, information on pages of thousands of web-sites. Technically, search engines conduct their search by website index. Commercial and governmental search engines differ from each other, depending on search parameters, performance of search process, and ways of search result representation. Most search engines use the so-called spider programs that scan web-resources and form indexed databases of search systems.

Sometimes, as a result of search, you can find a link to a document; however when you try to open it, access error is returned. In such cases OSINT personnel can find the needed content on such websites as www.archive.org.

Web-space, based on physical infrastructure of the Internet and data transfer protocol HTTP, unites hundreds of millions of web-servers, connected to the Internet (Fig. 2.4). In the beginning of the web-space data from several authors was published for a large number of visitors on a small number of websites. Today the situation has drastically changed; transition to second-generation web has taken place. Website visitors themselves are actively participating in content creation, which results in sharp increase of information volumes and web dynamics.

Today a huge freely-accessible information base exists in the web. Its volume is larger than anything one could imagine. Moreover, the scales of this base exceed everything that was available several decades ago by several orders of magnitude. In August 2005 Yahoo! announced that it had indexed about 20 billion documents. Google's achievement in 2004 amounted to something below 10 billion documents. And now Google has indexed more than a trillion online documents. According to Netcraft Web Server Survey (news.netcraft.com), presently, the number of web-servers exceeds a billion (Fig. 2.4).

Open sources and specialized data bases available in the web-space contain the "best part" of information needed for analytic research; however, the questions of finding and effectively using it are still open to discussion. As it has been already noted, during usage of web as a powerful information source, the main problems concern information volumes, navigation, information noise, and dynamic character of information in the Internet.

At first glance, potential of access to web-resources that attract users with their openness, volumes, and multi-dimensional content, seems endless. However, important events taking place in different domains indicate the opposite. It is in crisis situations that Internet seems to fail us most often. Problems are plentiful – from network structure overload to virus attacks, vulnerability, and service termination of separate web-servers. A whole bunch of issues is associated with volumes, versatility of representation, and dynamics of content segment of information space.

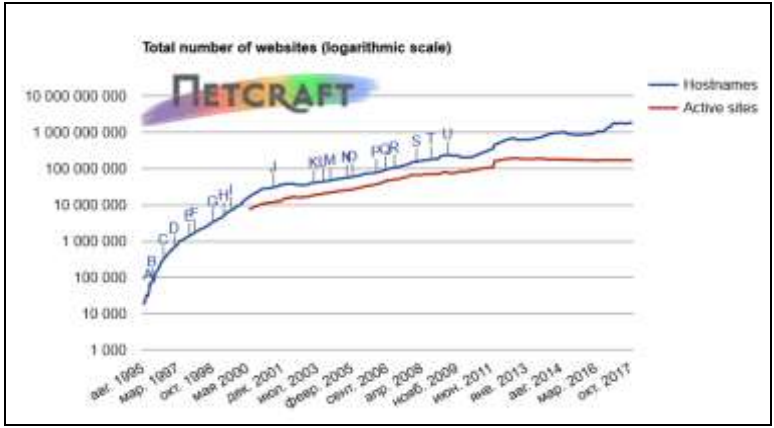

Fig. 2.4 Growth dynamics of the number of web-servers in the logarithmic scale (Netcraft, October 2017)

In spite of such qualities as openness and availability, existing web-space infrastructure cannot be considered reliable and credible. Let us enumerate a few more problems, inherent to web-space:

− The problem of providing users with access to heterogeneous web-resources from a "single frame" to let them get a general representation of information flows within the required topical scope is still unsolved;

− The opportunity for timely "reminding" and "promotion" of information, relevant for user, is still not provided by a large number of websites;

− The probability of service denial by crucial web-resources at the time when they are most needed is quite high.

We know that today there are content integration technologies allowing to partially solve the listed problems and ensure effective search and navigation in the web, as well as monitoring and aggregation of open web-resources.

For professional online search and information monitoring specialized software, as well as information search systems and services are used. Let us give some examples of the respective software products:

***Copernic Agent*** (www.copernic.com/en/products/ agent/) is a software program allowing its users to conduct meta-search (as it is stated on the company's website) involving 1000 search engines; unite results and eliminate duplications, block non-functional links, and display the most relevant results (Fig. 2.5).

***Avalanche*** (www.tora-centre.ru) is a family of web-monitoring software tools. Avalanche technology is based on three key solutions: Smart Folders concept, autonomous intelligent search robot, and built-in data base ("personal encyclopedia").

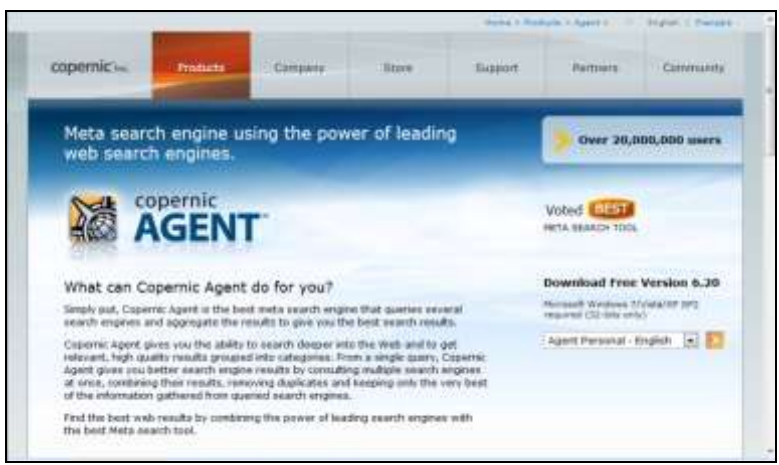

Fig. 2.5 – A fragment of Copernic Agent software website

***Newprosoft Web Content Extractor*** (www.newprosoft.com) is a software program for scanning and extracting data from websites.

***Portable Offline Browser*** from MetaProducts Corporation (www.portableofflinebrowser.com) is a software program allowing users to download the necessary website content and multimedia information, including Flash-animation, scripts, and active page content.

***Neiron Search Tools*** (neiron.ru/toolbar) is a software add-on uniting the information search results of two systems – Google and Яndex, and allowing users to perform competitive analysis based on evaluation of sites and context advertising.

***WebSite-Watcher*** (www.aignes.com) a software program allowing users to perform monitoring of websites, forums, local files, and providing filtering of information as well as convenient visualization of monitoring results.

Service solutions include the following ones.

***WatchThatPage*** (watchthatpage.com) is a free service allowing users to automatically gather new information from web-resources under monitoring.

***Diphur Monitor EveryThing*** (www.diphur.com) is a free service for monitoring of any web-sites, informing users about their updates and providing these updates to the users.

***Newspaper Map*** (newspapermap.com) is a service, combining geo-location and information search system by media-resources. While solving competitive intelligence problems, a user can select the target region, language, and list of online versions of newspapers and journals, and directly access the documents. The service supports Russian language and has a convenient interface.

***InfoStream*** (www.infostream.ua) is a service for content-monitoring of web-resources in Russia and Ukraine, providing access (in search mode) to information from 6000 sources, performing information classification, extraction of terms (persons, companies, toponyms), formation of event chains, evaluation of message tonality, and analysis of dynamics of publications on certain objects.

***Agregator.pro*** (agregator.pro) is an aggregator of information from news and media portals. It can be used for competitive intelligence to track objects of interest, get frequencies and contexts of references to the tracked object in the media, as well as analyze dynamics of these references in time.

**WebGround** (webground.su) is an aggregator of news information from Russian-language web segment. It can be used in competitive intelligence to track the themes of interest, get thematic events, and perform retrospective analysis of development of a particular topic in time (Fig. 2.6).

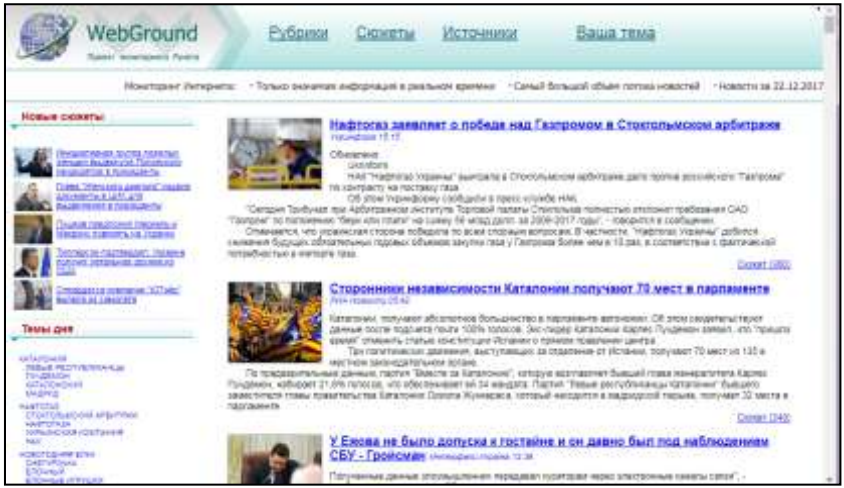

Fig. 2.6 A fragment of WebGround news aggregator

## 2.3.2. Deep web

Latest web-space research indicates that more than a trillion web-pages available through information search systems represent just the "iceberg top seen over the surface".

An important problem is information search in "hidden" or "deep" web space that, as mentioned above, contains much more data, that could be, potentially, useful for competitive intelligence, than open part of the Internet.

These data include, first and foremost, dynamic web-pages, information from numerous databases that can be of significant interest for analytical work. "Hidden" web also includes full-text information systems, such as LexisNexis or Factiva.

"Hidden" Internet resources also include peering networks, such as BitTorrent, EDonkey, EMule, Gnutella, Kazaa.

As it has been mentioned, the information volume (including information for competitive intelligence) available in the Web is far larger than the volume covered by search engines.

It is assumed that, in contrast to "knowledgeable" part of the Internet, its "hidden" part turns out to be much more voluminous.

A business analyst often faces the situation when he knows about the existence of a certain document in the web-space, but is unable to find it using traditional search systems, such as Google, Yahoo!, Bing, Yandex, Rambler or Meta. However, once he recalls the document's URL or finds it in the thumbnails he can effortlessly get to it. That is, the document is present in the web-space, but it cannot be found in a traditional way. The user trips upon some web resource, that is *invisible* for search systems.

Deep Web is a part of web-space and, according to some estimates, contains twice as much information as the remaining resources. In contrast to traditional web resources, Deep Web is not indexed by traditional search systems. Web-sites and data-bases, located in it, are not accessible when search is conducted through traditional search engines. Deep Web offers plenty of resources of the following content:

- Dynamic. Web pages, appearing when search is conducted;
- Response to the input query;
- Unrelated. Web-pages, unrelated to other pages and blocking access to their content when standard information search software is used;
- Hidden. Web-pages without reverse and inbound links;
- Limited access. Web-sites, that limit access to information;
- Not containing hyper-text documents;

- Text content, coded in the format of multimedia files that are not found by search engines.

So, the set of sources in the web-space, that are not accessible to traditional search engine users, forms the so-called "deep web", the concept, introduced by Jill Ellsworth in 1994. The deep web denotes the part of web-space that is not indexed by robots (web crawlers) of search systems. The term is based on the analogy: it is considered that the information unavailable for search "lies deep". At the same time, we shouldn't confuse deep web with resources, totally unavailable in the Internet – the so-called "dark web", that we are not going to address here. Some resources, accessible only to registered users, also fall within deep web category.

In 2000 an American company BrightPlanet (www.brightplanet.com) published a sensational report, according to which, web space included hundreds of times more pages than that time's most popular search systems could index. The company developed LexiBot software, allowing it to scan some dynamic web-pages, that were formed from data bases, and, obtained unexpected results using the software. It turned out that deep web included 500 times more documents than search engines could find. Of course, the numbers are not very accurate. Besides, it is known that an average page from deep web was 27% more compact than an average page from visible part of web space.

Today the situation has changed; for example, leading search engines are now capable of indexing documents, represented in formats, containing text. Naturally, these include pdf, rtf, and doc. In 2006 Google patented the deep web search method: «Searching through content which is accessible through web-based forms» (Fig. 2.7). According to different authors, visible web covers only 20-30% of web-space.

Fig. 2.7 – A fragment of WIPO web-resource, describing
Google's patent of deep web search technology

## *Pre-conditions of emergence*

Deep web contains resources not connected to other resources by hyper-links – for instance, pages, dynamically created as a result of queries to data bases, documents from data bases, available to users through search web-forms (not through hyper-links). Such documents remain inaccessible to a web crawler that is unable to input correct values into search form fields in real time (i.e. to formulate queries to data bases).

According to Price [Price, 2001], most pages from invisible Internet can, technically, be indexed but they are not, because search engines decide not to index them... Most "invisible" web-sites have high-quality content. These resources just cannot be retrieved by general-purpose search engines...

*… Some websites use database technologies, and that is, indeed, complicated for a search engine. Other sites, however, use a combination of files, containing text and multimedia, and that is why some of them can be indexed while others cannot.*

*… Some sites can be indexed by search engines, but they are not, because search engines consider it unpractical – for instance, due to high costs or because the data is so short-lived that it makes no sense to index it (for example, weather forecast, exact time of specific flight arrival to the airport, etc).*

Key limitations related to search engines' web crawlers can be explained by the following basic reasons: public search services consider it more important to ensure accuracy of search rather than thoroughness; to ensure provision of an answer to a query within acceptable time rather than exact matching of this answer to the query. This results in limitations on the depth of web-resource scanning, attempts to "filter" the content, "screening" of pages, containing redundant outbound hyper-links etc. In such situations the "baby might be thrown out with bath water". It is a common knowledge that the value of deep web resources is often higher than the value of resources from visible part of the web space.

We can mention another source of deep web content. Some owners deliberately do not want their web-resources to be retrieved by search systems. Most often, such web-resources represent something "semi-legal": hacker forums, unauthorized content archives etc. Many of such resources are, naturally, of high interest to business analysts.

Many companies start with connecting to the general Web, and only then spend significant amounts on security. Web-site owners may forbid indexing of certain pages of their resources by inserting the respective command into robots.txt, but search systems might as well ignore it. That is why either such resources themselves are deleted or the hyper-links are deleted and the resources are moved to the deep web. For instance, recently owners of business-catalogues Auto.ru and Drom.ru refused to share their ads with Yandex, i.e. the companies transferred their resources to deep web in order to protect their assets.

## *Types of deep web resources*

There are several types of deep web resources. For example, as it has been mentioned above, these can include *web-pages that become outdated* within a short time interval. *Besides, deep web includes resources, representing multimedia information. A*s we know, at present there are still no efficient algorithms for searching non-textual information. *Pages, generated dynamically in response to some query, also often fall within deep web category. Often such pages do not even exist until the respective query to respective databases is formulated.* So, the information is, technically, present in the web, but it becomes available only at the moment when the respective query is processed; however there is no universal algorithm for completion of search form by robots. Finally, if *there are no links, referring to a web-resource,* then there is no way that web-crawlers can learn about its existence.

Founder of Bright Planet company, Michael Bergman, managed to outline 12 types of deep web resources, belonging to online database category. The list includes both traditional databases (patents, medicine, and finance) and public resources – job offers, chats, libraries, reference books. Specialized search systems, servicing certain industries or markets, whose databases are not included into global catalogues of search services, were also classified by Bergman as deep web resources.

Deep web also includes multiple systems for interaction with users – support, consultancy, e-learning – that require human participation to form dynamic responses from servers. Another category can be described as closed (fully or partially) information, available only to users, accessing the network only from certain IP-addresses, groups of addresses, sometimes, cities or geographies. "Hidden" part of the web also features web-pages, registered on free servers that are indexed only partially (at best), as search systems do not try to crawl through them thoroughly in order to avoid ad-induced spam.

Finally, deep web also includes the so-called "grey" websites that function based on Dynamic Content Management Systems. Search systems usually limit the depth of indexing of such sites in order to avoid cyclic lookups of one and the same page.

## *Examples of deep web resources*

So, how do we find web-resources located in the deep web? If these resources require the user to complete special forms, appended, for example, by captures, it is necessary to access the database that, presumably, includes the necessary documents. Databases being the sources of deep web can be found using ordinary search systems through generalization of queries and input of follow-up clarifying words, such as "database", "data bank", etc.

Let us address a common example: a user needs to access statistics of airline crashes in Argentina. A natural query to a traditional search system returns a long list of newspaper headers. However, if you input the query «aviation database», you can instantaneously access the NTSB Aviation Accident Database (www.ntsb.gov/ntsb/query.asp).

Presently, certain specialized resources exist for search in deep web, particularly in its segment, constituted by databases. The leader among deep web navigators is CompletePlanet site (www.completeplanet.com) by BrightPlanet company. The site represents the largest catalogue of more than 100 000 links. BrightPlanet also developed a personal utility for search in online databases, LexiBot, that can perform search in several thousands of deep web search systems. A meta-search package DeepQueryManager (DQM) by the same company performs search across more than 70 thousand "hidden" web-resources.

Research conducted back in 2006 [He, 2007] indicated that deep web covered more than 300 thousand sites, associated with more than 450 thousand databases, not covered by traditional search systems. Resources representing the greatest interest for business-analysts include: databases of legal and physical persons; industry-specific databases; reputation databases (blacklists and white lists); criminological databases; databases of products and services; product catalogues etc. World-famous business resources, located in deep web include amazon.com, ebay.com, realtor.com, cars.com, imdb.com.

Let us provide a few more examples of databases and catalogues from deep web:

***FindLaw*** (www.findlaw.com) – one of the most popular legal-scope websites; a huge catalogue of law-related resources, including the annotated list of free-access databases of normative legal documents for which the resource serves as the point-of-entry (Fig. 2.8).

***About.com*** (www.about.com) – a portal, covering thousands of links to web-resources (complete with comments), including deep web resources (bearing «Invisible Web» flag). The portal provides the opportunity for search within the catalogue. The resource also features several articles on deep web problems, such as «What is the Invisible Web?», «Finding the Invisible Web», «Top Places to Search the Invisible Web», and others.

***Politicalinformation.com*** (www.politicalinformation.com) – a service providing operative search across 5000 of selected web-sites of political content and retrieving news from several dozens of credible sources.

***Infomine*** (infomine.ucr.edu) – this service retrieves information from databases, online journals (blogs), online bulletin boards, e-books, mailing lists, online catalogues etc., mostly of educational nature. It performs both general search and search according to topical categories.

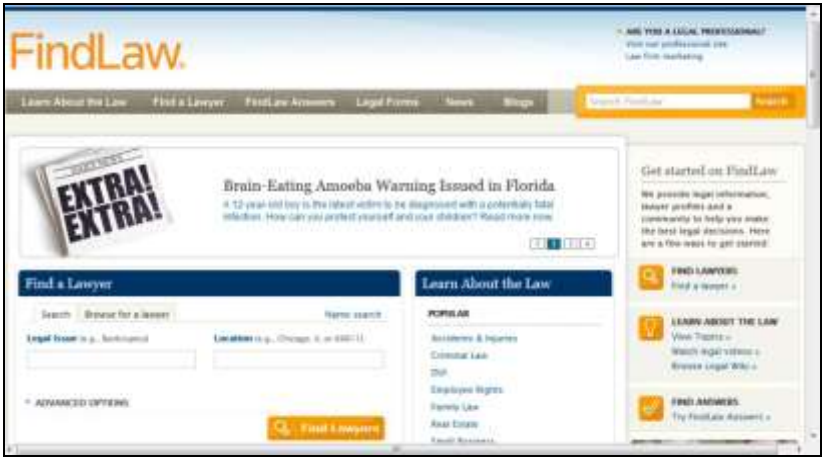

Fig. 2.8 – FindLaw service fragment

A peculiar feature of most of "hidden" resources is their narrow specialization. Mechanisms used for search in "hidden" and "surface" resources are the same, however, in most cases, search system robots for deep web include unique (for each resource) data access units.

A traditional search system, usually, can retrieve database address, however, it will not tell, which particular documents are contained in this database. Information and search systems for Ukrainian (zakon.rada.gov.ua) or Russian (www.kodeks.ru) legislation provide a typical example. Thousands of documents from these databases become available only after logging into the system, and standard search system robots are unable to index the content of these bases.

Paradoxically, archive of open web space resources can be considered as one of deep web resources. Such an archive – Internet Archive – was being compiled since 1996 by Alexa company (www.archive.org). Today, Alexa database volume exceeds 350 billion web-pages (Fig. 2.9).

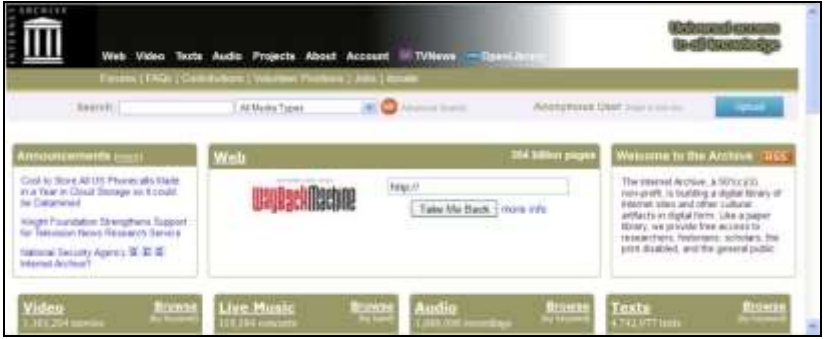

Fig. 2.9 – Title page of www.archive.org web-site

Alexa storage technology features a set of modern tools for management of a gigantic storage of documents. For example, Alexa technology performs clustering of web-resources, i.e. forms collections of documents on related themes. Particular interest of Alexa service for users is induced by Wayback Machine, opening access to different time projections of web space. One of the most interesting practical applications of the technology is restoration of documents that were published online at a certain moment, but deleted later. Growth of deep web threatens with serious storage problems related to increasing number of sites, exploiting different content management technologies, dynamic publication of documents from databases etc.

## Services for working with deep web

Traditional search systems try to narrow the deep web space, gradually occupying such niches as blogs, academic sites, and information agencies. For example, we can recommend the following complimentary deep web search services by Google: Google Book Search (books.google.com) – for book search, Google Scholar (scholar.google.com) – for academic publication search, Google Code Search (code.google.com) – for software code search.

Goldfire Research system by Invention Machine Corp. (inventionmachine.com) allows the user to process content, located on more than 2000 sites of governmental, academic, research, an business organizations in the USA.

Goldfire Research system has information on the mechanisms of access to databases of deep web and automatically generates queries to them.

Academic search system Infovell of University of California, Berkley (www.infovell.com) allows users to search deep web based on "key phrases", from passages to complete documents, or even sets of documents with a total volume of up to 25 thousand words. Infovell system does not depend on the language; users can search for pages in English, Arabic, Chinese, or even input mathematical equations and chemical formulas into search line.

Russian company «Р-Техно» (R-Techno) created the «it2b.интернетошпионопаук 3000+» (it2b.internetspyderspy 3000+), intended for downloading data from invisible Web segment. A search service Web Insight (www.r-techno.com/rtechno/online-services/webinsight) was built based on this system. It performs search across official web-sites and databases in Russia and near abroad, particularly, among documents of the Federal tax service (FTS), Federal court bailiff service (FCBS), Pension fund, Federal anti-monopoly service (FAS), Labor inspection, Federal registry service (FRS), Ministry of Russian Federation for civil defense, emergency situations, and liquidation of natural catastrophe consequences (MES), Arbitrary court, Ministry of internal affairs (MIA), Federal security service (FSS). Other well-known databases of "R-techno" include "Interpol search"; "US companies, accused for manipulations"; "Unethical FAS dealers"; "Debtors of metallurgic industry" etc.

Existing means of analysis and promotion of web-resources allow us to develop new approach to evaluation of ratio between scales of visible and deep web. For example, www.cy-pr.com site provides information on the actual number of documents on the web-site under consideration, presented in RUNet, and on the number of documents, indexed by different search systems, including Google and Yandex. Having received a representative set across web-sites, for example, according to Rambler's

top100 (top100.rambler.ru), one can get an estimate of the ratio between visible and deep parts of RUNet segment of web space.

According to calculations, volume of information located in the deep web segment, is approximately 3-5 times larger than the information volume in the visible part. It turns out that (with few exceptions) the larger the resource is the larger part of this resource belongs to deep web. From this perspective, smaller web-resources are more accessible. As the best part of news documents is located in the deep web, business-analytics tasks require special services for access to such information. Such service is provided by news content integration tools – online media archives. Russian and Ukrainian business-analysts are actively using the largest information archives from open sources "Integrum" (integrum.ru) and InfoStream (www.infostream.ua). It is the usage of open sources that allows competitive intelligence to stay within the limits of legal requirements and, at the same time, demonstrate high efficiency.

We can acknowledge that the faster the web-space grows the less reachable it becomes for traditional catalogues and search engines. Due to growth of the number of web-sites and portals using databases, dynamic content management systems, and emergence of new versions of information representation formats, deep web is growing rather intensively.  On the one hand, Internet, being a vast storage, increases the volume of information, that is "technically accessible", but on the other – informational chaos also increases, as does the entropy of network information space. Increasingly smaller part of information resources remains actually available to the users.

Leading search engines are still trying to find technical opportunities for indexing of database content and for access to closed web-sites, however, their tasks are objectively different from the tasks of business analysts – orientation of traditional search services towards mass services in this case does make sense. Thus, the niche for deep web search systems becomes increasingly larger.

### 2.3.3. Social media

Social media represent a set of online services and applications that allow users to communicate with each other, particularly, in real-time mode. The users can exchange opinions, news, and information (including multi-media information).

Social media are ideologically and technically based upon web 2.0 standard, allowing users to exchange content, generated by themselves (User-Generated Content), in contrast to the previous web concept, that, like traditional mass-media, envisioned centralized production of content that was later delivered to users/readers.

Naturally, social media are the most valuable information source for competitive intelligence that absolutely legally provides versatile information on people, events, companies, brands, and products. Such increasingly popular things as information operations, active information counteraction within competitive struggle, online mobilization, are, in many cases, mostly based on data manipulations in social media.

Seven categories of social media are outlined. They include social networks, forums, review sites, photo- and video-hosting servers, virtual dating agencies, and geo-social networks. We should note that the borders between these categories are a bit blurred.

An online social network (social networking service) is an online service, intended for building, representation, and organization of social inter-relationships, which provides a broad spectrum of information exchange features, and opportunities to create user profiles (allowing a user to present information about himself), build connections, find friends according to one's interests, add relatives, colleagues, classmates, etc.

A blog (blog is a derivative from web log) is a web-site, where most content is represented by records (text, images, or multimedia), posted by users on a regular basis. Blogs feature short posts (especially, the so-called "micro-blogs") of temporary significance; blogs are, usually, public, and open to outside readers, who can join public discussions with post authors (in their comments to blog records or on their own blogs). The set of all online blogs is called blogosphere.

Web-forums are web-applications intended for organization of communication among visitors of certain online resources (web-sites or portals). On web-forum resources users define topics of interest that are then discussed by other users through posting of messages within these topics.

Review web-sites are created in order to improve efficiency and quality of offered products and services (including products and services from outside web environment). Users, visiting review web-sites, leave their messages, complete questionnaires, and formulate opinions about some services or products.

A photo hosting is a web-site, allowing users to publish any images (usually, digital photos) in the Web. The key advantage of a photo hosting is the convenience of demonstration of posted photos. Respectively, a video hosting is a web-site, allowing users to upload and look through video information through their web-browsers. Video hosting concept gains popularity as a result of development of broadband access to the Internet.

A virtual dating service is an Internet service for virtual acquaintances among users, who want to communicate, create families, build relationships etc. While using virtual dating services, a user completes a form, in which (s)he specifies the selected nickname and other parameters, required by the service (gender, age, meeting purpose, interests, photos). Once registered, the user can communicate with other users, receive messages, and answer them.

A GeoSocial Network is a sub-category of social networks, where users leave data on their locations, which allows the network to unite and coordinate their activities based on information on other people located in the respective geographic locations and events taking place there.

The term "social network" denotes a concentration of social objects that can be considered a network (or graph), whose nodes are objects, while edges are social relations. The term was introduced in 1954 by a sociologist from "Manchester School" J. Barnes in his work "Class and committees in a Norwegian island parish". In the second half of the XX century the "social network" concept became popular among western researchers; at that time not only representatives of the society, but other objects with social connections were considered as nodes. Today the term "social network" denotes the concept that has become broader than just its social aspect; it incorporates, for instance, multiple social networks, including WWW. Not only static, but dynamic networks are considered. In order to understand their structure, one needs to consider the principles of their evolution.

Social network features are as follows:

1) Providing users with a broad spectrum of information exchange opportunities;
2) Creation of users' profiles, requiring users to provide some amount of personal information;
3) Friends in the social network are, mostly, real, not virtual.

A social network web-resource provides opportunities for:

1. Active communication.
2. Creation of public or closed user *profile*, containing personal data.
3. Organization and maintenance by a user of a list of other users, with whom (s)he has established some kind of social relationships.
4. Monitoring of connections between users inside the social networks.
5. Formation of groups of users, according to their interests.

6. Content management within one's profile.
7. Content syndication.
8. Usage of different applications.

## *Key social media*

The list of the largest social networks that might be of interest to competitive intelligence can include:

– Facebook;
– ВКонтакте (VKontakte);
– Google+;
– Мой Круг (Moi Krug, "my circle");
– LinkedIn;
– Badoo;
– Livejournal;
– Twitter.

***Facebook*** (www.facebook.com) is the largest social network, founded by M.Tsukerberg and his partners in 2004. Starting from September 2006, the network is accessible to Internet users. By October 4, 2012, *Facebook* audience numbered 1 billion users. Daily active audience exceeds 525 million people. Some 500 million people use *Facebook* mobile apps on a monthly basis. Every day network users leave about 3.2 billion "likes" and comments, and publish some 300 million photos. 125 million "friendships" are registered on the site. Monthly number of *Facebook* page views exceeds 1 trillion.

***Twitter*** (twitter.com) is a service, allowing users to send short text communications (up to 140 characters long) using web-interface, SMS, instant message exchange means, or third-party client software. By the start of 2011 the service, created by J. Dorsey in 2006, had more than 200 million users, while 50 million of them were using Twitter on a daily basis. 55 % of them were using Twitter on mobile gadgets. A peculiar feature of Twitter is public visibility of the "tweets" (i.e., so-called micro-blogging).

**Sina Weibo** (Chinese: 新浪微博) (weibo.com) – a Chinese microblog service, launched by Sina Corp company on August 14, 2009. The service is the most popular one in China; it is used by more than 30% of Internet users, and reaches the market similarly to Twitter in the USA. As of the 3rd quarter of 2017, there were 376 million of Sina Weibo users who were active on a monthly basis. Every day more than 100 million messages are sent through Sina Weibo. On September 8, 2017, Weibo required the users to bring their accounts into compliance with their real names before September 15.

**LinkedIn** (www.linkedin.com) is a social network for searching and establishing business contacts. More than 200 million users from 200 countries, representing 150 business branches, are registered on LinkedIn. LinkedIn social network, founded by R. Hoffman, was launched in 2003. The network provides registered users with an opportunity to create and maintain business contact lists. These contacts can be invited both from LinkedIn web-site and from outside, however the network requires preliminary acquaintance with these contacts. If a user does not have a direct connection with the contact, he can be introduced through another contact. LinkedIn contact list can be used for extension of connections, search for companies, people, and groups, according to their interests, CV publication, and job search; it allows us to recommend users (or be recommended by other users), publish vacancies, create interest groups etc. LinkedIn social network also allows one to publish information regarding business trips and conferences.

*«ВКонтакте» (VKontakte, i.e. "in contact")* (vk.com) is the largest social network in Russia and in Europe), created in 2006 and promoted as a "modern, efficient, and esthetic way of online communication". According to the data from August 2017, average daily audience amounts to 80 million visitors, while the total number of registered users exceeds 410 million. "VKontakte" users have access to the feature set that is common for many social networks: creation of a profile with personal information, content production and promotion, privacy settings management, interaction with other users (private and public), monitoring of activity of friends and communities through news feed. On May 24, 2013 Roscomnadzor (Russian surveillance service) entered vk.com domain and its IP-address into the "Unified registry of forbidden sites", but deleted it from the registry in a few hours and explained the error by human factor. On May 16, 2017, the President of Ukraine signed the order, ratifying the decree of the National security and defense council on the renovation of the list of sanctions against a set of Russian companies, including "VKontakte". Access to these resources is completely restricted since June 1, 2017.

*«Одноклассники»* (Odnoklassinki, i.e. "classmates") (OK.ru) is a Russian social network belonging to Mail.Ru Group. The project was launched on March 4, 2006. By January 1, 2016, more than 290 million users were registered in the network. Every day more than 42 million users visit the site (as of August 2017). On May 16, 2017, the President of Ukraine signed the order, ratifying the decree of the National security and defense council on the renovation of the list of sanctions against a set of Russian companies, including "Odnoklassniki" network. Restrictions on access to "Odnoklassniki" website in Ukraine entered into force on June 1, 2017.

**Google+** (plus.google.com) is a social network by Google that was officially put into operation in 2011. By the start of 2012 the number of users, registered in Google+, exceeded 90 million people. The service allows users to communicate through the Internet using special components: *Circles, Themes, VideoMeetings, Mobile version.* The basic principles of service operation are users, privacy, and live communication. Information shared by network users, influences personalized Google search results. The basis of Google+ operation is the concept of *Circles*, through which every user regulates his or her communication. Thanks to the circles concept, a user shares content and defines, which circle will have access to information, and which won't. Exchange of user materials takes place in the special feed (*Stream*), where you can monitor updates from circle members. Google also introduced the mobile social network version, featuring two unique functions: instant photo upload and chat (*Huddle*). Google+ social network allows users to get good positioning in Google search ratings.

*«Мой Круг» (my circle)* (www.moikrug.ru) is a Russian-language social network aimed at establishment of business contacts among people. The network architecture represents circles of users where the first circle includes close friends of the user, to whom (s)he discloses his(her) contact information, the second circle consists of friends of the user's friends, and the third – of their friends, respectively. "Moy Krug" network was created in 2005. In 2007 the project was purchased by "Yandex". Now it is one of Yandex services. In 2009 export of vacancies from the largest thematic web-sites hh.ru and rabota.mail.ru was connected to the service. In 2011 "Moy Krug" network completed the process of integration with social platforms and, thus, made it possible to find friends through Facebook, Twitter, LiveJournal and LinkedIn.

***Badoo*** (badoo.com) is a social network of acquaintances created in 2006 by a Russian businessman A. Andreyev. As of 2013, 180 million users were registered in Badoo. Having registered, the user can communicate through the chat, upload photos, communicate with friends in his(her) region or outside it. Paid premium services are also available. They are provided to those who want to increase their popularity, expand the circle of acquaintances. During the time of its existence Badoo released several freely licensable products, including different PHP updates, Pinba server for statistics collection in real time, free quick template processor Blitz for PHP.

*«Живой Журнал», ЖЖ, LiveJournal, LJ* (www.livejournal.com) is a platform for online-diaries (blogs) created in 1999 by an American developer B. Fitzpatrick. LJ allows users to publish their posts and comment on others' posts, create collective blogs (communities), add friends from among other users, and keep track of their posts in the feed. LJ servers are located in the USA, and the system belongs to an American company LiveJournal, Inc., although owned by SUP Media company from Russia. According to statistics from LiveJournal.com, as of the end of 2012, more than 40 million users were registered in LiveJournal. Settings and features of LiveJournal include: different types of records and commenting options; provision of extended user data; friends and news feed; user pictures; account security functions etc.

## *Social media monitoring*

Social media monitoring is the most important OSINT instrument. Through social media we can get the most thorough information on the audience of a particular product or service, and its opinion of the company's activity.

Let us consider several services for efficient social media monitoring, focusing on the most easily accessible ones:

***Seesmic*** (seesmic.com) is a free service for social media monitoring service. It supports monitoring of such resources as: **Twitter, Facebook, LinkedIn, Chatter, Google Buzz, Ping.fm**. It has applications for both web and PC, iPhone, Android, Windows Mobile.

***Socialmention*** (www.socialmention.com) is a platform for free search and analysis of information in social networks. The system searches for references in the selected networks, or in all networks at once. It also conducts analysis of reference tonality, related keywords, popular sources, and many other aspects. The system covers more than 100 social media items, including social networks, thumbnails, blogs, forums, and others.

***Hootsuite*** (hootsuite.com) is a multifunctional service for working with social media. Hootsuite system allows us to work with Twitter, Facebook, LinkedIn, MySpace and Foursquare accounts, as well as with WordPress blogs. HootSuite service is a certified partner of Twitter. It performs scheduled posting, provides an opportunity to track posts by keywords and references. HootSuite system is also fully integrated with Facebook. HootSuite system is conditionally free, it has a free version (analytics, 5 social profiles, 2 RSS/Atom feeds). The system is available through mobile platforms: iPhone, Android, Blackberry. All mobile software is free.

*YouScan* (www.youscan.ru) is a system for monitoring of Russian-language social media. YouScan system monitors references in blogs, forums, social networks (Facebook, VKontakte), Twitter, YouTube, and presents monitoring results through analytical interface that supports simultaneous work of several operators (employees). The system provides reports by number of posts, referring to keywords, by authors, sources, and tonality. The system offers 5 tariff plans, including one free plan. (Freemium – number of topics – 1; number of users – 2).

*BuzzLook* (buzzlook.ru) is a service for monitoring of such social media as VKontakte, Facebook, Livejournal, Flickr, YouTube and Twitter, allowing users to monitor brand reputation; study the activities of competitors; answer customer questions in social networks; gather suggestions from customers; support online communities.

*IQBuzz* (www.iqbuzz.ru) is a service for monitoring of social media – a large number of sources and platforms, such as LiveInternet, LiveJournal, Twitter, Yandex, blogs, video hosting services, such as RuTube and YouTube, different news, entertainment, specialized, topical, and regional portals. The system performs around-the-clock monitoring, allowing users to get information in real-time mode. IQ Buzz system allows you to define the tonality of user messages, analyze socio-demographic characteristics of their authors based on social network profile information. There is an opportunity to add new sources for monitoring, according to user queries.

*Socialbakers* (www.socialbakers.com) is a service for collecting statistics on social networks' functioning. The service provider calls it "the heart of Facebook statistics". Socialbakers is famous for its Facebook ratings of brands compiled in different categories. Beside Facebook, Socialbakers service provides an opportunity for free monitoring of information in other social networks, such as Twitter, Google+, LinkedIn.

***SocialSeek*** (socialseek.com) is an easy-to-use free service for real-time monitoring of several social media. It conducts search in news feeds, blogs, Twitter, Facebook, Youtube.

***Socialpointer*** (www.socialpointer.com) is an easy-to-use service for information monitoring in social networks, newsfeeds, blogs. It includes basic analytics.

***PeerIndex*** (www.peerindex.net) is a free service for analysis of social media, first and foremost, Twitter, Facebook, LinkedIn. It defines the size of "social capital" or impact of a company, a professional, a publication etc.

***PostRank*** (www.postrank.com) is a Google service for real-time analysis of data by themes, trends, events, related to a person or a business.

***Topsy*** (topsy.com) is a free service conducting search across social media in real time mode.

***HowsSciable*** (www.howsociable.com) is a free instrument for monitoring of brands and keywords in 32 social networks.

***Twitalyzer*** (www.twitalyzer.com) is an analytic client software for Twitter, allowing users to monitor the number of clicks, analyze positive and negative comments, segment the audience. The software is integrated with Google Analytics system; it can build interactive diagrams and graphic tools.

***WildFire*** (monitor.wildfireapp.com) is a multi-functional online service for commercial media-marketing, including Wildfire Messages tool for creating, monitoring, and managing messages. It allows you to configure scheduled message posting in social networks. The service provides complete feature set for brand promotion in different social networks.

***Kurrently*** (www.kurrently.com) is a free search system targeting Twitter and Facebook, allowing users to monitor and disseminate information in the social networks.

**Trackur** (www.trackur.com) is a commercial online tool for monitoring and analysis of social media. It allows users to monitor brand reputation across news web-sites, blogs, forums, and social networks, such as Twitter, Google+, and Facebook.

**Babkee** (www.babkee.ru) is a system for monitoring the references in the social media. It allows users to solve such problems as brand reputation evaluation; efficiency analysis of advertizing campaigns in the Web; marketing research of offerings, competitors, and target audience; responding to user requests and respective support activities. The system is positioned as a unique service for evaluating the significance of comments and messages. A free system version is available.

**Buzzware** (www.buzzware.ru) is a social media monitoring service, **allo**wing users to study consumer opinions of the brands, expressed in blogs and social networks. The service can be used for reputation analysis, competitor study, and for gaining understanding of user experience, expectations, and, of course, for online promotional campaign efficiency evaluation.

**SemanticForce** (www.semanticforce.net) is a service, conducting monitoring of non-structured sources, such as comments in online media and online stores (Fig. 2.10). It compiles more than 20 types of analytic reports. SemanticForce service is integrated with external systems: Klout, Copiny, GoogleAnalytics.

**Brandspotter** (www.brandspotter.ru) is a system for monitoring of social media, offering a standard service package, including definition of emotional tonality of comments, monitoring of message statistics by topics, platforms, and authors.

**Крибрум** (www.kribrum.ru) is a social network monitoring system, allowing users to monitor and analyze references to brands, products or services, key persons, events, geographical locations. It includes tools for automatic evaluation of emotional tonality of posts and for compiling of interactive reports. The system filters out the posts, in which the brand is mentioned as something unimportant. The data is displayed by the system 2–4 hours after being published.

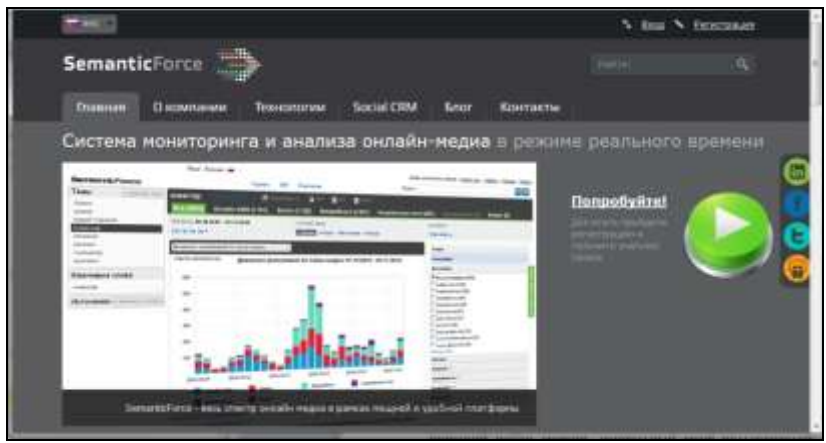

Fig. 2.10 – A fragment of SemanticForce **service web-site**

*Wobot* (wobot.ru) is a service, allowing users to monitor the opinions in social networks in retrospect. It features a broad spectrum of metrics and a social graph of users. It includes a self-learning mechanism, allowing the user to define the comment tonality.

*TweetDeck* (tweetdeck.com) is a free cross-platform application, providing a tool for managing and monitoring of messages in social networks, such as Twitter, Facebook, MySpace, LinkedIn. It supports multi-channel column interface, various filters, including filtering by keywords. No analytics.

## 2.3.4. Databases of open sources

In the process of OSINT access to databases and other Internet resources is provided in order to facilitate efficient processing, storage, search for, and exchange of open information. These databases are located in different computational networks, both local and global, web-resources, "Deep Web" and others. At the same time, OSINT systems use processed information from open sources and support the following information databases:

- Database with operational information that provides interconnection between information needs and queries, indicates the status of material collection and processing, and contains statistical data (graphs, tables);
- Database with technical information that supports information collection operations and consists of unprocessed textual, audio- and video-materials, translations and recordings of transcripts.

As a rule, for successful competitive intelligence a databank, including the following key databases, has to be created and supported [Ландэ, 2005]:

1. Competitors (existing and prospective);

2. Market information (trends; information on product ranges, prices; targeted information);

3. Technologies (products, exhibitions, conferences, standards, quality);

4. Resources (raw materials, human and information resources);

5. Legislation (international, central, regional, and departmental normative and legal acts);

6. General trends (policy, economy, region-specific features, sociology, and demographics).

Today, the key information sources for competitive intelligence are: Internet, press, as well as open databases. However, while access to ordinary online resources can be considered conditionally free, access to databases, in most cases, requires both registration and payment for the services. Besides, almost all of them can be attributed to the so-called "hidden" web space.

The most popular databases used by competitive intelligence specialists are: databases of customs, taxation, statistics, legal offices, courts, chambers of commerce, privatization bodies, stock markets, information, rating, analytic and other agencies etc. Certain available databases of other controlling bodies and organizations are also very helpful.

Traditionally, competitive intelligence is based on such information sources as published open-access documents that include overviews of product market, information on new technologies, formation of partnerships, mergers and acquisitions, announcements of job vacancies, exhibitions, conferences etc. That is why lately databases based on media archives (including online media) started gaining popularity.

The "Great Three" world services providing users with access to business and analytical information include ***LexisNexis, Factiva,*** and ***Internet Securities.***

The largest full-text online information system ***LexisNexis*** (www.lexisnexis.com), containing more than 2 billon documents from 45 thousand sources with more than 30-year-deep archive of business information and more than 200-year-deep archive of legal information, belongs to "hidden" web category (Fig. 2.11). 14 millions of new documents are added to the archives on a weekly basis. In contrast to unstructured data sets of "surface" web, LexisNexis users can enjoy the benefits of powerful search tools to get credible and well-structured information.

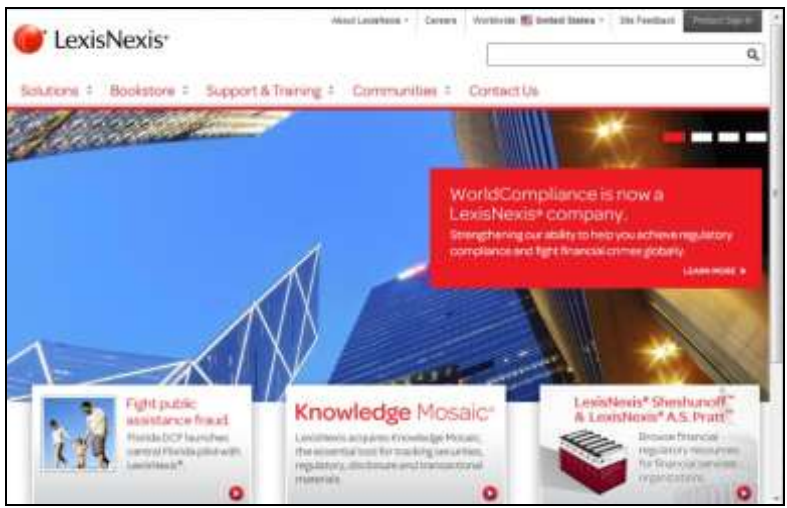

Fig. 2.11 – A fragment of LexisNexis service website

**Factiva** (global.factiva.com) service is a subdivision of Dow Jones company. Presently, it belongs to News Corporation and provides access to business and analytical information. Factiva service is based on more than 35 thousand primary sources from 159 countries of the world. Factiva database contains materials on more than 36.5 million companies, as well as complete Investext information set.

**Internet Securities** company (www.internetsec.com), ISI Emerging Markets brand, covers 80 thematic information sections, based on 16 thousand sources – article texts, financial and analytic reports, corporate information, macro-economic statistics, market data (Fig. 2.12). The key products of ISI Emerging Markets are: CEIC Data, Emerging Market Information Service (EMIS), Islamic Finance Information Service (IFIS), IntelliNews, ISI Compliance Edition, ISI DealWatch.

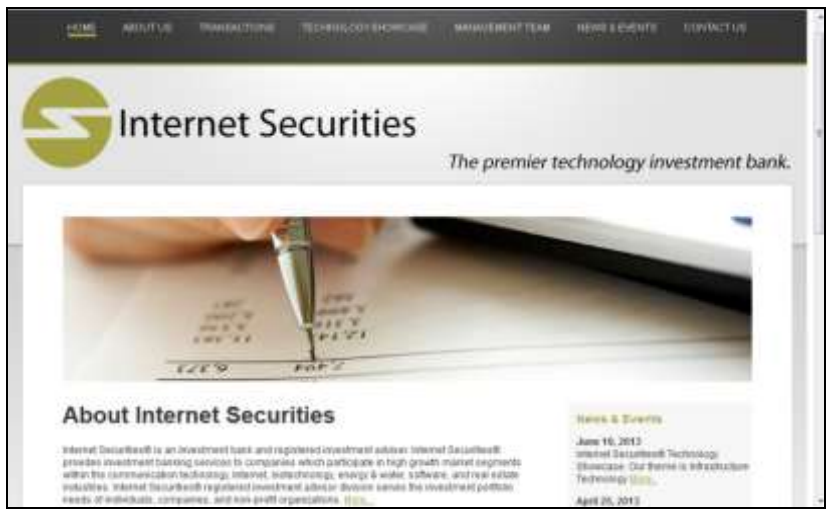

Fig. 2.12 – A fragment of InternetSecurities service website

In Russia the most popular services include «**Интегрум**» (**"Integrum"**) (more than 10 thousand sources, particular services are: "media analysis", "media archive", "media feed", "Companies" DB, "Connections" DB), «*Медиология*» (*"Mediology"*) (13 thousand sources: media, TV, radio, newspapers, journals, information agencies, Internet, blogs, database of 30 thousand objects: companies, persons, and brands), «*Яндекс.Новости*» (*"YandexNews"*) (a service for automatic processing and systematization of news; more than 4000 sources; the service does not allow for usage of materials unrelated to news), Public.Ru – a large online library of Russian-language media. Since 2000 *Public.Ru* is building its database that stores archive materials of Russian publications since 1990. More than 70 million articles of Russian-language media articles collected from 4600 sources are available through the database archives. The key types of sources contained in the database are: federal publications; regional publications; information agencies; TV channels; radio broadcasting stations; Internet-publications.

In Ukraine the niche is occupied by online media content-monitoring system *InfoStream* (more than 6 thousand information sources, more than 100 million documents in the archive).

Ukrainian corporation «*Медіа-простір*» (*"Media-space"*) (550 information sources, 25 regional information bureaus) that performs Ukrainian information space analysis, provides media-reviews. Systematization of information messages is performed by four objects of reference: subject domains; political subjects; personalities; territories.

Information-monitoring system *Web-Observer* in its "basic configuration" covers 500 sources. The system is implemented in UNIAN information agency; "UNIAN-monitor" service is based on it.

Ukrainian Internet-monitoring system *MonitorIX* covers online sources (as a network information search system), media (130 publications), TV, blogs, and forums. It supplies users with operational and archive monitoring results.

By the end of 2017 Ukrainian information analytical platform for business analytics, competitive intelligence, and verification of counteragents, YouControl (https://youcontrol.com.ua/) allowed users to get relevant information about a company from 42 national registries and learn about changes in them on a daily basis. Trough the system of automated monitoring YouControl follows every company that is added to the list, verifies information in the most important registries, and collects the detected differences into a unified document. YouControl service allows users to detect, how a company is related to other companies; in this process such factors as common founder, director, address etc. are taken into consideration. The interface allows users to review corporate connections of an organization, portfolio of which is currently analyzed, in a graphic step-by-step mode.

Let us provide another example of a foreign database from "hidden" web. LexisNexis corporation provides **Auto TrackXP** service, featured in the list of twenty world's largest "hidden" websites (according to BrightPlanet rating). Auto TrackXP is a 30-therabyte database, covering almost all aspects of US civil life. Auto TrackXP database contains information on almost every US citizen. TestProfiles.com – ChoicePoint Online part – contains personal characteristics and data on competencies of US citizens. For example, if you need to verify whether some individual possesses other person's documents, the system provides a paid service ProCheck, allowing you to compare information from different sources and state catalogues. For individual "portfolio-compiling fans" ChoicePoint has a more modest, but no less interesting set of services (www.choicetrust.com). Patients in doubt can use the Doctor Check service to select doctors from 40 different specialties and verify their qualifications. The report, compiled by the system, can be used, for example, by an insurance company as grounds for insurance policy application rejection.

The system is widely used as a legal resource for competitive intelligence tasks. At the same time, today

Americans are, generally, expressing their indignation, claiming that existence of such services violates their civil rights.

*Insight Profiles* (www.insightprofiles.com) service contains personal characteristics and data on skills and competencies of US citizens. For example, if you need to verify whether some individual possesses other person's documents, the system provides a paid service *ProCheck* (procheck.com), allowing you to compare information from different sources and state catalogues.

In Russia and Ukraine the following databases are popular:

«*Лабиринт*» *("Labyrinth")* (www.labyrinth.ru) – a database containing publications of leading business sources and intended to assist people performing analytical and research works, writing articles, comments, report notes, etc. The database features biographies of Russian public persons, notes on organizations and companies, information on Russian Federation subjects, and other reference materials;

*«Компасс» ("Compass")* (www.kompass.com/ru) – a database, positioned as an international information B2B (business-to-business) search system, performing search across companies, products and services, managers, in order to form databases of targeted marketing and distribution, potential customers;

*«КАРЕ»* (kare.pulscen.com.ua), a database of Ukrainian enterprises including 384000 companies, DB of Ukrainian agro-industrial complex – 218000 companies;

*Dun & Bradstreet* (*D&B*) database. The local office of *D&B* in Ukraine is "Business-monitoring" company that belongs to "Avesta-Ukraine" group and represents business information. The company provides connection to the System of Professional Analysis of Markets and Companies (SPARK-Russia) and to Dun & Bradstreet (D&B) database. In Russia connection to this database is provided by Interfax information agency (www.dnb.ru);

Databases of an international corporation **Creditreform**, represented in Russia by "Creditreform RUS" (www.creditreform-rus.ru) and in Ukraine by "Creditreform Ukraine" (www.creditreform.ua). The companies provide access to the international network of information business facilitation BIGNet (Business Information Group Network). This network unites independent organizations providing online business consultations across the world (more than 8 million a year), as well as access to BIGNet and to its own database (www.crefoport.ru) containing data on 30 millions of companies.

*Europages* (www.europages.eu) – European business-directory – is an information search B2B-suystem covering more than 2 million suppliers, manufacturers, and distributors in Europe and across the world.

It is virtually impossible to enumerate absolutely all information sources, as this market is very dynamic, new databases are constantly emerging, existing sources are uniting, and weaker ones are taken over by the stronger. At the same time, one of the key competitive intelligence rules is formulated as follows: "the more independent sources confirm the information, the more credible it is".

Beside databases, one of the most efficient information sources is represented by reports and notes of outsource companies that perform professional competitive intelligence and gather data on business structures and markets. Their actual products are the results of competitive intelligence.

A lot of such special companies exist in the world. One of such large companies, owning 80% of western market, is **Dun & Bradstreet (D&B)**, whose database was mentioned above. Information on any company in this service is estimated at the level of $100 and higher. More profound analysis of a market or a competitor might cost you $10,000. Timeframes range from several hours (if the information is available in the database) to several days (for reference notes) to several months (for analytic work).

At the European market the most well-known players include the above-mentioned Irish company *Creditreform*, German *Schufa Holding AG* (479 million documents in the DB, including 66 million on physical persons), Australian *Intercredit Information Holding*, Latvian *Coface IGK* (IGK System is a database of debtors, containing data on current debts, lawsuits, and insolvencies), as well as many others. Some of these companies combine the competitive intelligence functions with other activity types, such as responsibilities of credit bureaus.

There are many commercial software applications, tools, and databases in the Web, providing opportunities for full-text search. Here are some examples:

- Google Academy (scholar.google.com) provides easy ways of broad academic literature search. This search system can perform search across all existing subjects and sources, such as articles, monographs, textbooks, dissertations, abstracts, preprints etc. from one entry point. Google Academy allows the user to find the necessary academic work among all available international research publications.
- Spokeo. This service specializes on collecting information about people (their names, addresses, phone numbers) from phonebooks, social networks, marketing lists, business sites, and other open sources.
- Blog Pulse is an automated system to detect trends in blogs through application of computational learning and natural language processing methods.
- Pipl. The system allows users to find Deep Web resources that cannot be retrieved using standard search systems.
- Monitter. The search system, created based on Twitter browser. Monitter displays the results of direct search through search systems by keywords in your browser.
- Maltego. This system is an analytic application that performs collection and analysis of data, as well as forms semantic groups based on gathered information. Maltego allows its users to detect existing and previously unknown connections between information objects.

The common problem while trying to obtain information through western agencies that have offices in Russia and Ukraine is that, as a rule, provided information on western non-residents, is much more profound and concise than the information on local firms. Consequently, in such cases it is more relevant to address local information companies; their results cost less and have better quality.

At the Russian competitive intelligence market the most popular and high-quality reports are provided by such companies as «Р-Техно» ("R-Techno"), «Медиология» ("Mediology"), "Special Information Service", «Интегрум» ("Integrum"), «Кронос-Информ» ("Cronos-Inform"), and many others.

In Ukraine there is a whole set of similar companies, including «Авеста-Украина» ("Avesta-Ukraine"), «Сидкон» ("Sidcon"), Interbank security service «Скиф» ("Scythian") and others.

All national and foreign information companies have online offices and receive orders through the Web, so they can be attributed to a specific type of Internet-sources.

We should also note that if a client orders services from an outsource company, it performs most of the information work for this client, but final conclusions and solutions, as well as recommendations for managerial decisions remain the responsibility of the client himself. Only the client can possess all the insider information in its thoroughness.

## 2.4. OSINT technologies

OSINT is a very versatile form of information analysis and collection. In the process of OSINT one should use security measures while collecting information from the Internet. For example, VPN can be used to ensure anonymity and avoid detection of the fact of information collection. Assessment of the sources becomes important for the general process of collection and analysis during OSINT. OSINT analyst needs intellectual analysis to detect true or false process that impacts prediction of the future. Finally, analysts should apply assessment analysis so that its results could be included into the ready-made classified, non-classified, or patented intellectual product.

As a rule, information collection in OSINT differs from data collection in other intelligence disciplines, where acquisition of unprocessed information subject to analysis might be the major problem. The major problem in OSINT is detection of relevant, credible sources within vast quantities of publicly available information. However, the task is not very difficult for those who know how to get access to local knowledge, and how to use experts, capable of producing new, individual knowledge "along the way".

## *OSINT phases*

OSINT process includes four phases: planning, preparation, collection, and production of the final material – analytics and four key processes: analysis, accumulation of intelligence data, evaluation and distribution among directions. The process of intelligence, as well as processes of preparation of information countermeasures (planning, preparation, execution and drawing of conclusions), intersect and repeat themselves according to requirements of practice.

According to the Instruction on field intelligence, OSINT increases the efficiency and provides support for intelligence process and other operations.

Fig. 2.13 illustrates the phases of OSINT process.

Gathering of intelligence data synchronizes and integrates the processes of planning, usage of efforts and means, processing and distribution of system elements for support of military operations, uniting intelligence and operational functions.

After analysis the information, obtained from different sources, becomes intelligence data, containing the information on the enemy, threats, climate, weather conditions, landscape, etc.

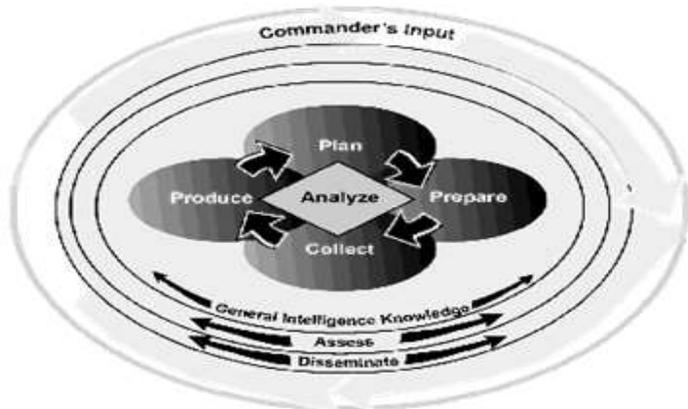

Fig. 2.13 – Collection of information through OSINT allows the management to take correct decisions [APT, 2012]: Plan ⇒ Preparation ⇒ Collection ⇒ Production. General intelligence knowledge. Assessment. Dissemination

It is established that such elements of OSINT structure as steady information flows, technical means, software, security of communication channels and databases include the means of:

- Ensuring availability of intelligence data. Ensuring availability of intelligence data is a process allowing intelligence organizations to get access to the data actively and swiftly;
- Development and maintenance of automated intelligence network. The main task is provision of information systems that facilitate connection, simultaneous analysis and processing, dissemination of materials, and creation of conditions for availability of intelligence data;
- Provision and maintenance of access. This task entangles provision, support, and maintenance of access to classified and non-classified software, databases, networks, and other Internet resources for troops of ally countries, joint forces, and international organizations;
- Building and maintenance of databases. This task envisions building and maintenance of classified and

non-classified databases. Building and maintenance of a database contributes to fast analysis, preparation of reports, processing, dissemination, long-term military action.

## 2.5. Legal aspects

Undoubtedly, OSINT as a particular activity should be performed within the limits of the national legislation. The basis for this is provided by the constitutional rights for searching, obtaining, transferring, and usage of information in all civilized states. At the same time, it should be noted that in some countries the legislation, limiting the activities on information collection and processing, virtually, puts OSINT under a ban.

In Ukraine everyone "has the right to freely collect, store, use, and distribute information in an oral, written, or any other way – according to his(her) choice" (Constitution of Ukraine, chapter 2, article 34).

So, in Ukraine legal regulation of the information sphere, to which OSINT, certainly, belongs, is based on the following principles:

1) freedom of search for, acquisition, transfer, production, and distribution of information in any legal way;

2) limitations of access to information are established only by laws of the state;

3) openness of information on activity of national government and local self-government bodies and free access to such information, except for cases provided by laws of the state;

4) according to access category, information is sub-divided into open (publicly available) and limited-access information. Limited-access (classified) information, in its turn, is also subdivided into two categories according to its nature: state-secret information and confidential information.

Although there is no legally established concept of «OSINT» in Ukraine, activity on collection, storage,

processing, and distribution of information is regulated by a whole set of legislative and normative acts:

Law of Ukraine "On print media (press)" of 16.11.1992, # 2782-XII, articles 6, 25.

Law of Ukraine "On security activity" # 4616-VI of March 22, 2012, articles 9, 13, 19;

Civil code of Ukraine (article 505), Criminal code of Ukraine (articles 231, 232), Code of Ukraine on administrative offences (article163);

Law of Ukraine "On information" of 02.10.1992, # 2657-XII (with changes of 13.01.2011), articles 5-7.

We should not forget that activities on ensuring business security (even through OSINT) can, sometimes, be considered operational and search activities, that can be performed (according to the Law of Ukraine "On operative and search activity" of 18.02.1992, #2135-XII) only by subjects, specified in the respective articles of the listed Laws. The list of these subjects is exhaustive and other legal and physical persons are forbidden to perform operative and search activities.

The Cyber-security strategy of Ukraine, approved by the Order of the President of Ukraine # 96/2016 of January 27, 2016, provides the key tasks of power structures, including: "intelligence bodies of Ukraine – implementation of intelligence activity on detection of threats to national security of Ukraine in cyber-space, other events and conditions, concerning cyber-security spheres", and provides for "creation of a system for timely detection, counteraction, and neutralization of cyber-threats, particularly, with engagement of volunteer organizations", and all these points, surely, concern implementation of OSINT means in the specified domain.

At the same time, the legal and normative documents currently in force provide for criminal responsibility for illegal collection of data for purposes of usage, and for usage of data itself, if the data is a commercial secret, as well as for disclosure of a commercial secret.

Under such broad interpretation of legislative norms all procedures of collection, processing, and storage of information about competitors become, on the one hand, legitimate, and, on the other, quite problematic. In Ukraine access to a large set of data (for instance, on owned and mortgaged property, land plots, bank accounts etc), freely available in most countries, is closed. In these countries much of the data can be obtained only through consultations with the respective specialists.

Moreover, activities of companies performing open-source intelligence attract even greater attention of state control bodies. Such interest results from several groups of legal problems, respectively, related to:

1) commercial secret protection;

2) personal data protection;

3) observance of authorship rights (copyright laws);

4) potential competition on OSINT market.

We can also outline three classes of key copyright law-related problems concerning competitive intelligence: problems related to legality of incoming information usage (information sources), based on which the reports (OSINT results) are published; problems associated with authorship rights for OSINT results; problems related to rights for usage (application) of specialized software necessary for OSINT.

Besides, one of the problems faced by Ukrainian OSINT services is almost complete absence of anti-damping legislation. Although access of large international players to this market is obstructed by lack of necessary connections, databases, archives, and even linguistic and legal training, they can, potentially, provoke damping at OSINT services market.

The situation might change if a clear legislative base for activities on collection and analytical processing of information, particularly, on open-source intelligence, is developed.

Certain articles of the Law of Ukraine "On protection from unfair competition" # 236/96-BP of 07.06.1996 had important implications. The Law (article 15-1) forbids "Unlawful collection of commercial information", "Disclosure of commercial information", "Unlawful usage of commercial information" (chapter 4, articles 16, 17, 19, respectively).

The decree of the Cabinet of Ministers of Ukraine of August 9, 1993, # 611 "On list of data that do not represent commercial secret" provides a whole class of documents, related to activity of business structures, that are, in fact, open, particularly, constituent documents, reporting forms, information on participation of founders and executive officers in other companies, etc.

Often the purpose of business intelligence efforts is to discover the commercial secret of competitors. We can say that OSINT activities are, sometimes, targeted at information, that is not publicly available and is protected by law. These activities violate multiple articles of the Criminal Code of Ukraine, including article 231 "Illegal collection with purpose of usage, or usage of data, representing commercial or bank secret".

Thus, open-source intelligence can legitimately use only those methods and ways of information collection and processing that do not violate the legislation, i.e. the key OSINT functions include high-quality collection, systematization, and, mostly, analysis of information, and not spying, bribing, or illegal hacking.

For the first time the right for preservation of commercial secret was provided by the Law of the USSR of June 4, 1990 "On enterprises in the USSR". Article 33 of the Law clarified the commercial secret concept as data, not belonging to state secret, and related to manufacturing, technological information, management, finances, and other activities of enterprises, disclosure (transfer, leak) of which can damage their interests.

The Civil code of Ukraine defines commercial secret (article 505, point 1) as information that "is a secret in a sense, meaning that in general or in some particular form and in the aggregate, is unknown and is not easily accessible to persons that, generally, deal with type of information it is attributed to; as a result it has commercial value and was subject of measures appropriate for existing conditions, related to preservation of its secrecy, undertaken by person that legally controls this information".

According to these definitions, as soon as the information (as a result of some actions) gets, for example, to the pages of any website, it ceases to be a commercial secret, as it becomes easily accessible.

Although many articles of the Criminal code of Ukraine (231, 232, 232-1, 361, 363) provide criminal responsibility for both disclosure of commercial secret and illegal collection and usage of data, attributed to it, the existing normative and legal base does not clearly specify, which particular data on financial and economic activity of an enterprise are its commercial secret (except for definition of bank secret, provided in article 60 of the Law of Ukraine "On banks and banking activity").

Government structures, banks, large corporations are not always able to ensure protection of personal data bases they store, and as a result an enormous flow of confidential information gets to the market. Ensuring the security of personal data is and objective necessity. Today personal data, information on people, becomes the most expensive commodity. Such information in the hands of a malefactor is a powerful weapon. So, personal data needs to be protected.

Personal data is an important component of a broader concept of privacy. That is why personal data protection is a component of ensuring privacy. Alongside freedom of speech and other rights, privacy is one of the basic values of humanity.

As of today, the key European documents regulating personal data protection are the Convention of the Council of Europe for the Protection of Individuals with Regard to Automatic Processing of Personal Data and the Directive of Euro-parliament ETS #108, 1981, that is mandatory for all EU member countries, and can provide a legislation example to be followed by other states (including ours). EU countries are gradually harmonizing their legislations with the Directive. The UK adopted their "Data Protection Act" back in 1998. Its technical implementation is the draft standard «Specification for the management of personal information in compliance with the Data Protection Act 1998» (BS 10012, 2009). The US developed the local version of the standard simultaneously with the UK. The draft document on personal data protection for US governmental structures («Guide to Protecting the Confidentiality of Personally Identifiable Information (PII)» (SP 800122)) regulates observance of «The Privacy Act of 1974» and «Privacy Protection Act of 1980». Canada adopted «Privacy Code» – a set of documents for implementation of the legislation on protection of personal data (The Privacy Act and PIPEDA).

In EU member countries definitions of personal data tend to be as broad as possible, so the citizens often do not follow the respective legislation because of its excessive "load". And respective state authorities, usually, do not take serious action, except in some special cases. Such issues as collision between privacy requirements and freedom of speech remain extremely relevant. As a rule, modern European laws forbid collection, storage, usage, and distribution of critical personal data without the subject's consent.

The right for privacy is provided by the Constitution of Ukraine. According to Article 32 of the Constitution of Ukraine, no one can be subject to meddling in his personal and family life except for cases provided by the Constitution of Ukraine. Besides that, the Constitution of Ukraine also regulates protection of some other privacy aspects. For instance, Article 30 protects the security of residence (territorial privacy), Article 31 – confidentiality of mailing, telephone conversations, telegraph, and other correspondence (communication privacy), Article 32 forbids collection, storage, usage, and distribution of confidential personal data without the subject's consent, while article 28 forbids to make an individual the subject of medical, scientific, or other research, against the individual's free consent (protection of certain elements of physical privacy).

Constitutional norms provide an exhaustive list of grounds for privacy intervention and conditions for such intervention. However, our countries have many brunch legal norms that contradict provisions of the Constitution. It is these norms that do not comply with the international standards and European legal practice.

In accordance to Ukrainian legislation, personal data in Ukraine includes second name, first name, and patronymic, accompanied by any other identification data, such as address, phone number, or educational status.

In order to clarify the relation of a physical or legal entity to personal data protection it is important to define the subjects of relationships associated with personal data: according to Article 3 of the Law of Ukraine # 2297-VI, the subjects of relations associated with personal data are:

– Subject of personal data;
– Owner of personal data base;
– Manager (operator) of personal data base;
– Third party;
– Authorized governmental body for personal data protection;
– Other governmental bodies and local self-government bodies, whose competency embraces the issues of personal data protection.

Ukrainian legislation provides for notification character of personal data processing. Prior to processing of personal data, owner and manager (operator) are obliged to notify the competent personal data protection body of their intention to process personal data. Then the data on owners or managers (operators) are included in the special registry of operators. Information contained in the registry of operators becomes commonly accessible.

Laws on personal data concern the majority of population as participants of data "processing". And as every person is a personal data subject, the Law bears general character and concerns everyone.

Both legislative acts, Ukrainian and Russian, are immediately related to the spheres of information technologies and telecommunications; and both contain provisions that are subject to debate, contradict common practice, and seem to be impossible to observe. Provisions of the law concern all legal and physical entities, and Internet sphere is not an exception. The law on personal data protection can change the operation principles of Ukrainian internet resources: e-mail services, dating sites, online stores, social networks, although the owners of these resources themselves hope to remain "untouched" by the law. In order to comply with all the provisions of the law on personal data, owners of internet-resources must carefully think through organization of their operation. Presently, there are many web-services that do collect, store, and use personal data. Compliance with provisions of the law is not an easy task for owners of these resources; particularly, government officers have an opportunity to oblige internet-companies to obtain written consent for personal data usage from every user. It is not a secret that many sites post information containing personal data of people (who are, say, looking for vacancies or acquaintances), including special data categories, such as nationality or faith. The task of those who provide such services is to legitimately process the data and, at the same time, protect it according to legal requirements.

Particularly, personal data is very widely used in social networks and e-mail services. For example, it is especially difficult for web-resource owners to comply with the provisions of the law regarding written consent of every user for processing of his(her) personal data. At the same time, it is the operator that is obliged to provide proof of the fact that such written content has been obtained.

A modern internet-company collects and processes different categories of personal data of its employees,

contractors/counteragents, and certain data of its service users. People who place their personal information in social networks or dating services consciously make it open for all users of the resource, and it can be legally considered "publicly available", so it is not mandatory to maintain privacy of this information, but social networks also feature information that is concealed by the user, making it accessible only for a limited group (of "friends"). In such cases the internet-resource had to envision special protection means for this information.

OSINT practitioners have to face multiple contradictions and non-compliances within the existing legislation; for instance, in the Law of Ukraine "On protection of personal data" (part 9, article 6) provides, that usage of personal data for historical, statistical, or academic purposes can take place only in de-personified format. That is, the respective records in OSINT reports should look as follows: "Person A held negotiations with person B", "The flag over Reichstag was pitched up by individuals I, II, and III". In scientific reports one should not refer to colleagues, even if they provided written consent. Certain difficulties are caused by the necessity to inform the authority "about every change of data, needed for registration of the respective database", which, among other data, contains information on all its operators (users).

Besides that, many business intelligence services, that absolutely legally compile a base of personal data to solve a certain given problem, have to destroy all the fruits of their work once their task is completed. However, if the main purpose is, for example, providing services to clients through satisfying their requests, then a secondary purpose for any respectable organization is to develop a customer database. And such a base, usually, has its own commercial value. We know many cases of absolutely legal reselling of customer databases, for example, when the activity of database-owning company is terminated. Still, Ukrainian legislation provides conditions for destruction of personal data, including (article 15), "termination of law relations between subject of personal data and owner or operator of base...". This means that the operator (for instance, the service provider), has to destroy the database, compiled while the service was provided.

That is why owners and operators of such databases reformulate their goals in a special way, for instance, "service provision with an opportunity to store personal data during the guarantee term...". Thus, the legislative norms are observed and the interests of the contractor (owner or manager/operator of personal data base) are taken into consideration.

OSINT departments perform processing of personal data that are stored in open sources in the Internet, i.e., are publicly available. No consent of personal data subject is required for data processing. However, in this case it is the responsibility of the owner or operator to prove that personal data being processed are publicly available. This means that he should either collect proofs, certifying that the data is gathered from open sources, or obtain personal data subject's consent, allowing him to store the respective document. Moreover, one should have a document certifying the public availability of the source of personal data. However, the question whether the information resource (web-site) owner has the written consent of the subject, remains unanswered.

The problem of criminalization of certain OSINT services is as relevant as ever. Many security services use databases of persons. These bases are used for quite decent purposes, for example, for verification of data on employees, partners, and competitors. Obviously, they will use such databases in future as well, but they will have to do this illegally and "go underground". Technical facilities for usage and maintenance of such databases are provided by numerous systems like Cronos (software shells, distributed absolutely legally). Thanks to such tools, any interested user can get access to multiple databases working under these shells.

At the same time, today owners and operators of personal data bases find it more profitable to comply with the legislative requirements and protect their businesses from threats represented by controlling authorities and unfair competition. Obviously, for certain organizations it will be problematic to comply with the legal requirements. In such cases, they can turn for help to integrator companies, operating at the information security market.

In the USA the key legal mechanism for open source intelligence under the Department of Defense is the Defense Open Source Council (DOSC). It serves as a forum for coordination and facilitation of programs and activities on open source intelligence for all services and military command. The council consults and reports to the deputy Minister of Defense for intelligence about open source intelligence issues, new OSINT unit efficiency improvement initiatives, and the activity of the Department of Defense in general. Particular tasks of the Council include:

- Coordination of OSINT unit actions and approval of its open source intelligence plan;
- Definition of the set of requirements to open source intelligence process.

The US army standard "ATP 2-22.9" provides the key terms, basic concepts and methods for collection of intelligence data from open sources for the US Army. The document underlines the characteristics of OSINT as an intelligence domain, its connection with other intelligence domains, and opportunities of its application in unified operations.

Usage of publicly available information is an important aspect of technical intelligence (TECHINT). Although intentions, opportunities, and vulnerability factors of the enemies are subject to secrecy, OSINT results (for instance, data from open service "Google Earth") contribute to obtaining of information on the most "secretive" states and organizations. Such examples indicate the high degree of responsibility associated with OSINT activities.

Copyright law is one of protection forms for published and unpublished works, provided by chapter 17 of the US Code, where the authors of "original works" (literary, drama, musical, and artistic) are defined.

National copyright laws represent limitations for OSINT. Violation of rights, particularly those provided by chapter 17 of the US Code, and copyright laws, still leaves a loophole for legitimate OSINT usage that can be defined by four factors:

- Purposes and character of usage;
- Properties of used original works;
- Volume and parts of the original work being used;
- Impact of original works usage upon potential market or value of these works.

# 3. Elements of non-linear dynamics for information operations recognition

In this chapter we focus on information operation recognition based on study of dynamic properties of information streams in global computer networks, particularly, in the Internet.

Study of information streams in the Internet, i.e. the flow of messages published on pages of web-sites, in social networks, blogs etc, calls for application of modern tools. Particularly, existing methods of generalization of information arrays (classification, phase integration, cluster analysis etc) are not always suitable even for adequate quantitative reзpresentation of the processes taking place in the information space [Lande, 2007].

Quantitative analysis of dynamics of information streams generated in the Internet is becoming one of the most informative methods allowing us to research the relevance of various topical flows. This dynamics depends on various qualitative factors, and many of them are not subject to accurate description. However, the general character of temporal dependence of the number of online topical publications still allows for development and research of mathematical models and for prediction of their behavior. Observation of temporal dependences of online information flow volumes persuasively shows that the mechanisms of their generation and distribution are, obviously, related to complex nonlinear processes. This is the subject of the current chapter.

Various approaches are applied for analysis of time series reflecting the dependence of information stream volumes on time. It turns out that all these approaches are interconnected, and, moreover, the concept of correlation plays the key role. Further discussion is built around the structure, shown on Fig. 3.1, while special attention is given to the interconnection issues.

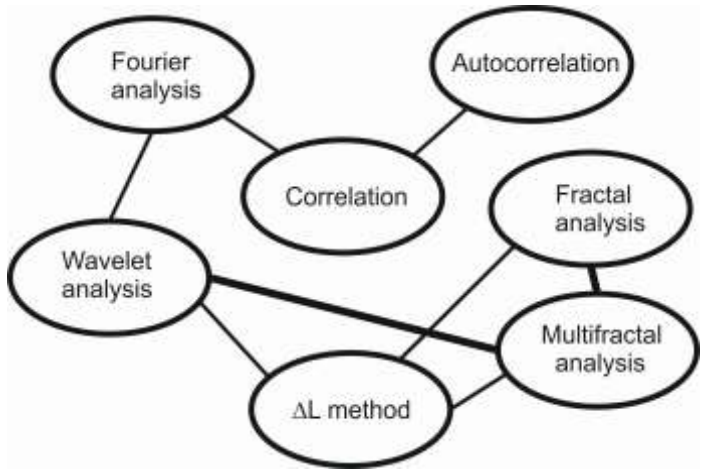

Fig. 3.1 – Interconnections between approaches to analysis of time series

## 3.1. Time series

A time series is a set of observed values, sorted in chronological order. Further on we are going to address discrete time series, in which values are recorded at regular time intervals. Let us denote such a time series as $x_1, x_2, ..., x_T$ or shortly $\{x_t\}_{t=1}^{T}$ assuming that series value are recorded at regular time intervals (with regularity $h$): $t_0, t_0 + h, t_0 + 2h, ..., t_0 + (T-1)h$.

If values of a time series are uniquely defined by some mathematical relation (such as, for instance, $x_t = A \cdot sin(vt)$), then the series is a determined one. If the time series values can be described only in terms of a probability distribution, then we are talking about a statistical time series; it is the time series of this type that will be addressed below. While analyzing a time series, we are going to consider it a realization of a stochastic process.

To provide examples, we are going to use three time series, obtained using a popular online service Google Trends. These time series reflect the level of interest to Donald Trump, Hilary Clinton, and "Russian hackers" between August 2016 and April 2017. Time series obtained using Google Trends display the popularity dynamics of a search query. The maximum value on the diagram equals 100 and represents the date when the query was most popular, while other points are defined as percentages of this maximum value. All three time series are shown on Fig. 3.2. In order to simplify further reference to the three time series, let us denote them as T (D. **T**rump), C (H. **C**linton), and H («Russian **h**ackers»).

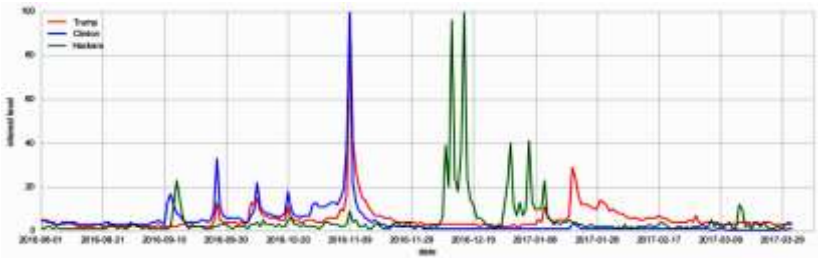

Fig. 3.2 – Time series reflecting interest to Donald Trump (T), Hilary Clinton (C), and «Russian hackers » (H) between August 1, 2016, and April 1, 2017, according to Google Trends

In some cases it is useful to consider a smoother version of the initial time series. The smoothing helps us to detect significant trends in the row dynamics, eliminating the noise and different peculiar features that manifest themselves under smaller scales. There are different smoothing methods. The simplest smoothing method is the calculation of a moving (sliding) average. Simple moving average equals the average mean of the series elements within the interval of a given length:

$$SMA_t = \frac{1}{w}\sum_{i=0}^{w-1} x_{t-i},$$

where $w$ is the width of the smoothing interval (the number of elements, across which the average mean is calculated), $SMA_t$ is the moving average value in point $t$.

The obtained $SMA_t$ value lies in the middle of smoothing interval, so the smoothed series $y_t$ can be defined as $y_t = SMA_{t+\left[\frac{w}{2}\right]}$.

If the moving average is used for smoothing, then the wider the smoothing interval is, the smoother function we will get. Fig. 3.3 shows how the smoothed series T looks under larger values of $w$.

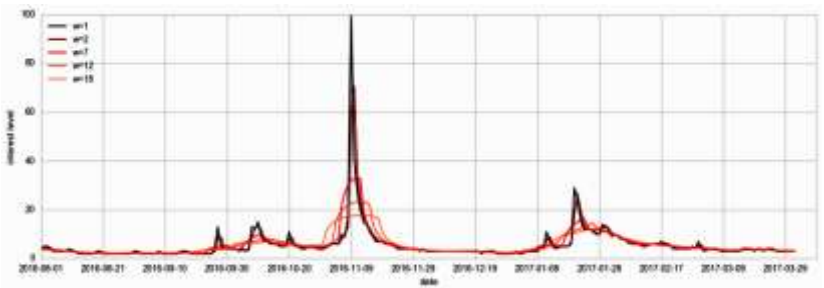

Fig. 3.3 – Initial time series T and the time series, smoothed by simple moving average under smoothing interval widths of 2, 7, 12, and 18

Results of time series smoothing can be demonstrated by a diagram, where the abscissa denotes the time axes, while the ordinate axes represents the smoothing interval width. The diagram shows the values of $y_t^{(w)}$, i.e. the smoothed series elements in point $t$ when the interval of width $w$ is used (Fig. 3.4).

When simple moving average is used, all points within the smoothing interval are assigned equal weights. Naturally, unequal weights might be used as well. In this case we will get the weighted moving average

$$WMA_t = \frac{1}{w} \sum_{i=0}^{w-1} a_i x_{t-i},$$

where

$$\sum_{i=0}^{w-1} a_i = 1.$$

Another frequently used method used for smoothing of time series is the ***exponential smoothing***. Previous values of the series are assigned exponentially decreasing weights. Let us denote the smoothed series elements as $y_t$ and assume that $y_0 = x_0$.

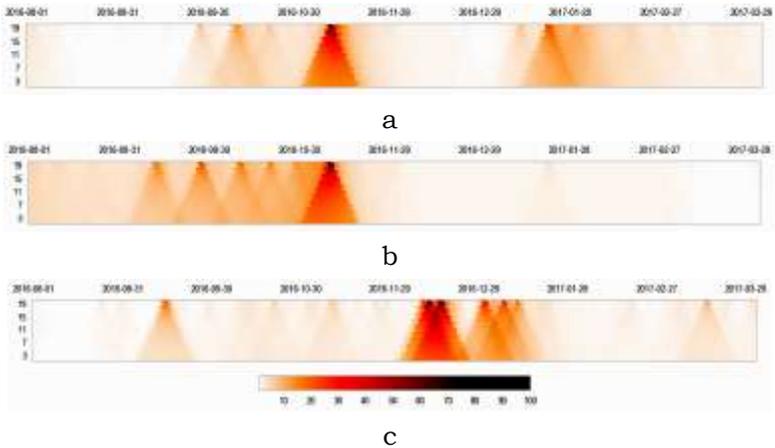

a

b

c

Fig. 3.4 – Values of time series T (a), C (b), and H (c), smoothed by simple moving average, depending on the smoothing interval width. Abscissa axes denotes the time while ordinate axes denotes the interval width

Next elements of the series $y_t$ can be obtained according to the recursive formula

$$y_t = \alpha x_t + (1-\alpha) y_{t-1},$$

where $0 < \alpha < 1$ is the smoothing coefficient. Naturally, under $\alpha = 1$ the new series $y_t$ coincides with the initial one $x_t$. Thus, if the value of $\alpha$ is close to 1 then, when $y_t$ is calculated, maximum weight is assigned to the respective $x_t$, while the previous series history does not play a significant role. On the other hand, if $\alpha$ equaled 0, the whole series $y_t$ would be smoothed to as single value $y_t = y_0$. That is, under $\alpha$ close to 0 the previous history of the series is considered with larger weights than the current value.

Fig. 3.5 shows the series T, as well as the respective smoothed series under different values of parameter $\alpha$.

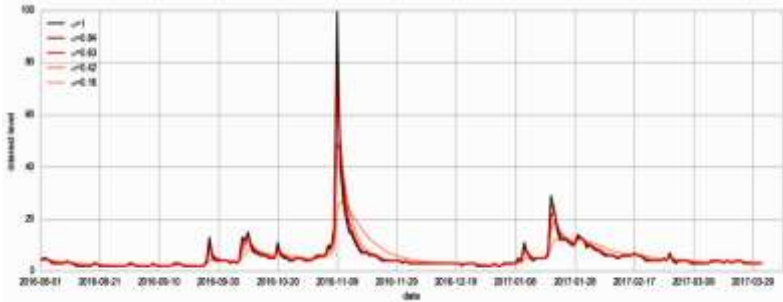

Fig. 3.5 – Initial series T and exponentially smoothed series under smoothing parameter values of 0.84, 0.63, 0.42, and 0.16

Just like in the case of simple moving average, let us display the results of the series smoothing on a diagram. In this case we will use the ordinate axes to represent parameter $\alpha$ (Fig. 3.6). The diagram shows the values of $y_t^{(\alpha)}$ (value of a smoothed series in point $t$ under smoothing parameter $\alpha$).

As examples we are considering time series T, C, and H that have a weekly regularity period. This is a characteristic feature of many processes in the information space. It is known that news reports are often published on a weekly basis, while user activity levels are different on weekdays and weekends.

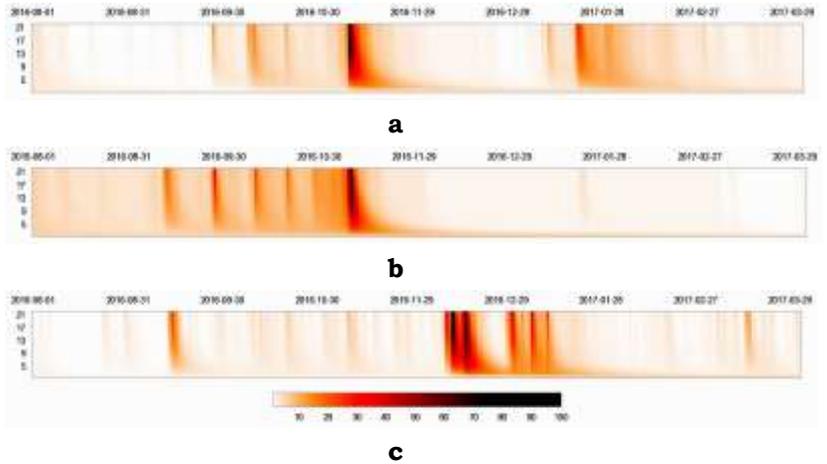

Fig. 3.6 – Values of exponentially smoothed time series T (a), C (b), and H (c), depending on parameter $\alpha$. Abscissa axes denotes the time; ordinate axes denotes parameter $\alpha$

In order to remove the periodical component from the series, let us smooth them using the simple moving average method with interval width of 7 (the number of days in a week) in accordance to the formula:

$$x_t^{New} = \frac{x_{t-3} + x_{t-2} + x_{t-1} + x_t + x_{t+1} + x_{t+2} + x_{t+3}}{7},$$

where $x_t$ denote the initial series values, and $x_t^{New}$ are new series values at the moment $t$. Fig. 3.7 shows the smoothed time series.

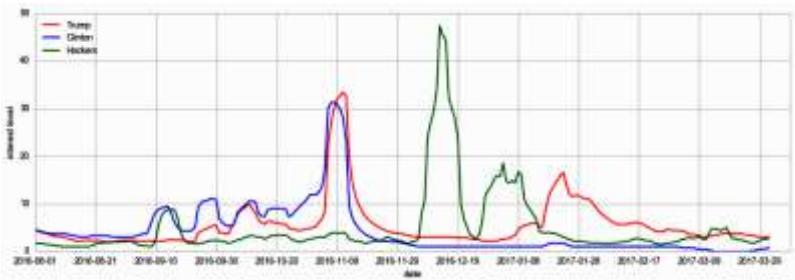

Fig. 3.7 – Time series T, C, and H, smoothed using simple moving average with the interval length 7

## 3.2. Correlation analysis

Many time series research methods are based on some assumption of statistical balance or permanence. One of such useful assumptions is the stationarity condition [Box 2015].

A time series is *strictly stationary or stationary in a narrow sense* if its statistical properties do not change with time. Formally, if the joint distribution of random variables $x_t, x_{t+1}, ..., x_{t+n}$ coincides with the distribution of $x_{t+k}, x_{t+k+1}, ..., x_{t+k+n}$ under any integer values of $k$, then the time series $\{x_t\}_{t=1}^T$ is strictly stationary. Stationary time series has a constant mathematical expectation (average mean)

$$\mu = Ex_t$$

and dispersion

$$\sigma^2 = Var(x_t) = E(x_t - Ex_t)^2.$$

Values of $\mu$ and $\sigma^2$ can be estimated as sample average

$$\hat{\mu} = \bar{x} = \frac{1}{T} \sum_{t=1}^T x_t \tag{1}$$

and sample dispersion

$$\hat{\sigma}^2 = \frac{1}{T} \sum_{t=1}^T (x_t - \bar{x})^2. \tag{2}$$

Stationarity property also plays an important role during comparison of time series. Linear dependence between two random variables is measured by co-variation (covariance). For time series a cross-covariance function is defined. By definition, *cross-covariance with delay $k$ between random (stochastic) processes $\{x_t\}_{t=1}^T$ and $\{y_t\}_{t=1}^T$ equals*

$$\gamma_{xy}(k,t) = Cov(x_t, y_{t+k}) = E\big[(x_t - \mu_x)(y_{t+k} - \mu_y)\big].$$

As a result of the assumption of stationarity in the narrow sense, distribution of pairs of variables $x_t$, $y_{t+k}$ is the same for any values of $t$. Consequently, co-variation between $x_t$ and $y_{t+k}$ does not depend on $t$, and depends only on the value of $k$, i.e. $\gamma_{xy}(k,t) = \gamma_{xy}(k), \forall t$. The set of values $\{\gamma_{xy}(k)\}$ forms a cross-covariance function.

Having normalized the cross-covariance coefficient, we will get the ***cross-correlation coefficient***

$$\rho_{xy}(k) = \frac{Cov(x_t, y_{t+k})}{\sigma_x \sigma_y} = \frac{\gamma_{xy}(k)}{\sigma_x \sigma_y}.$$

Cross-correlation function is the measure of similarity between two time series.

Most often, cross-covariance and cross-correlation coefficients are estimated according to the formulas

$$\hat{\gamma}_{xy}(k) = \frac{1}{T} \sum_{t=1}^{T-k} (x_t - \bar{x})(y_{t+k} - \bar{y}), \qquad \hat{\rho}_{xy}(k) = \frac{\hat{\gamma}_{xy}(k)}{\hat{\gamma}_{xy}(0)}.$$

We should note that such estimates apply to series that are stationary in the narrow sense, as the respective coefficients do not depend on time, while in the general case this may be untrue. Often a weaker requirement is applied – stationarity in the broad sense.

*A time series $\{x_t\}_{t=1}^{T}$ is **stationary in the broad sense**, if its mathematical expectation does not change with time, i.e. $\forall t \exists Ex_t = const$ and covariance function depends only on the difference between arguments $Cov(x_t, x_s) = K(t-s)$.*

As the definition states, mathematical expectation is constant, and it is easy to see that dispersion also does not change with time $Var(x_t) = Cov(x_t, x_t) = K(0) = const$, estimates (1) and (2) apply, just as in the case of strictly stationary series.

Fig. 3.8 illustrates correlation calculation. Two centered time series are considered. In order to calculate the correlation coefficient, we should multiply respective elements of the series and calculate their average value. Multiplication (product) result is marked on Fig. 3.8 with a line. The square of darkened region under the line (with sign taken into consideration) equals the covariance between the two series.

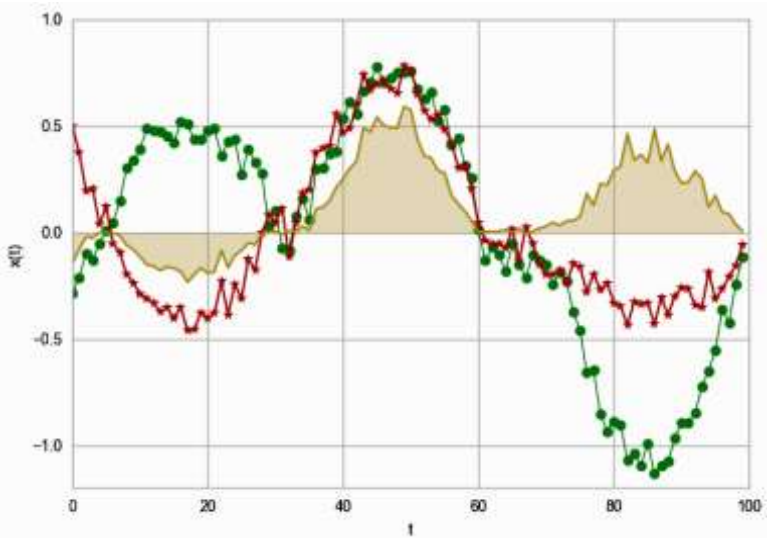

Fig. 3.8 – Illustration of correlation definition. Two time series are displayed. The darkened region indicates the contribution into the value of correlation between the two series.

To provide an example, let us calculate an estimate of cross-correlation functions for series T, C, and H. Fig. 3.9 shows the correlation function for series T and C. Maximum value (of approximately 0.8) is reached under delay 0. That is, the two series, representing interest to Donald Trump and Hillary Clinton, are strongly correlated.

Fig. 3.9b illustrates correlation function for series T and H. Maximum value (of approximately 0.7) is reached under delay of 34 days. This reflects the fact that, starting from December 13, 2016 (34 days after the election in the USA, held on November 8) the number of news reports about "Russian hackers" significantly increased.

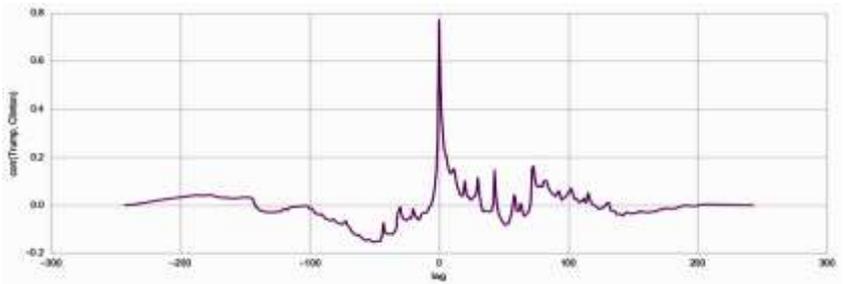

a

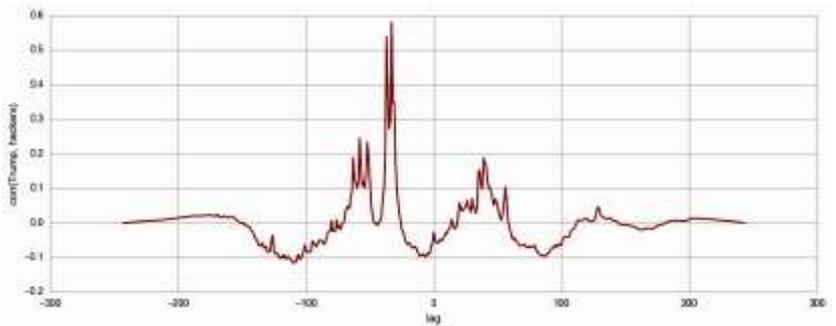

b

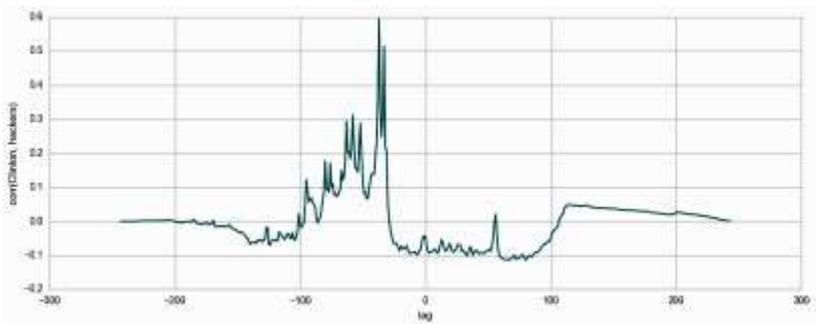

c

Fig. 3.9 – Correlation functions for pairs of series T and C (a), T and H (b), C and H (c). The abscissa axis shows the time delay (lag), while the ordinate axis shows the estimate of correlation coefficient

### *Autocorrelation*

We can calculate covariance for a single time series as well, not for two different series. Such covariance is called **auto-covariance** with delay or lag $k$

$$\gamma_k = Cov(x_t, x_{t+k}) = E[(x_t - \mu)(x_{t+k} - \mu)].$$

A set of values $\gamma_k$, $k = 0, 1, 2, \ldots$ is called an auto-covariance function and their normalized value $\rho_k$, $k = 0, 1, 2, \ldots$ – an **autocorrelation function**

$$\rho_k = \frac{E[(x_t - \mu)(x_{t+k} - \mu)]}{\sqrt{E(x_t - \mu)^2 E(x_{t+k} - \mu)^2}} = \frac{Cov(x_t, x_{t+k})}{Var(x_t)} = \frac{\gamma_k}{\gamma_0}.$$

An autocorrelation function describes the dependence between the values of a stochastic process in different moments of time (Fig. 3.10).

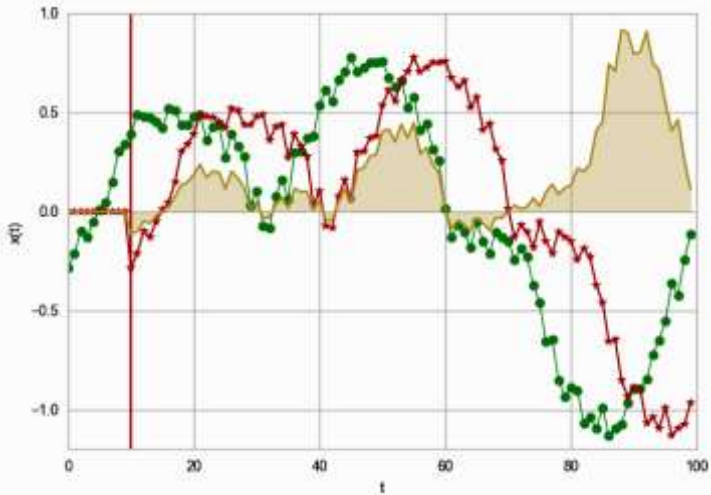

Fig. 3.10 – An illustration to definition of auto-correlation. One and the same time series is displayed, while its values are shifted by 10 units to the right. The darkened area indicates the contribution into the autocorrelation coefficient value with lag 10

Most often, auto-covariance and autocorrelation coefficients are estimated according to formulas

$$\hat{\gamma}_k = \frac{1}{T} \sum_{t=1}^{T-k} (x_t - \bar{x})(x_{t+k} - \bar{x}), \qquad \hat{\rho}_k = \frac{\hat{\gamma}_k}{\hat{\gamma}_0}.$$

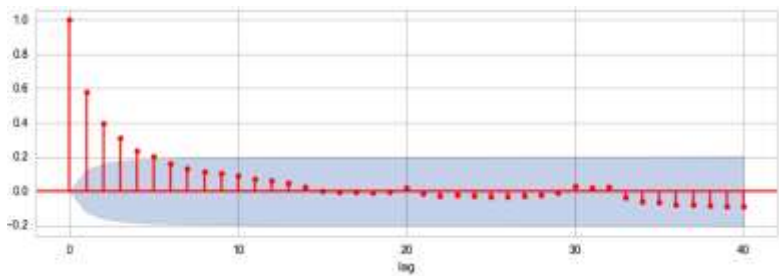

**a**

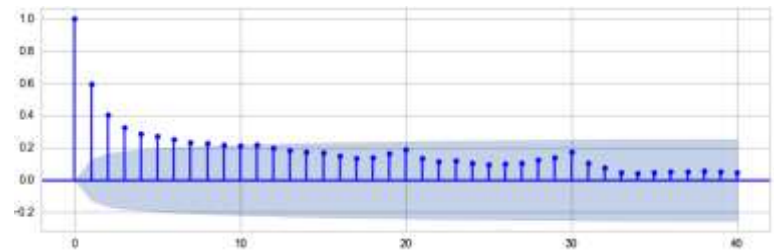

**b**

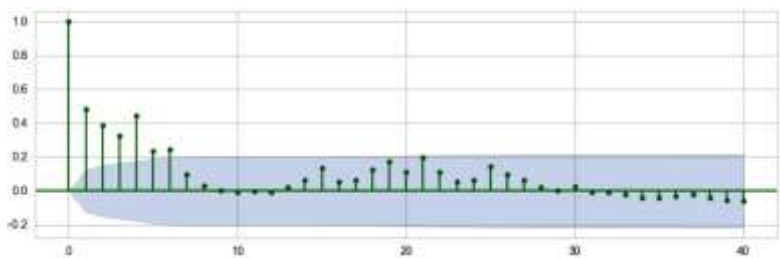

**c**

Fig. 3.11 – Autocorrelation functions for series T (a), C (b), and H (c). Abscissa axis indicates the time delay (lag), while the ordinate axis shows the autocorrelation

coefficient. The darkened area indicates standard deviation for estimation of the autocorrelation coefficient

When the estimates of autocorrelation coefficients are calculated, a question arises: do coefficients $\rho_k$ equal zero starting from some value of $k$? In order to answer this question, we need to compare the value of the estimate $\hat{\rho}_k$ with its standard deviation. If we assume that $\rho_k = 0$, then the standard deviation of the estimate $\hat{\rho}_k$ is

$$se(\hat{\rho}_k) \cong \frac{1}{\sqrt{T}}.$$

In practice an empirical rule is often used. According to the rule, autocorrelation coefficients are estimated for delays, not exceeding $T/4$. Autocorrelation functions for series T, C, and H, are shown on Fig 3.11.

Definition of autocorrelation function was introduced for stationary time series, however, its value can be estimated for any time series. For non-stationary time series such autocorrelation function is decreasing very slowly.

## 3.3. Fourier analysis

The classical Fourier analysis presents an opportunity to study a function in time and frequency domains. The essence of transition to frequency domain is that the function is expanded into a series of components, representing harmonic oscillations with different frequencies. Each frequency is assigned a coefficient representing the amplitude of oscillations of the given frequency. If we represent the function graphically in time domain, we can get information on how it changes with time. If we display the function in frequency domain, we will get information on frequencies of oscillations it contains. For these purposes direct and inverse Fourier transforms are used:

$$\hat{x}(\nu) = \int\limits_{-\infty}^{\infty} x(t)e^{-i2\pi\nu t}dt,$$

$$x(t) = \int\limits_{-\infty}^{\infty} \hat{x}(\nu)e^{i2\pi\nu t}d\nu.$$

Figure 3.12a shows an example of a function that is, in fact, a sum of three sinusoids with different periods. It is rather difficult to realize that the function consists of three harmonic oscillations and define their periods just by looking at the diagram in time domain. Figure 3.12b shows Fourier transform for the function. This diagram clearly indicates that the function contains oscillations of three different frequencies.

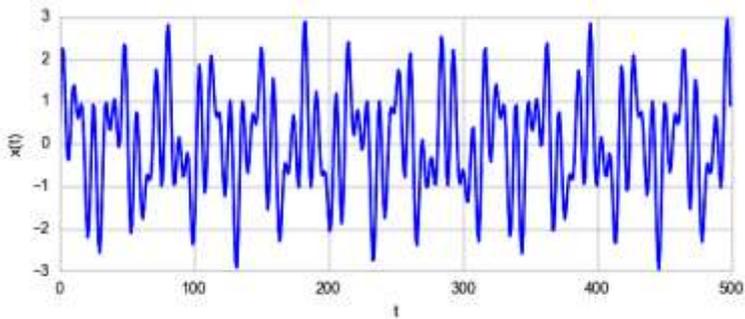

a

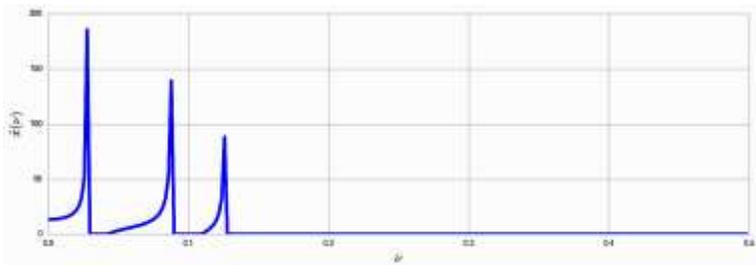

b

Fig. 3.12 – Function, representing the sum of three sinusoids with different periods (a) and estimated Fourier spectrum for this function (b)

Today Fourier transform and Fourier spectrums are applied in different contexts in machine learning systems. Fourier spectrums are often used as learning parameters. For instance, [Rodrigues 2014] suggested a model of time series prediction in which Fourier spectrum and some other values were used as input parameters of a neuronal network.

Fourier transforms and spectrums are often used for speech recognition. In [Alam 2014] specific indicators, based on Fourier transform, are used in the system of speech recognition with different learning conditions. Fourier spectrums are also used as learning parameters for neuronal networks in the systems of automated detection of certain events within speech or against noise background [Sazonov 2010, Wang 2014]. Another recognition-related problem is definition of emotional tonality of speech. [Wang 2015] suggested a model based on certain Fourier parameters, and demonstrated efficiency of usage of these parameters for identification of different emotional states in voice signals.

In machine learning algorithms based on kernel usage, such as support vector machine, random Fourier features are often used for approximation of kernels of high dimensionality. Such an approach was suggested in [Rahimi 2008] and based on Bochner's theorem from harmonic analysis. It guarantees that under some kernel properties, its Fourier transform is a probability distribution.

Fourier transform can be considered as definition of correlation between initial signal and harmonic functions with different oscillation frequencies. Fig. 3.13 shows an illustration, similar to Fig. 3.8 and Fig. 3.10.

In spite of its advantages and numerous applications, Fourier transform is a bad method if you need to study functions that evolve with time. These functions require a way of spectrum estimation not across all time series length, but across its specific parts. An example of such an approach is the window Gabor transform:

$$G(v, \tau, s) = \int\limits_{-\infty}^{\infty} x(t)e^{-\frac{(t-\tau)^2}{s^2}}e^{-i2\pi vt}dt.$$

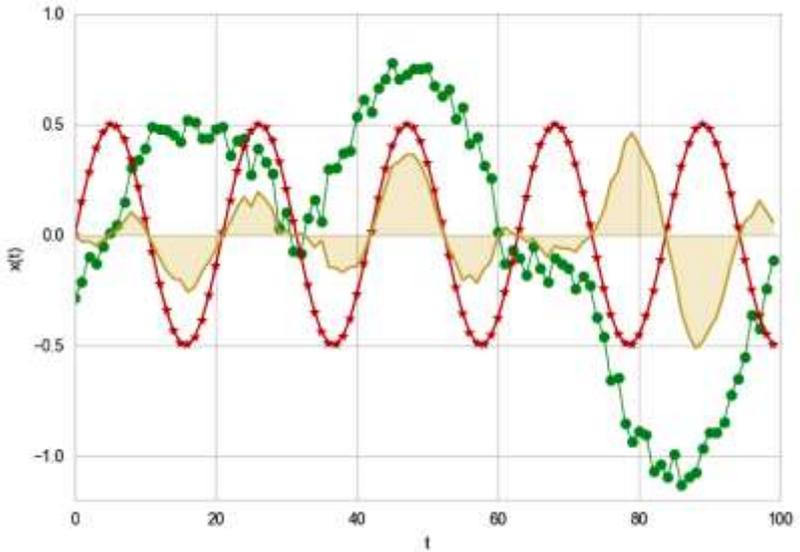

Fig. 3.13 – An illustration for Fourier transform definition as calculation of correlation between the initial signal and harmonic oscillation. The darkened area indicates the contribution of Fourier transform (or the amplitude of the oscillation of the given frequency) into the value

Time window $e^{-\frac{(t-\tau)^2}{s^2}}$ selects the fragment of the time series with a centre $\tau$ and the width of $s$, allowing us to single out a part of kernel under consideration.

When Gabor transform is used the problem of window width selection becomes relevant. Another class of transforms, namely, wavelet transforms, allows us to make window function frequency-dependant, so that the window could become wider for low frequencies and narrower for high frequencies. The key advantage of a wavelet transform is that the selected time series fragment is analyzed with the degree of detail, suitable for its scale.

## 3.4. Wavelet analysis

Wavelet transform has correlative nature. In this case we are considering correlation between an initial function and the wavelet function in different scales. In order for the procedure to be applicable and for correlation coefficients to be informative, the wavelet must have certain mathematical properties. Literally, the word "wavelet" means a small wave, so, as the name suggests, it should be easily localizable in time. From mathematical viewpoint, a wavelet is a function $\psi(t)$ that satisfies the following conditions:

1. $\psi(t)$ is square-integrable ($\psi \in L^2(\mathbb{R})$) or, in other words, it has finite energy

$$E = \int\limits_{-\infty}^{\infty} |\psi(t)|^2 dt < \infty.$$

2. Let us denote $\hat{\psi}(\lambda)$ as Fourier transform of function $\psi(t)$, then

$$\int\limits_{0}^{\infty} \frac{|\hat{\psi}(\lambda)|^2}{\lambda} d\lambda < \infty.$$

Fig. 3.14 shows examples of wavelets that are often used in practice.

### Continuous wavelet transform

Wavelet $\psi(t)$ properties of which have been described above is often called mother or basic wavelet. Mother wavelet provides the basis for construction of a family of functions through extension or compression and parallel shift. These operations are necessary for studying different areas of the initial signal with varying degree of detail.

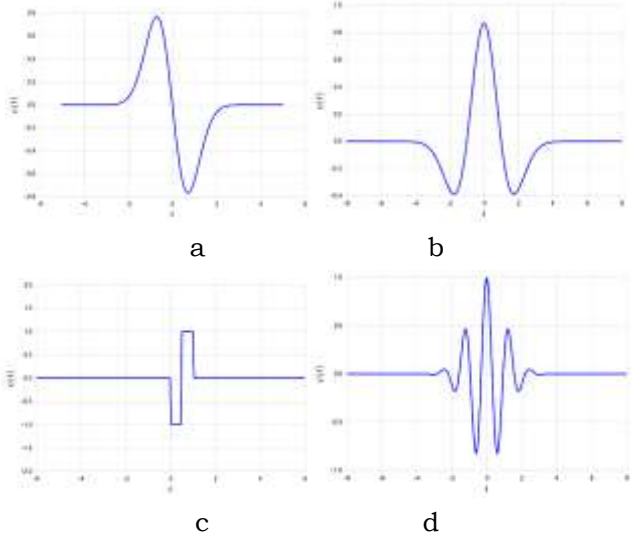

<div style="text-align:center">a       b</div>

<div style="text-align:center">c       d</div>

Fig. 3.14 – Examples of wavelets that are often used in applications: (a) Gaussian wave (first derivative of Gaussian function), (b) Mexican hat, (c) Haar wavelet, (d) Morlet wavelet (real part).

Let us introduce parameters $s$ (scale) and $l$ (location); then we will get the following transformed version of mother wavelet

$$\psi_{s,l}(t) = \frac{1}{\sqrt{|s|}}\psi\left(\frac{t-l}{s}\right).$$

**Continuous wavelet transform** of function $x(t) \in L^2(\mathbb{R})$ is the expression

$$W(s,l) = \frac{1}{\sqrt{|s|}}\int\limits_{-\infty}^{\infty} x(t)\psi^*\left(\frac{t-l}{s}\right)dt = \int\limits_{-\infty}^{\infty} x(t)\psi_{s,l}^*(t)\,dt,$$

where $l, s \in \mathbb{R}$, $s \neq 0$; $\psi^*$ is a function that is complex conjugate to $\psi$, and variables $\{W(s,l)\}_{l,s \in \mathbb{R}}$ are called coefficients of wavelet transform.

The formula for definition of continuous wavelet transform clearly indicates that the essence of such transform is calculation of correlation coefficients of a specific kind. Fig. 3.15 shows how Mexican hat wavelet is superposed against the initial time series, and correlation between the part of the series and the "template", represented by the wavelet, is defined.

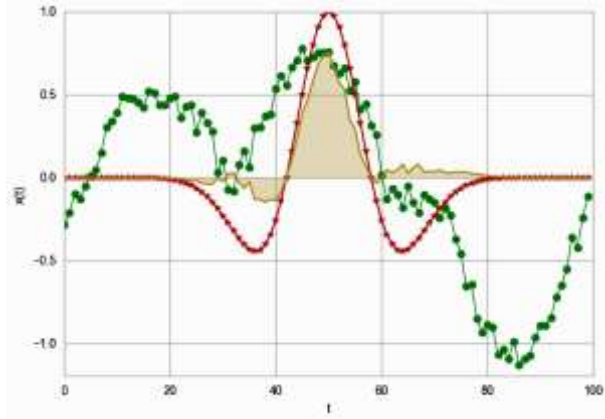

Fig. 3.15 – An illustration to calculation of wavelet transform as correlation between the initial signal and the wavelet function. Darkened area indicates the contribution into the wavelet transform value for a given location and scale

We should note that continuous wavelet transform is a reversible operation. Inverse wavelet transform is performed as follows:

$$x(t) = \frac{1}{C_g} \int\limits_{-\infty}^{\infty} \int\limits_{0}^{\infty} W_x(s,l)\psi_{s,l}(t)\frac{dsdl}{s^2}.$$

Let us illustrate the results of wavelet transform on several simple examples (Fig. 3.16). The first example is a sum of two oscillation processes. Mexican hat wavelet transform coefficients are shown under the graph. Time changes along the horizontal axis (location parameter $l$), while the scale changes along vertical axis (parameter $s$). Two periodic processes can be seen on the wavelet coefficient graph. On Fig. 3.16 we can see how periodic processes with different amplitudes and frequencies, as well as separate signal peaks are reflected by wavelet coefficient values.

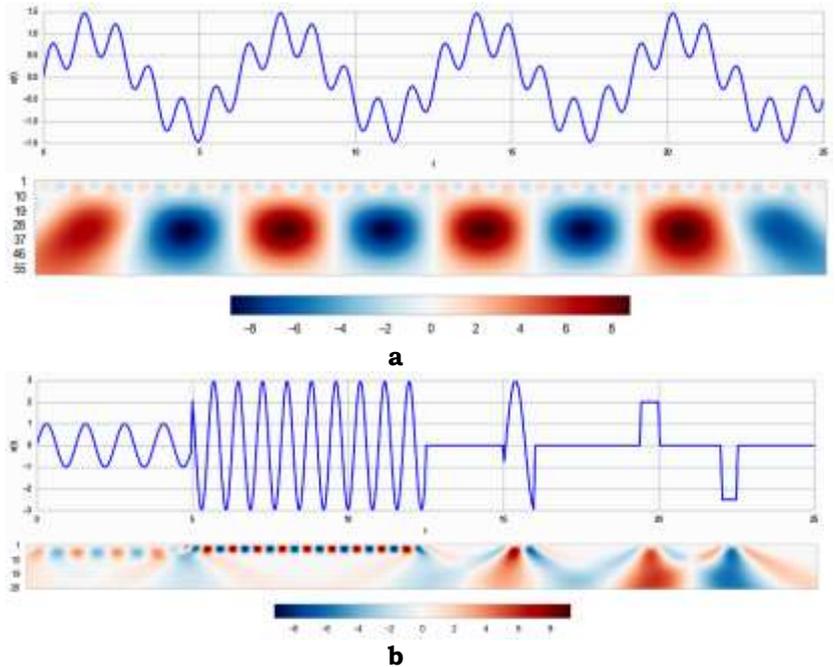

Fig. 3.16 – Wavelet transform examples. A sum of two sinusoids and its wavelet transform (a). A function comprised of oscillation processes with different frequencies and amplitudes as well as separate peaks (b).

Let us take a look at wavelet transform coefficients for time series T, C, H. Fig. 3.17a, b show the wavelet transform results for time series T obtained using Mexican hat wavelet (a) and Gaussian wave (b). Fig. 3.17c and d indicate the wavelet transform results for time series C and H, obtained using Mexican hat wavelet.

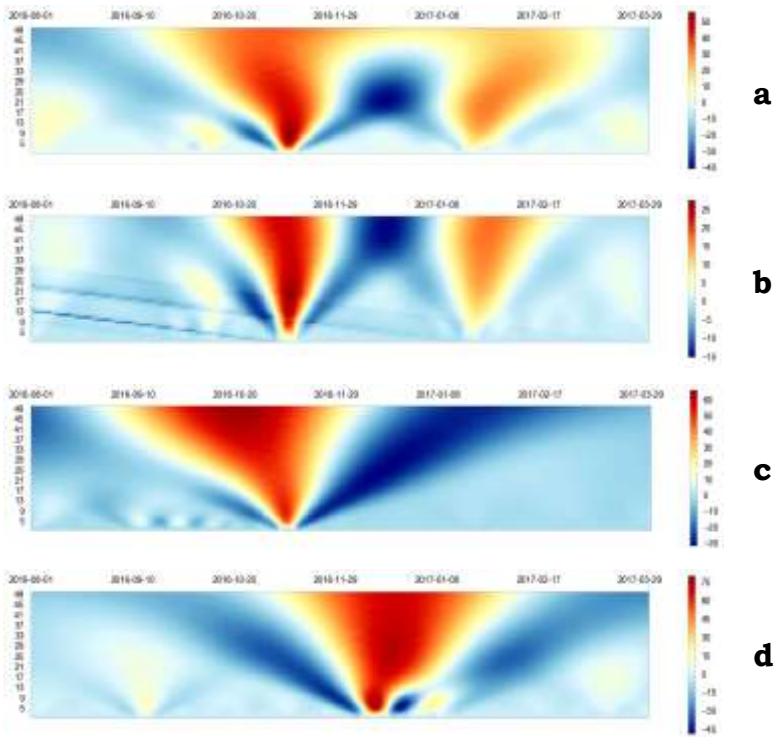

Fig. 3.17 – Wavelet transform for series T using Mexican hat wavelet (a); real part of wavelet transform for series T using Morlet wavelet (b); wavelet transform for series C (c) and series H (d) using Mexican hat wavelet

Let us stress that both continuous wavelet transform and Fourier transform can be considered in correlation terms. Fourier transform is the correlation between the initial time series and the wave $\varphi(t) = e^{-i2\pi\nu t}$. The wave covers the whole time axis and is characterized only by frequency $\nu$, so Fourier transform depends only on frequency. Wavelet transform is the correlation between the initial time series and the wavelet $\psi(t)$. Thus, wavelet transform depends only on the wavelet's position on the time axis and on its scale, which are determined by parameters $l$ and $s$, respectively.

### *Energy of the signal*

By definition, the total energy of the signal $x(t)$ equals

$$E = \int\limits_{-\infty}^{\infty} |x(t)|^2 dt.$$

Using wavelet transform coefficients, we can calculate the energy of the signal for given location and scale

$$E(s, l) = |W(s, l)|^2.$$

Values $E(s, l)$ can be shown on a diagram just like wavelet transform coefficients. Such a diagram is, usually, called a scalogram. Besides that, we can also define the relative contribution of energy that corresponds to a specific scale, into the total energy, or, in other words, energy distribution by scale

$$E(s) = \frac{1}{C_g} \int\limits_{-\infty}^{\infty} |W(s, l)|^2 dl.$$

We should note that, in contrast to a diagram (graph) with wavelet transform coefficients, on a scalogram all values are positive, while the areas with maximum energy are most distinctive (Fig. 3.18).

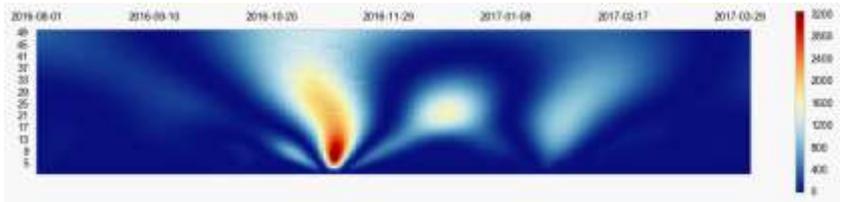

Fig. 3.18 – A scalogram for time series T

### *Comparing time series with the help of wavelet transform*

Above, we demonstrated a way to detect dependence between two time series based on cross-correlation. Now we are going to consider some ways of time series comparisons based on wavelet transform coefficients.

These ways can also be applied for detection of a certain type of relation or interconnection between time series. Metrics for comparing wavelet transform coefficients, as well as examples of their application to real practical problems are described in detail in [Addison 2017].

Let us consider two time series $x_t$ and $y_t$, and denote wavelet transform coefficients of these series as $W_x(s, l)$ and $W_y(s, l)$. Let us start with the simplest way of comparison, i.e. the difference between absolute values of the respective coefficients

$$DiffMOD_{x,y}(s, l) = |W_x(s, l)| - |W_y(s, l)|.$$

On Fig. 3.19 the values of $DiffMOD_{x,y}(s, l)$ for two pairs of time series are shown (for time series T and C above, and for time series T and H below). This simple procedure allows us to select the areas in which wavelet transform coefficients are similar, and, consequently, the initial time series contain similar fragments.

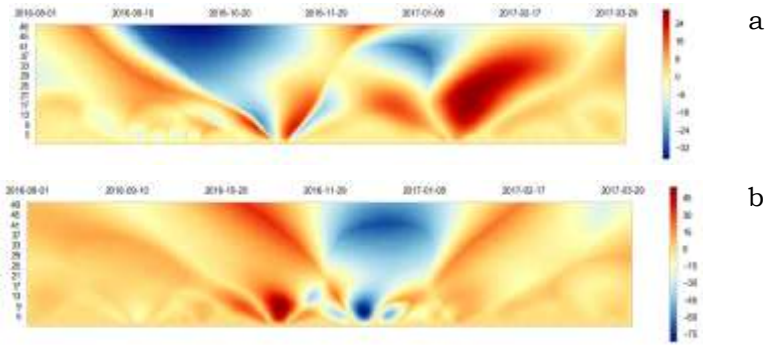

Fig. 3.19 –$\boldsymbol{DiffMOD_{x,y}(s, l)}$ for series T and C (a), T and H (b)

Another simple way of comparison is the ratio of absolute values of wavelet transform coefficients.

$$RatioMOD_{x,y}(s, l) = \frac{|W_x(s, l)|}{|W_y(s, l)|}.$$

Such a metric should be used with care, as $W_y(s,l)$ can assume values equal or close to zero.

Additional information can be obtained if we use a complex wavelet (for instance, Morlet wavelet). Then, beside absolute values of wavelet coefficients, we will have phase. A complex wavelet can be always represented in the following way

$$W(s,l) = |W(s,l)|e^{i\phi(s,l)},$$

where, as we know, the absolute value equals

$$|W(s,l)| = \sqrt{\text{Re}\big(W(s,l)\big)^2 + \text{Im}\big(W(s,l)\big)^2},$$

and phase equals

$$\phi(s,l) = \tan^{-1}\left[\frac{\text{Im}\big(W(s,l)\big)}{\text{Re}\big(W(s,l)\big)}\right].$$

Consequently, we can also compare phases of coefficients

$$\Delta\phi_{x,y}(s,l) = \phi_x(s,l) - \phi_y(s,l)$$

Cross-wavelet transform is used to single out areas with the same energy between signals in the transform area, and to determine the relative phase

$$CrWT_{x,y}(s,l) = W_x^*(s,l)W_y(s,l).$$

On figures they, usually, display the value $\left|CrWT_{x,y}(s,l)\right|$, similarly to scalogram. In such a case, if time series $x$ is identical to $y$, we will get the scalogram for series $x$.

Calculation of cross-wavelet transform represents particular interest, when a complex wavelet (such as Morlet wavelet) is used. In such a case

$$CrWT_{x,y}(s,l) = W_x^*(s,l)W_y(s,l)$$
$$= |W_x(s,l)|e^{-i\phi_x(s,l)}|W_y(s,l)|e^{i\phi_y(s,l)} =$$
$$= |W_x(s,l)||W_y(s,l)|e^{i(\phi_y(s,l)-\phi_x(s,l))}.$$

Thus, through cross-wavelet transform calculation we can determine the difference of phases between coefficients of wavelet transforms for two time series.

Fig. 3.20 illustrates the values of $CrWT_{x,y}(s,l)$ for series T and C. For series T and C we select an area that represents the peak of interest during elections, so this is the high-energy area for both time series.

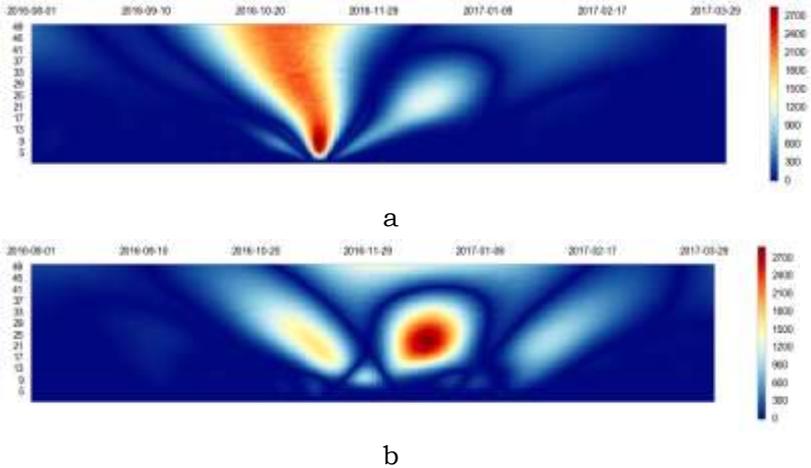

a

b

Fig. 3.20 – Cross-wavelet transform using Mexican hat wavelet for series T and C (a) and for series T and H (b)

If we integrate wavelet transform coefficients by time, we will get Wavelet Cross-Correlation Measure that depends on scale

$$W_{x,y}(s) = \frac{\left|\int W_x^*(s,l) W_y(s,l) dl\right|}{\sqrt{\int |W_x(s,l)|^2 dl \int \left|W_y(s,l)\right|^2 dl}}$$

$$= \frac{\left|\int CrWT_{x,y}(s,l) dl\right|}{\sqrt{\int |W_x(s,l)|^2 dl \int \left|W_y(s,l)\right|^2 dl}}.$$

Such a measure allows us to detect correlation between signals that contain oscillations with different amplitudes or phases, but are still correlated with each other (3.21).

Besides, wavelet cross-correlation measure definition can be extended if we introduce dependence on location shift between series

$$W_{x,y}(s,k) = \frac{\left|\int W_x^*(s,l) W_y(s,l-k) dl\right|}{\sqrt{\int |W_x(s,l)|^2 dl \int \left|W_y(s,l)\right|^2 dl}}$$

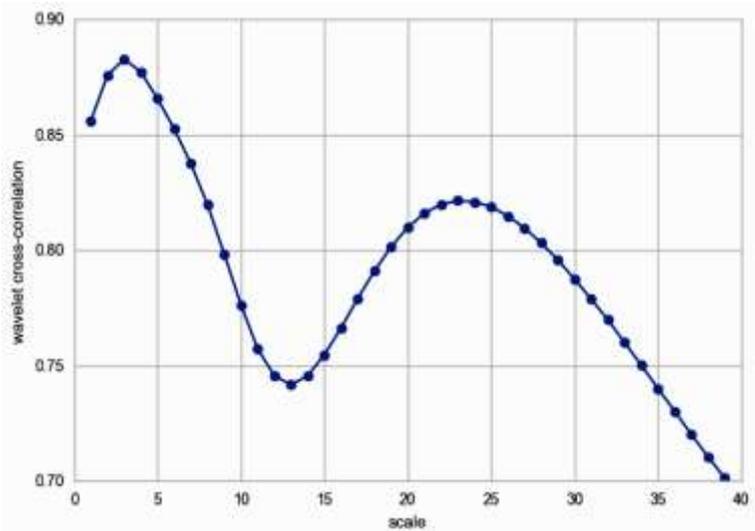

Fig. 3.21 – Dependence of wavelet cross-correlation measure on scale for time series T an C

In order to study the connection between time series components locally on the transform plane we can define square estimate of **wavelet coherence** as follows:

$$WCH_{x,y}^2(s,l) = \frac{\left|\langle W_x^*(s,l)W_y(s,l)\rangle\right|^2}{\langle|W_x(s,l)|^2\rangle\langle\left|W_y(s,l)\right|^2\rangle},$$

where $\langle\cdot\rangle$ defines a local smoothing operation in terms of both time and scale, while the smoothing takes place on transform components.

Cross-wavelet analysis methods are used for research of properties of several time series that depend on each other in some non-trivial way. For instance, in geophysics there is a problem of detecting cause-and-effect connection or correlation between meteorological and other natural events that happen far away from each other. In [Maraun, 2004] specific features of cross-wavelet transform application and wavelet coherence estimation for two-dimensional time series of this kind are analyzed. Besides that, wavelet analysis methods, including cross-wavelet transform, were used in [Adamowski, 2008] for studying meteorological time series and data on river flow levels. Components were selected from the time series of two types to be further used in the model for flood forecasting. The paper indicates that application of cross-wavelet analysis makes sense when there is a relatively stable phase shift between flow and meteorological time series. Based on cross-wavelet transform, the difference of phases between flow and meteorological data was determined; as a result the flood forecasting model was improved.

Another example is [Labat, 2010], where cross-wavelet analysis was conducted for climate indices and indicators of fresh water discharge in Africa. In this case cross-wavelet transform and coherence estimation were used to visualize and analyze periodic fluctuations in data within the time slot of 2-8 years. [Kelly, 2003] confirms that cross-wavelet analysis-based methods can be an efficient instrument for detection of quasi-periodic nature of a time series.

Cross-wavelet analysis methods also found their application in healthcare. A paper by [Li, 2007] describes application of cross-wavelet transform, coherence estimation, and some other wavelet-based methods for studies of dynamics of mutual impacts between oscillations, generated by two different neuron groups. Research results can be used for analysis and quantitative definition of terminal interaction between neural oscillators, as well as for studying the mechanisms of epilepsy.

[Aguiar-Conraria, 2008] uses cross-wavelet analysis instruments to show that connection between monetary policy variables and macro-economic variables changes with time, and these changes are not homogenous for different frequencies.

Data obtained using cross-wavelet transform can also be used as initial data for classification algorithms. In [Dey, 2010] cross-wavelet transform coefficients were input into an artificial neural network and Fuzzy classifier.

## *Discrete wavelet transform and approximation of functions by time series*

Discrete wavelet transform is defined in such a way that the initial signal could be thoroughly reconstructed using infinite sums of discrete wavelet coefficients. Such an approach provides an opportunity for fast calculation of the wavelet transform and its inverse [Addison, 2017].

Let $x(t)$ belong to space of $2\pi$ −periodic square-integrable functions. Then $x(t)$ can be represented as Fourier series

$$x(t) = \sum_{n=-\infty}^{\infty} c_n e^{int},$$

where coefficients $c_n$ are calculated as follows

$$c_n = \frac{1}{2\pi} \int_0^{2\pi} x(t)e^{-int}dt.$$

The set of functions $w_n(t) = e^{int}$ represents an orthonormal basis in the space $L^2(0, 2\pi)$, constructed using scale transform $w_n(t) = w(nt)$ from the basic function $w(t) = e^{int}$.

Now let us assume that $x \in L^2(\mathbb{R})$. The basic function in the space $L^2(\mathbb{R})$ should be a function that is rapidly decreasing towards 0 on $\pm\infty$. That is why for construction of basis they use wavelets – well localized soliton-like functions. In order to cover all the real axis with wavelets they use shifts along the axis. For the sake of simplicity integer shifts $k$ can be used, as well as analogues of sinusoid frequency, as powers of 2: $\psi_{jk} = \psi(2^j t - k)$. Wavelet $\psi \in L^2(\mathbb{R})$ is called orthogonal if the family of functions $\{\psi_{jk}\}$ forms an orthonormal basis in $L^2(\mathbb{R})$.

## 3.5. Correlation with a pattern

Using a continuous wavelet we can detect the most "wavelet-like" parts of the time series under consideration (Fig. 3.22). The idea is to compare parts of the series with a certain pattern under different scales (Fig. 3.23). At the same time, wavelet as a function should have certain mathematical properties, such as quick decrease towards zero when argument tends to infinity. In some cases it is useful to apply the pattern that does not meet wavelet-specific requirements. For this purpose, instead of the wavelet transform, let us calculate correlation between the part of the time series and some pattern $p$

$$C(l,k) = \frac{\sum_{i=1}^{k}(x_{l+i} - \bar{x})(p_i - \bar{p})}{\sqrt{\sum_{i=1}^{k}(x_{l+i} - \bar{x})^2 \sum_{i=1}^{k}(p_i - \bar{p})^2}},$$

The obtained coefficient $C(l,k)$ depends on values $x_{l+1}, ..., x_{l+k}$. That is, parameter $l$ reflects location of the pattern, while parameter $k$ reflects the number of points in the pattern and the part of the series under consideration. In this case, parameter $k$ is the analogue of scale $s$ that was used in wavelet transform.

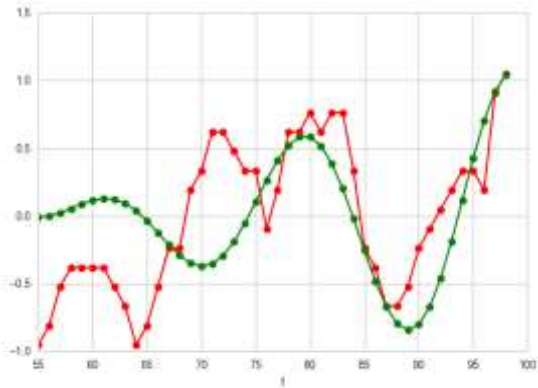

Fig. 3.22 – Part of the time series superposed against the pattern

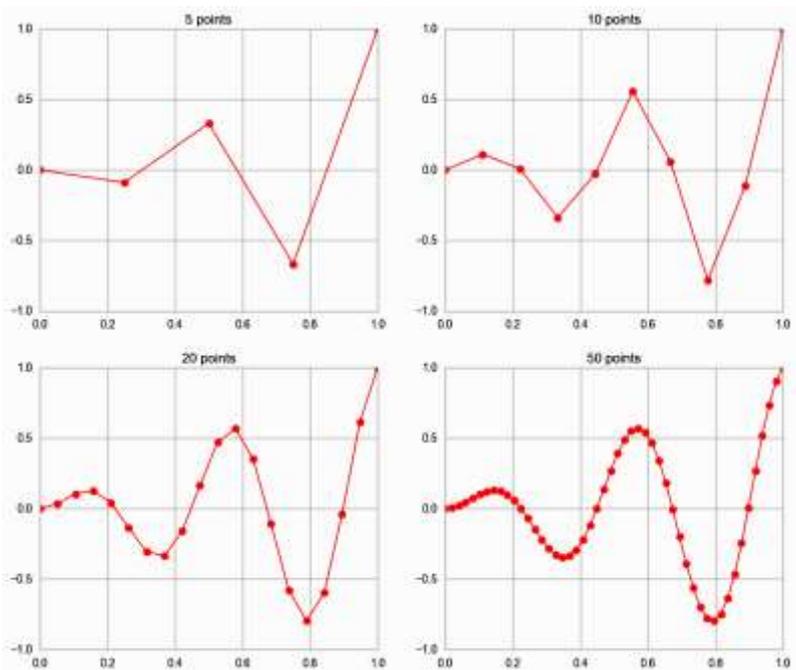

Fig. 3.23 – "Snake" pattern with different numbers of points

While during wavelet transform coefficient calculation we have been constantly using the whole time series, in this case, in order to calculate $C(l,k)$ only $k$ points from the time series and the pattern of length $k$ are used.

Let us display the obtained correlation coefficients $C(l,k)$ on a graph, similar to a scalogram (Fig. 3.24).

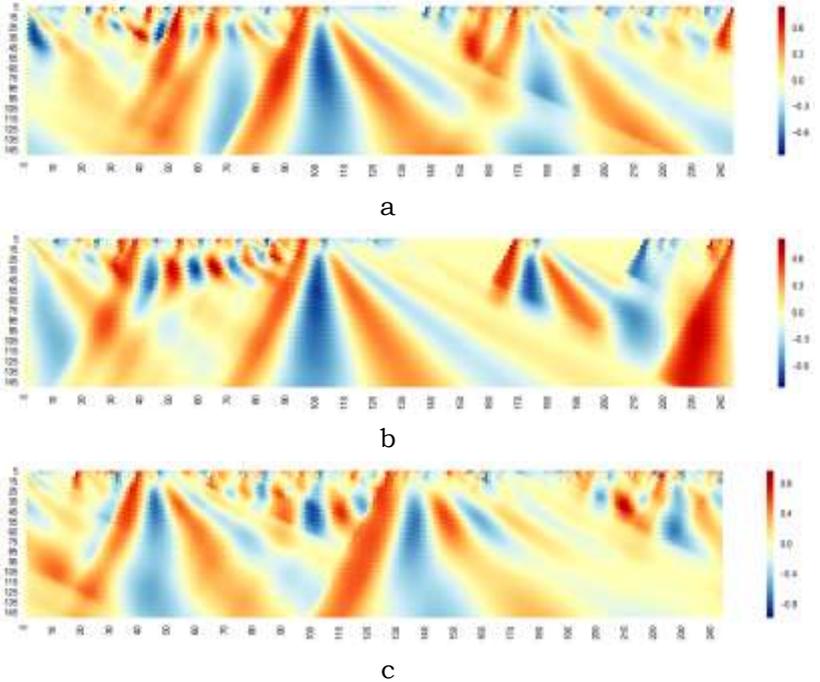

a

b

c

Fig. 3.24 – Correlation coefficients $C(l,k)$ calculated for series T (a), C (b), and H (c), using the template, shown on Fig 3.22

## 3.6. Fractal analysis

The term "fractal" was introduced and popularized by Benoit Mandelbrot. Most often fractals are defined as geometrical objects that have strongly jagged (indented) shape the property of self-similarity.

At present, there is no strict and generally acknowledged definition of a fractal, although Benoit

Mandelbrot used several trial definitions. One of them, introduced in [Mandelbrot, 1982], sounds as follows:

A fractal is a set whose Hausdorff-Besicovitch dimensionality strictly exceeds its topological dimensionality.

The strict definition of Hausdorff-Besicovitch dimensionality will be introduced later. The essence of this definition, however, is to select a class of strongly jagged objects, for description of which it is not enough to use just topological dimensionality. For instance, there are curves whose topological dimensionality always equals 1, but they are curved in such a complex way that they fill a plane or a space. For instance, Peano curves pass through every point of a unit square. Another example is the trajectory of a Brownian particle that is not smooth in any point.

The first definition, while being strict, does not cover many physical fractals, and, consequently, is not used. So, another fractal definition was suggested:

A fractal is a structure, consisting of parts that are, in some way, similar to the whole.

The second definition focuses on self-similarity as a distinguishing feature of a fractal. Let us provide a strict definition of a self-similar set that is used in mathematics. For this we will need several preliminary definitions.

Let the set $E \subset \mathbb{R}^d$ be a closed one. Then the mapping $S: E \to E$ is called a similarity mapping on $E$, if $\exists t : 0 < t < 1 : |S(x) - S(y)| = t|x - y|, \forall x, y \in E$.

That is, the similarity mapping $S$ turns the set $E$ into a geometrically similar set.

Let us consider a set of similarity mappings $S_1, \ldots, S_m$. The set $F \subseteq E$ is invariant in relation to transforms $S_i$, if

$$F = \bigcup_{i=1}^{m} S_i(F).$$

A set that is invariant in relation to a set of similarity mappings, is called a self-similar set.

The definition of a self-similar set can be perceived intuitively. Indeed, by definition, a set is self-similar if it can be "assembled" from fragments that are similar to the entire set. The simplest example of a self-similar set is an interval $[0,1]$. Lest us assume $S_1 = \frac{x}{2}$ and $S_2 = \frac{1}{2} + \frac{x}{2}$. Then $[0,1] = S_1([0,1]) + S_2([0,1])$.

Obviously, ordinary self-similarity is not sufficient for an object to be called a fractal. We wouldn't consider an interval on a straight line or a piece of crated paper as fractals. We are going to consider fractal objects that have the self-similarity property and complex structure.

Let us provide a basic example of a fractal set that is going to prove handy for illustration of the key ideas further on. It is Cantor set (or Cantor dust). A classical process of Cantor set construction starts with a unit interval $C_0 = [0,1]$. Let us now eliminate the middle third of $C_0$, which will leave us with the set $C_1 = \left[0, \frac{1}{3}\right] \cup \left[\frac{2}{3}, 1\right]$. $C_1$ consists of two intervals; now let us eliminate middle thirds from both of them and get the set $C_2$. If we continue this operation, we are going to get a sequence of sets $\{C_i\}_{i=1}^{\infty}$. Cantor set is the intersection of all these sets

$$C = \bigcap_{i=1}^{\infty} C_i.$$

We should note that the set $C$ is self-similar. Let us consider similarity mappings $S_1(x) = \frac{x}{3}$ and $S_2(x) = \frac{x}{3} + \frac{2}{3}$, then $C = S_1(C) \cup S_2(C)$. On the other hand, we know that the Lebesgue measure of a Cantor set equals 0, just as the measure of a point or any enumerable set. However, it is obvious that the structure of a Cantor set is much more complex, and we are facing the need for a special measure for description of such a set. This measure is going to be the fractal dimensionality of a set.

## *Fractal dimensionality*

Let us get back to Hausdorff-Besicovitch dimensionality definition. For this we are going to need the concept of set covering.

Let $U$ be a non-empty set within $\mathbb{R}^d$. Diameter of the set $U$, by definition, equals

$$|U| = sup\{|x - y| : x, y \in U\}.$$

If $F \subset \bigcup_{i=1}^{\infty} U_i$ and $0 < |U_i| \leq \delta$ for every $i$, then the collection of sets $\{U_i\}$ is called $\delta$ – covering for the set $U$.

Let $F$ be a subset of some closed set within $\mathbb{R}^d$. For any $s \geq 0$ and $\delta > 0$ let us define

$$\mathcal{H}_\delta^s(F) = \inf\left\{\sum_i |U_i|^s : \{U_i\} - \delta - \text{covering for } F\right\},$$

where the infinum is taken across all possible $\delta$ – coverings of the set $F$. By definition, s-dimensional Hausdorff measure is

$$\mathcal{H}^s(F) = \lim_{\delta \to 0} \mathcal{H}_\delta^s(F).$$

Such a limit exists for every set $F \subset \mathbb{R}^d$, providing it can equal zero or infinity.

Hausdorff-Besicovitch dimensionality of a set $F$ is defined as

$$D_H(F) = \inf\{s : \mathcal{H}^s(F) = 0\} = \sup\{s : \mathcal{H}^s(F) = \infty\}.$$

Or (the same)

$$\mathcal{H}^s(F) = \begin{cases} 0, s < D_H(F); \\ \infty, s > D_H(F). \end{cases}$$

For example, let us calculate the dimensionality of Cantor set $C$. Above we described the procedure of the set construction. According to this procedure, on the $n$-th iteration we have $2^n$ intervals of length $1/3^n$ each, and on further iterations the set is only further diminished. That is why we can select the value $1/3^n$ as covering diameter $\delta$ and use $2^n$ sets in the covering. By definition

$$\mathcal{H}^s(C) = \lim_{\delta \to 0} \mathcal{H}_\delta^s(C),$$

and now we can make a transition from the limit under $\delta \to 0$, to the limit under $n \to \infty$

$$\lim_{\delta \to 0} \mathcal{H}_\delta^s(C) = \lim_{n \to \infty} \sum_{i=1}^{2^n} \left(\frac{1}{3^n}\right)^s.$$

Now we only have to calculate the limit value

$$\lim_{n \to \infty} \sum_{i=1}^{2^n} \left(\frac{1}{3^n}\right)^s = \lim_{n \to \infty} \left(\frac{2}{3^s}\right)^n = \begin{cases} 0, \dfrac{2}{3^s} < 1, \\ 1, \dfrac{2}{3^s} = 1, \\ \infty, \dfrac{2}{3^s} > 1. \end{cases}$$

Consequently, the limit $\mathcal{H}_\delta^s(C)$ under $\delta \to 0$ does not equal zero or infinity under $\frac{2}{3^s} = 1$, so

$$D_H(C) = \frac{\ln 2}{\ln 3} \approx 0,63.$$

## Statistically self-similar processes and Hurst exponent

Many objects of the surrounding world are statistically self-similar (a classical example is a shoreline), which means that parts of these objects maintain statistical characteristics under scale changes. During the research of evolution of information flows, the structures of document sets in the Web, and the processes taking place in the information space, we are often witnessing the emergence of self-similar structures, particularly, time series.

Let us provide a definition of a self-similar process.

*A real-value process* $\{x(t), t \in \mathbb{R}\}$ *is a **self-similar process with Hurst exponent** $H > 0$, if for every $\alpha > 0$ finite-dimensional distributions $\{x(\alpha t), t \in \mathbb{R}\}$ are identical to* *finite-dimensional distributions $\{a^H x(t), t \in \mathbb{R}\}$, which can be written out in the following abbreviated form:*

$$\{x(\alpha t), t \in \mathbb{R}\} =^d \{a^H x(t), t \in \mathbb{R}\}.$$

That is, by definition, for a self-similar process, change of time scale is equivalent to change of process value scale. This means that implementations of such a process (member functions) look similarly under different scales. At the same time, naturally, the process is not an exact copy of itself on different scales, only statistical properties are maintained.

Hurst exponent is a measure of persistency, i.e., the tendency of a process towards trends. The value of H = 0.5 corresponds to uncorrelated behavior of time series values, like in Brownian movement. Values within the range 0.5 <H <1 indicate that the dynamics of the process directed to a certain side in the past will, most probably, result in continuous movement in this same direction in future. If H<0.5, then the process will, most probably, change its trend.

Let us describe some properties of self-similar processes that are important for application. First, these auto-covariance function of such processes is decreasing and looks as follows

$$\rho_k \approx k^{(2H-2)} L(t) \text{ under } k \to \infty,$$

where $L(t)$ is a function, that slowly changes on infinity, i.e.

$$\forall x > 0: \lim_{t \to \infty} \frac{L(tx)}{L(t)} = 1.$$

Consequently, for self-similar processes the series of covariance coefficients is divergent

$$\sum_{k=1}^{\infty} \rho_k = \infty.$$

This endless sum indicates the long-term dependence within the series.

Second, dispersion of the selective average decreases slower than the value, inverse to the set size

$$\sigma^2\left(x_t^{(m)}\right) \sim m^{2H-2},$$

where the sequence $\left\{x_t^{(m)}\right\}$ is obtained by splitting of the initial sequence $\{x_t\}$ into non-intersecting blocks of length $m$ and taking of the average within each block.

## Hurst exponent evaluation methods

The method suggested by Hurst himself is called the method of normalized span or $R/S$ analysis. For the time series $\{x_t\}_{t=1}^T$ standard deviation $S$ is calculated according to the formula

$$S = \sqrt{\frac{1}{T}\sum_{t=1}^{T}(x_t - \bar{x})^2}, \quad \text{where } \bar{x} = \frac{1}{T}\sum_{t=1}^{T}x_t,$$

while the span of the series

$$R = \max_{1 \le t \le T} x^{(t)} - \min_{1 \le t \le T} x^{(t)}, \text{ where } x^{(t)} = \sum_{i=1}^{t}(x_i - \bar{x}).$$

Relation $R/S$ is, in fact, the normalized span. Hurst found that for many observed time series the normalized span is well described by an empirical relation

$$\frac{R}{S} = \left(\frac{T}{2}\right)^H.$$

Hurst exponent values can be evaluated if we calculate the values of statistics $R/S$ depending on $T$ and build a diagram of such dependence in the double logarithmic scale. Hurst exponent estimate will be the estimated angle of inclination of the line, which approximates the dependence of $\log R/S$ on $\log T$ in the best way.

Let us use $R/S$ method to calculate Hurst exponent for series T, C, and H. On Fig. 3.25 the results of evaluation for series T and C are shown. Obtained values of Hurst exponent are 0.62 and 0.68, respectively. They speak in favor of the data's inclination to trends, although not a very high one.

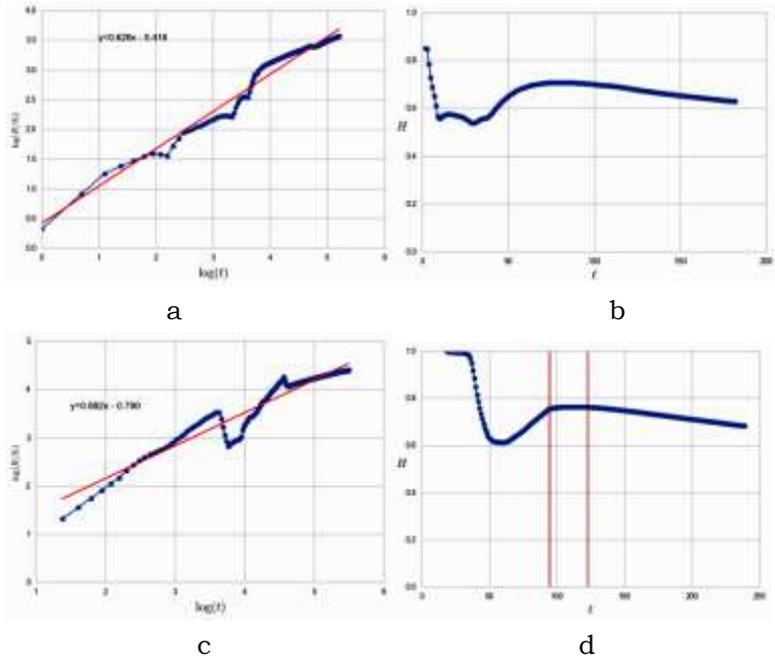

a        b

c        d

Fig. 3.25 – Evaluation of Hurst exponent for series T and C. Dependence of $R/S$ statistics on time in logarithmic scale for series T (a) and C (c). Dependence of Hurst exponent on time for series (b) and C (d)

In case of series C on Fig. 3.25 we can see that dependence of $\log R/S$ on $\log t$ is poorly approximated by linear dependence, because the graph has sharp twists. If we build dependence of Hurst exponent on time (Fig. 3.26), we can define the moment, when the value of the indicator starts to decrease. If we mark this moment of time on the diagram for the series H, we can see that this is the moment of sharp increase of time series values; before it the values had significantly lower dispersion.

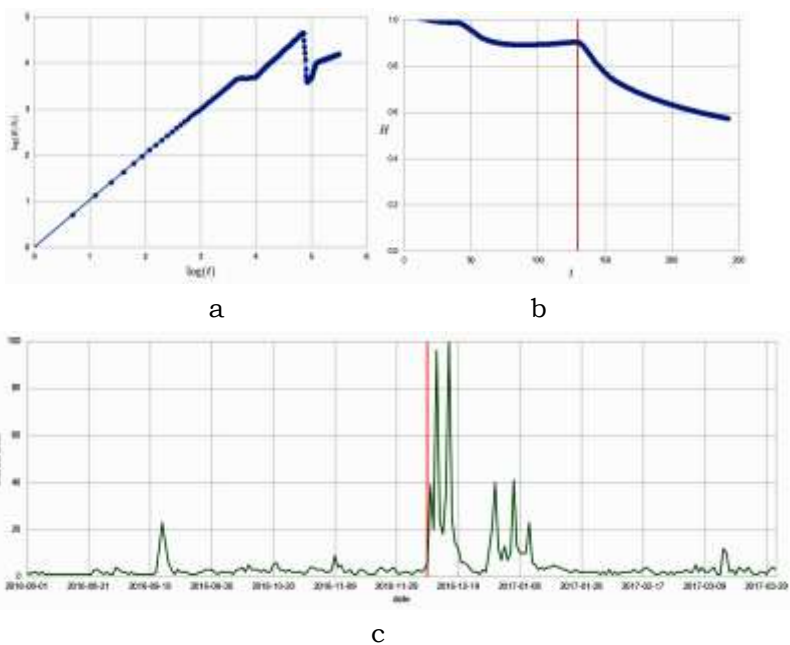

a            b

c

Fig. 3.26 – Evaluation of Hurst exponent for series H. Dependence of $R/S$ statistics on time in the logarithmic scale (a). Dependence of Hurst exponent on time (b). The vertical line marks the moment, when Hurst exponent value starts to decrease. Time series H with the marked moment of time, after which Hurst exponent starts to change (c)

Behavior of the series H, starting from December 2016 (the moment when peak values are witnessed) can be considered separately. Evaluation of Hurst exponent for the second fragment of the series is shown on Fig. 3.27. At the same time, we should keep in mind that in the given example the time series becomes too short, because only series with at least 200 elements usually qualify for $R/S$ analysis. Nevertheless, abrupt "step-like" changes in the dependence of Hurst exponent on time, confirm that the process under consideration consists of different processes that should be addressed separately.

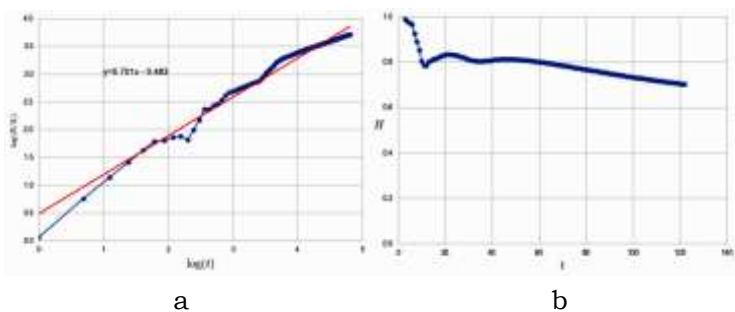

<div align="center">a          b</div>

Fig. 1.27 - Evaluation of Hurst exponent for the second part of series H. Dependence of statistics $R/S$ on time in the logarithmic scale (a). Dependence of Hurst exponent on time (b)

### $\Delta L$–method

Scalograms, obtained though application of continuous wavelet transform are used to visualize the properties of a time series. In [Ландэ, 2009] another visualization method is suggested, allowing to detect trends, periods, and local properties of a time series. The suggested approach is much easier to implement than wavelet analysis.

The method, which the authors called $\Delta L$-method, is based on DFA (Detrended Fluctuation Analysis) method, which we are going to address below. The essence of the approach is determination and display of absolute deviation of accumulated value series points from the respective linear approximation values.

Let us describe the $\Delta L$-method in greater detail. For a start, let us set a certain window width $s$ (i.e. the scale in which the series is considered). Let us consider the point $x_l$ and select a window of width $s$ for it so that the point number $l$ would be in the center of the window (or one value to the left or right if $s$ is even). Let us construct a linear approximation by the points within the window and define the local approximation value in point number $j$ for the interval with center in $l$ as $\Delta_{l,j,s}$. Then let us calculate absolute deviation $x_j$ (Fig. 3.28) from the approximation line $\Delta_{l,j,s} = \left| x_j - L_{l,j,s} \right|$.

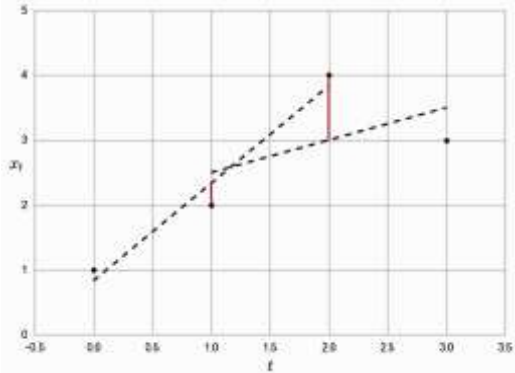

Fig. 3.28 – Four points of the time series with linear approximation for two windows three points wide each. Besides that, the figure shows the deviation $\Delta_{l,j,s}$ of the central point of the window from the respective linear approximation

The method envisions calculation of $\Delta_{l,j,s}$ values for all points $l = 1, \ldots, T$ and window widths $s = 1, \ldots, [T/4]$. For the fixed window width we calculate the average square deviation

$$E(l,s) = \sqrt{\frac{1}{s} \sum_j \left| x_j - L_{l,j,s} \right|^2} = \sqrt{\frac{1}{s} \sum_j \Delta_{l,j,s}^2}.$$

Then obtained values are displayed on a scalogram-like diagram. Examples of such diagrams are shown on Fig. 3.29.

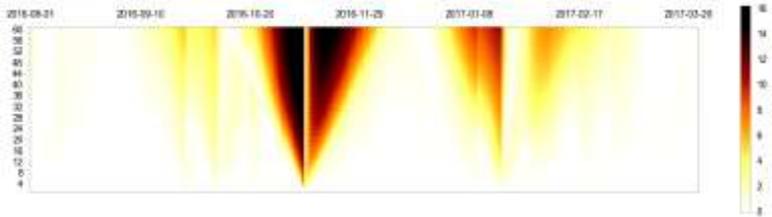

a

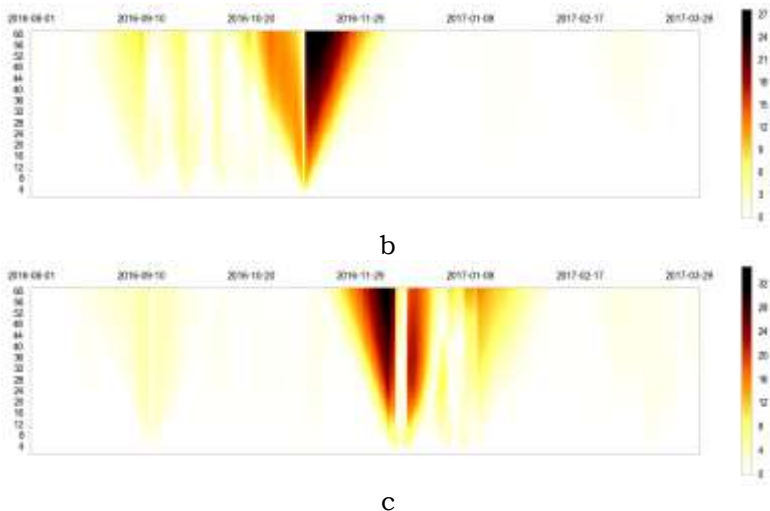

b

c

Fig. 3.29 – Coefficients obtained using the $\Delta L$-method for series T (a), C (b), and H(c).

The suggested method of visualization of absolute deviations $\Delta L$, just like the wavelet transform method, allows us to detect isolated and irregular spikes, sharp changes of qualitative indicator values in different time periods, as well as harmonic components within a series.

## 3.7. Multifractal analysis

A single fractal dimensionality is often insufficient for description of self-similar objects witnessed in natural environments. In many cases such objects are not homogenous. The most general description of such objects is provided by multifractal theory, according to which an object is characterized by an endless hierarchy of dimensionalities, allowing us to tell homogenous objects from heterogenous ones.

A multifractal set (signal) can be understood as a certain union of different homogenous fractal subsets (signals), while every one of them has its own fractal dimensionality. Values of such fractal dimensionalities are reflected in the multifractal spectrum whose formal definition will be introduced further. It is important that multifractal spectrum can be used as similarity measure. Such an approach can be used, for instance, to form representative sets from arrays of documents, in addition to traditional methods based on detection of content similarity. Practical applications of such an approach are: providing the user with perceptible    search results, reflecting all the document array spectrum, or selection of document subsets for further research [Ландэ, 2009a; Ландэ, 2009b].

In order to consider the multifractal set and multifractal spectrum ideas in greater detail, we are going to need several additional concepts and definitions. As an example, we are going to use the generalized Cantor set, as well as measure on this set. Multifractal spectrum in this example emerges in a natural and simple way, so it helps us to get an intuitive understanding. Besides, Hölder exponent (that will be defined a bit later) is also crucial for multifractality analysis.

Let us start with building of a generalized Cantor set and defining measure on it. A classical Cantor set can b e generalized in several ways. For instance, we can use the compression function $F_i(x) = rx + (1-r)i$, where $i = 0,1$ and $r \in \left(0, \frac{1}{2}\right)$, instead of $F_i(x) = \frac{1}{3}x + \frac{2}{3}i$. Let us define this set as $C(r)$. On the generalized Cantor set we can introduce an evenly distributed measure, or a $p$-measure with the same compression function, but with probabilities $p$ and $1 - p$, where $p \in [0,1]$.

A more interesting generalization is the Cantor set with variable separation coefficients. Let us assume, we have a sequence $\{r_j\}$, $r_j \in \left(0, \frac{1}{2}\right)$. We are going to build the set using the following iterative procedure. Let $C_0 = [0,1]$. Let us eliminate from the middle of $[0,1]$ interval an open interval of length $1 - 2r_1$. What remains is two closed intervals of length $r_1$. We will define union of these two intervals as $C_1$. On $j - th$ iteration the set $C_j$ will consist of a union of $2^j$ closed intervals of lengths $r_1 \cdot r_2 \cdot \ldots \cdot r_j$. Thus, we will obtain the set

$$C(\{r_j\}) = \bigcap_{j=1}^{\infty} C_j.$$

Closed sets that emerge under iterative construction of Cantor set can be coded by finite words from the alphabet $\{0,1\}$. At the first step we obtain the left interval $I_0$ and right interval $I_1$. On $n-th$ step the name of interval $w$ has the length of $n$ while on the next step the interval $I_w$ will be divided into intervals $I_{w0}$ and $I_{w1}$.

On the set $C(\{r_j\})$ we can introduce $p$-measure, that will have the following property

$$\mu(I_{w0}) = p\mu(I_w), \qquad \mu(I_{w1}) = (1-p)\mu(I_w).$$

We will get an even more generalized measure on $C(\{r_j\})$ if we use a sequence of weights $\{p_j\}, p_j \in [0,1]$ and define the measure using the following rule $\mu(I_{w_1 w_2 \ldots w_n}) = p_{w_1 1} \cdot p_{w_2 2} \cdot \ldots \cdot p_{w_n n}$, where $p_{0j} = p_j$, and $p_{1j} = 1 - p_j$.

In order to draw certain conclusions, concerning the structure of the measure, we are going to need Hölder exponent. First, let us consider the definition and the sense of Hölder exponent for functions and measures, and then let us apply them to the just-constructed generalized Cantor set, and the measure on it.

### Hölder exponent and multifractal analysis for measures

Hölder exponent, containing information on behavior of a function in the vicinity of a point, is a characteristic of the function's smoothness. The smaller the Hölder exponent value, the less smooth the function is.

Let $x$ be a limited function on $\mathbb{R}$ и $t_0 \in \mathbb{R}$, then local Hölder exponent of the function $x$ in point $t_0$ is defined as

$$h_x(t_0) = \sup_{\Delta t \to 0}\{\alpha \geq 0 : |x(t + \Delta t) - x(t)| = O(\Delta t^\alpha)\}.$$

In other words, the local Hölder exponent characterizes the behavior of the function in the vicinity of a point as follows

$$|x(t + \Delta t) - x(t)| \sim \Delta t^{h_x(t)}.$$

The last relation should be compared with a similar relation for monofractal processes

$$|x(t + \Delta t) - x(t)| \sim \Delta t^H,$$

Where $H$ – is Hurst exponent. That is, for multifractal processes, the local Hölder exponent $h_x(t)$, is, in essence, on $t$.

In order to calculate the regularity of measure in the vicinity of a point, Hölder exponent is also introduced.

Hölder exponent or local dimensionality of measure $\mu$ on $\mathbb{R}$ is defined as follows

$$h_\mu(x) = \lim_{r \to 0^+} \frac{\log[\mu(B(x,r))]}{\log r},$$

Where $B(x,r)$ – is a sphere with the center in point $x$ and radius $r$.

Let us get back to examples with Cantor set. For evenly distributed measure on the classical Cantor set:

$$h_\mu(x) = \frac{\log 2}{\log 3}, x \in C.$$

However, if we consider the set $C(\{r_i\})$ with measure $\mu$, defined using weights $p$ and $1 - p$, then the local measure dimensionality will vary across different points. For instance,

$$h_\mu(0) = \lim_{j \to \infty} \frac{\log p^j}{\log \prod_{i=1}^{j} r_i} = \frac{\log p}{\lim_{j \to \infty} \frac{1}{j} \sum_{i=1}^{j} \log r_j} = \frac{\log p}{\log r_0}, h_\mu(1)$$
$$= \frac{\log(1-p)}{\log r_0},$$

where $\lim_{j \to \infty} \frac{1}{j} \sum_{i=1}^{j} \log r_j = \log r_0$. If we assume that $\sup_i r_i < 1/2$, and without loss of generality assume that $p < 1 - p$, then we can show that

$$h_\mu(x) \in \left[ \frac{\log p}{\log r_0}, \frac{\log(1-p)}{\log r_0} \right], x \in C(\{r_i\}).$$

(proof can be found in [Aldroubi 2016]). Thus, for *p*-Cantorian measure $\mu$ we know the possible values of local dimensionality. In order to describe certain properties of the measure $\mu$ we should consider the level sets of local dimensionality values, particularly such sets as

$$E_h = \left\{ x \in \mathbb{R} : h_\mu(x) = h \right\}.$$

Further on we can compare the sizes of sets $E_H$ under different values of $h$. In many practically important cases for comparison of such sets we will need to use fractal dimensionality. So, we are approaching the definition of a multifractal spectrum.

A multifractal spectrum of measure $\mu$ on $\mathbb{R}$ is a mapping $d_\mu(h) = D_H(E_h)$.

That is, using a multifractal spectrum, we can show, which values of Hölder exponent are present in the heterogenous object (measure, set, signal), and in what proportion. Every Hölder exponent value corresponds to fractal dimensionality of a set of points, in which the Hölder exponent value equals the given one (Fig. 3.30).

For *p*-Cantorian measure $\mu$ we can show that (as proved in [Aldroubi 2016])

$$d_\mu(h) = \inf_{q \in \mathbb{R}} \left( qh - \frac{\log(p^q + (1-p)^q)}{\log r_0} \right).$$

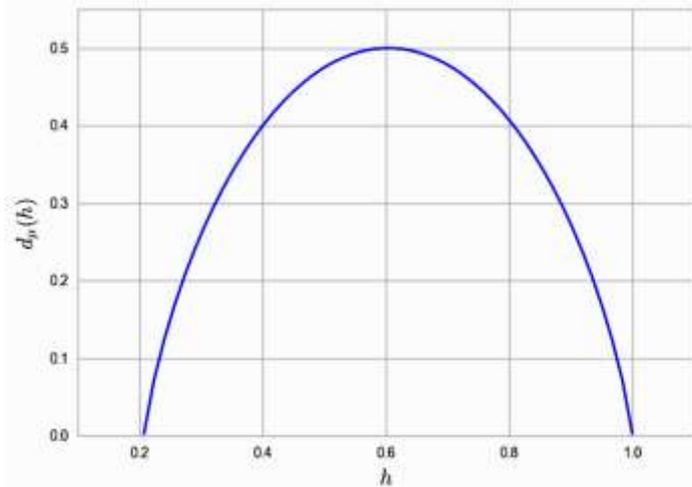

Fig. 3.30 – Multifractal spectrum for *p*-Cantorian measure.

## *An approach to multifractal spectrum estimation*

Above we described a theoretical approach to definition of a multifractal spectrum for a measure. For practical purposes direct calculation of Hölder exponent in every point and calculation of fractal dimensionalities of level sets of this indicator is impossible. The key to practical approach lies in the following definition of the structure function of measure $\mu$

$$Z(q,j) = \frac{1}{2^j} \sum_i \mu\left( B\left( \frac{i}{2^j}, \frac{1}{2^{j+1}} \right) \right)^q,$$

where the sum is taken only on such intervals where the measure does not equal 0. Besides that, let us define the scaling function

$$\tau(q) = \liminf_{j \to +\infty} \left( \frac{\log\left(Z_\mu(q,j)\right)}{\log 2^{-j}} \right).$$

It is known that in order to cover the set $E_h$ we need approximately $2^{-d_\mu(h)j}$ of balls, and, from the definition $h_\mu(x) = \lim_{j \to \infty} \frac{\log[\mu(B(x,2^{-j}))]}{\log 2^{-j}}$ it follows that $\mu\left(B(x, 2^{-j})\right) \sim 2^{-h_\mu(x)j}$, so the scale function can be estimated as follows

$$Z(q,j) \sim 2^{-j} \sum 2^{-h_\mu qj} \sim 2^{-j} 2^{-d_\mu(h)j} 2^{-hqj} = 2^{-(1+d_\mu(h)+hq)j}.$$

On the other hand, the scale function is defined in such a way that $Z(q,j) \sim 2^{-\tau(q)j}$, so

$$2^{-(1+d_\mu(h)+hq)j} = 2^{-\tau(q)j}.$$

And under $j \to \infty$

$$d_\mu(h) = \inf_{q \in \mathbb{R}} (1 - \tau(q) + hq).$$

Thus, we get the expressions for multifractal spectrum through scale function. This approach allows us to numerically define multifractal spectrum for time series. First, we define the structure function, then (based on it) – the scale function, and after that – using the Legendre transform we come up with the multifractal spectrum.

### *Multifractal analysis for functions*

For functions certain definitions and statements apply, which are similar to the respective definitions and statements for measures.

A multifractal spectrum of a locally limited function $x \colon \mathbb{R} \to \mathbb{R}$ is a mapping

$$d_x \colon \mathbb{R}^+ \cup \{+\infty\} \to [0; d] \cup \{-\infty\},$$

such that

$$d_x(h) = D_H\left(E_x(h)\right)$$

(assuming $D_H(\emptyset) = -\infty$).

The following statement concerning Hölder exponent applies to functions (similar to Hölder exponent definition introduced for measures above). We should remind that

$$|x(t + \Delta t) - x(t)| \sim C_t \Delta t^{h_x(t)},$$

Then, based on this expression, it is not difficult to formulate the following statement.

<u>Lemma.</u> Let $x: \mathbb{R} \to \mathbb{R}$ be a limited function, for which $h_x(t) = H \in [0,1]$, then

$$h_x(t) = \lim_{j \to +\infty} \inf \left( \frac{\log\left(R_x\left(B(t, 2^{-j})\right)\right)}{\log 2^{-j}} \right),$$

where $R_x(A) = \max_{t \in A} x(t) - \min_{t \in A} x(t)$ – is the span of function $x$ on set $A$.

It follows from the lemma that the multifractal formalism for functions can be based on the structure function

$$Z(q,j) = \frac{1}{2^j} \sum_i R_x \left( B\left( \frac{i}{2^j}, \frac{1}{2^{j+1}} \right) \right)^q \qquad (1)$$

and the corresponding scale function

$$\tau(q) = \lim_{j \to} \inf_{+\infty} \left( \frac{\log(Z_x(q,j))}{\log 2^{-j}} \right) \qquad (2)$$

*which results in the following definition of a multifractal spectrum*

$$d_\mu(h) = \inf_{q \in \mathbb{R}} (1 - \tau(q) + hq).$$

## *Multifractal processes*

A stochastic process is called ***multifractal*** if it has stationary increment and satisfies the equality

$$\mathbb{E}[|x(t)|^q] = c(q) t^{\tau(q)+1},$$

for certain positive $q$, where $\tau(q)$ is the scale function.

If the scale function $\tau(q)$ linearly depends on $q$, then the process is called monofractal. If the process $x(t)$ is self-similar with Hurst exponent $H$, then $\tau(q) = Hq - 1$.

## *The DFA method and its application to multifractal spectrum estimation*

[Peng, 1994] proposed the Detrended Fluctuation Analysis (DFA) method to determine long-term correlations in noisy and non-stationary time series. The key feature of the DFA method is that it is based on the theory of random walks. The time series is not analyzed in the original form. Instead, the series is centered and transformed into accumulated sums

$$y_t = \sum_{k=1}^{t} x_k.$$

In this case, $y_t$ can be considered the location of the random walk after $t$ steps. Further, the DFA method involves analysis of the mean square deviation of series values from the trend in various non-intersecting fragments of the series.

Many modifications and practical applications of the DFA method were suggested. An overview of such methods can be found, for example, in [Kantelhardt 2009]. An important step was the development of an approach to numeric estimation of multifractal spectrum based on DFA method. Such a method, called Multifractal Detrended Fluctuation Analysis (MF-DFA), was suggested in [Kantelhardt 2002]. Efficiency of MF-DFA method was analyzed for various model time series (Brownian motion, fractional Brownian motion, binomial cascades) [Oswiecimka 2012]. Besides that, the method is actively used for analysis of real time series, often in economic sphere [Suarez-Garcia 2013].

Detailed step-by-step description of MF-DFA algorithm is set forth in [Thompson 2016]. Let us describe all these steps.

**Step 1.** Bringing of the time series to the aggregated form. Data sets can be classified as aggregated or disaggregated. An example of aggregated data is the number of new messages, related to a certain topic, in the Web. Respective disaggregated data is the increment of the number of messages in comparison to the previous day. If the initial time series $\{z_t\}_{t=1}^{T+1}$ is aggregated, then we need to make a transition to the disaggregated series $\{y_t = z_{t+1} - z_t\}_{t=1}^{T}$. The time series that we are going to use in the algorithm, during centering and calculation of accumulated sums is as follows

$$x_t = \sum_{k=1}^{t} (y_k - \bar{y}), \qquad t = 1, \dots, T.$$

This series processing step is necessary for correct work of the method, as it is based on random walk theory.

**Step 2.** Let us construct a set $S = \{3, [N/4]\}$. For every value $s \in S$ let us split the time series $\{x_k\}_{k=1}^{N}$ into $N_s = \left\lfloor \dfrac{N}{s} \right\rfloor$ non-intersecting parts of length $s$. If $N$ cannot be divided by $s$ with integer result, then we need to repeat the procedure, starting from another side of the time series. As a result we will get $2N_s$ parts.

**Step 3.** For each value $j = 1, \dots, N_s$ $j-th$ time series part consists of values $\left\{x_{(j-1)s+1}\right\}_{i=1}^{s}$, and, similarly, under $j = N_s + 1, \dots, 2N_s - \left\{x_{N-(j-N_s)s+i}\right\}_{i=1}^{s}$. For each $j$-th time series part we need to define the trend $X_j(i)$. In many cases it is enough to use linear approximation of time series values, obtained through least squares method. For longer series polynomial approximation of degree $m$ is used (further we are going to provide an algorithm for optimal $m$ value selection). Now let us calculate the average square deviation of the time series part from the trend

$$F^2(j,s) = \frac{1}{s}\sum_{i=1}^{s}\left[x_{(j-1)s+i} - X_j(i)\right]^2, \text{ under } j = 1,...,N_s,$$

and, similarly, under $j = N_s + 1,...,2N_s$.

<u>Step 4.</u> Let us define the set of momentum orders as $Q$. The set $Q$ should include 0, positive and negative values. Usually, they select a set, which is symmetrical in relation to 0. For fixed values $s \in S$ и $q \in Q$ let us calculate the norm $l_q$ for the vector, consisting of estimated dispersions $\left\{F^2(j,s): j = 1,...,2N_s\right\}$.

$$F_q(s) = \left(\frac{1}{2N_s}\sum_{j=1}^{2N_s}\ln\left[F^2(j,s)\right]^{q/2}\right)^{1/q}, \quad q \in Q \setminus \{0\},$$

$$F_0(s) = \exp\left\{\frac{1}{4N_s}\sum_{j=1}^{2N_s}\ln\left[F^2(j,s)\right]\right\}.$$

<u>Step 5.</u> For each $q \in Q$ we should perform linear approximation of dependence of $\ln\left[F_q(s)\right]$ on $\ln(s)$. At the same time, the inclination of the obtained linear function is the estimate for $h(q)$. The scale function $\tau(q)$ can be obtained from expression $\tau(q) = qh(q) - 1$.

<u>Step 6.</u> Let us estimate the derivative of the obtained estimate of function $\tau(q)$

$$\alpha_0 = \frac{d\tau(q)}{dq}\Big|_{q=q_0}, \quad q_0 \in Q,$$

and, as a result, we will get the estimate of a multifractal spectrum

$$f(\alpha) = q_0\alpha_0 - \tau(q_0).$$

Examples of scale functions and multifractal spectrums for series T, C, and H, obtained using MF-DFA method, are provided on Fig. 3.31.

For correct work of the described algorithm we need to select the set $S$ of intervals into which the series is split, in advance. Let us clarify, what this set $S$ should look like. On the one hand, if we use the values of $s \geq N/4$, we are going to split the series into a small number of parts, and the estimate $F_q(s)$ will be calculated across a small number of dispersion estimates $F^2(j,s)$. On the other, if we use polynomial regression, then under $s \leq 10$ on step 3 the number of points used for regression will be too small. Thus, a reasonable limitation is: $10 \leq s \leq N/4$. In case of linear regression we can also use smaller values $s = 3, 4, \ldots$.

For long time series including more than two thousand values we recommend using such limitations as

$$s_{\min} = \max\{10, N/100\}, \quad s_{\max} = \{20 s_{\min}, N/10\},$$

and select the step in such a way that the set $S$ would involve no more than 100 values. Besides that, the degree $m$ should be selected in such a way that regression polynomial $X_j(i)$ would adequately reflect the trend in each part of the time series.

In case of short time series, such as T, C, and H, it is enough to use linear regression.

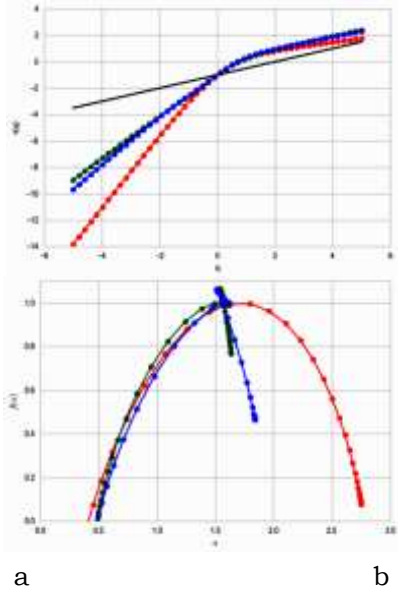

<div align="center">a        b</div>

Fig. 3.31 – Scale functions (a) and multifractal spectrums (b) for series T, C, and H, obtained using the MF-DFA method

## *Filtration*

During estimation of a multifractal spectrum the periodical component in the series might significantly influence the result. Influence of a sinusoid trend upon the estimate of spectrum is illustrated in [Nagarajan 2004]. Besides that, a smoothing filter is suggested for minimization of trend impact (at the same time, the filter does not influence the multifractal properties of the time series). We should note that traditional series smoothing methods (such as ordinary sliding average) significantly influence the multifractal properties.

Let us consider a smoothing filter:

1. Let us calculate Fourier transform for the series under consideration $x_t$ and denote it as $f(v)$.

2. Let us assume that sinusoid trend corresponds to frequency $v_k$. Let us define the modified Fourier transform

$$\left|\tilde{f}(v_k)\right| = 0.5(\left|f(v_{k-1}) + f(v_{k+1})\right|)$$
$$\left|\tilde{f}(v)\right| = \left|f(v)\right|, v \neq v_k$$

3. Using the inverse Fourier transform let us make a transition to the filtered series $\tilde{x}_t$.

Let us demonstrate the usage of such a filter on a certain time series $x_t$, as well as its modified version $y_t = x_t + s_t$, where $s_t$ is a sinusoid (Fig. 3.32). Let us calculate Fourier transform for series $y_t$ and denote it as $f(v)$. Fourier transform for series $x_t$, $y_t$, as well as $\tilde{y}_t$ differs in one point $v_k = 0.01$, which corresponds to oscillation frequency $s_t$ (Fig. 3.34).

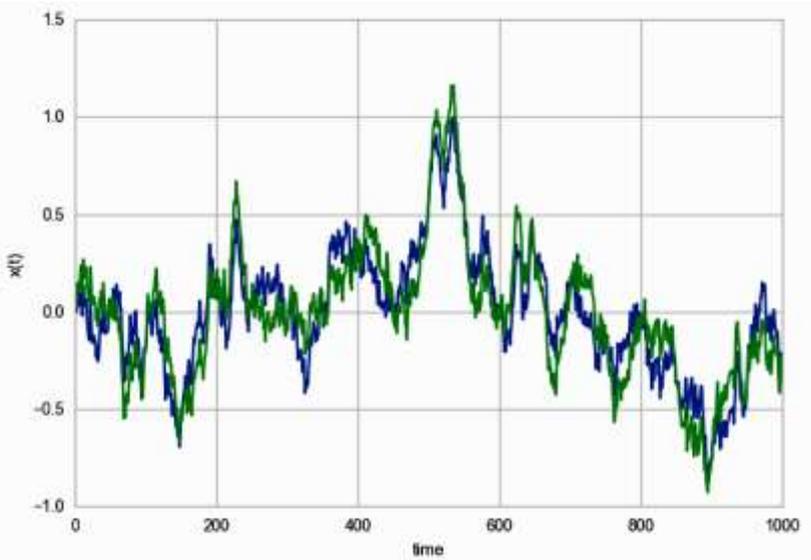

Fig. 3.32 – Multifractal time series (blue) and the same time series plotted against sinusoid trend (green).

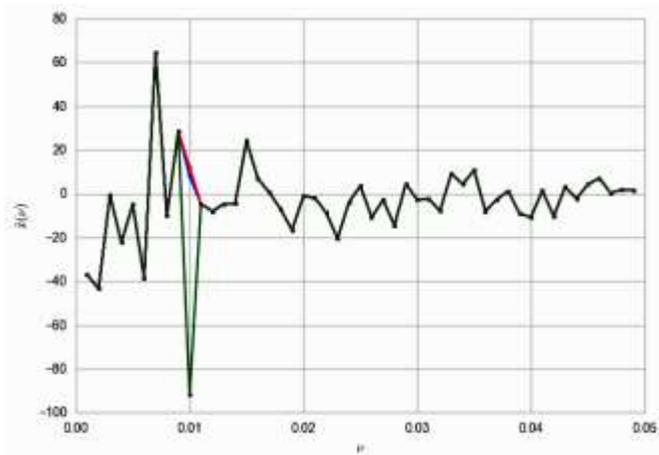

Fig.3.33 – Fourier transform for the initial series $x_t$ (blue), series $y_t = x_t + s_t$ (green) and filtered series $\tilde{y}_t$ (red)

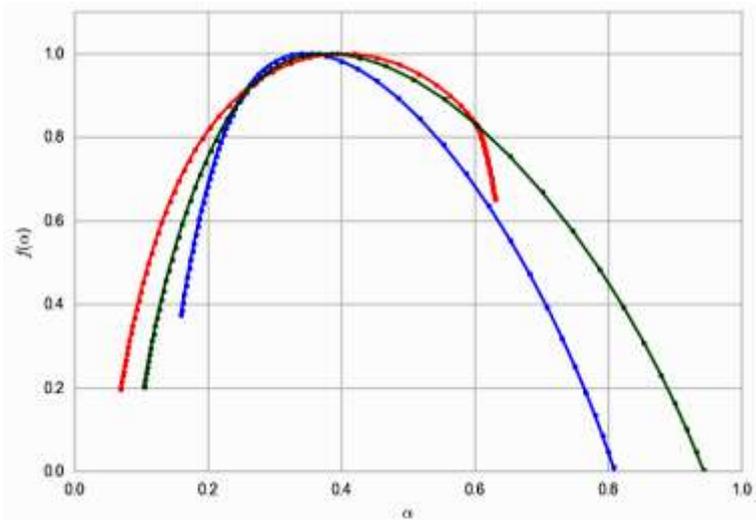

Fig. 3.34 – Multifractal spectrums for the initial series $x_t$ (blue), series $y_t = x_t + s_t$ (red), and filtered series $\tilde{y}_t$ (green)

## Usage of wavelets for estimation of multifractal spectrum

Wavelets are a natural instrument for analysis of fractal characteristics of an object [Aldroubi, 2016], primarily, because Hölder exponent can be estimated through wavelet coefficients. The following statement, connecting, wavelet transforms with Hölder exponent, applies.

Statement. Let the function $x$ in point $t$ have Hölder exponent of $h_x(t)$. Let us assume that in point $t$ the function $x$ does not have an oscillating property (the singularity is not oscillating). Then

$$W_\psi[x](s,t) \sim s^{h_x(t)}, \quad s \to 0^+.$$

under condition that first $n$ moments of the wavelet $\psi$ equal 0 and $n > h(t_0)$. On the other hand, if we select a wavelet, for which $n < h(t_0)$, then

$$W_\psi[x](s,t) \sim s^n, \quad s \to 0^+.$$

It follows from the statement that we can characterize the behavior of the function $x$ in the vicinity of point $t$ as follows: the more smooth the function is, the faster its wavelet transform value $W_\psi[x](s,t)$ decreases under decreasing scale $s$. For instance, if function $x$ is continuously differentiable in point $t$ (which means that $h_x(t) = +\infty$), then the wavelet transform behaves according to the last formula above. That is, values of $W_\psi[x](s,t)$ depend only on the shape of the wavelet (parameter $n$). Conceptually different case is witnessed when function $x$ in point $t$ has the value of Hölder exponent between 0 and 1 (this is a rather common case in practice). Then asymptotic behavior of wavelet coefficients is defined by the above-mentioned formula, i.e. depends on Hölder exponent, for any wavelet with zero momentum. The specified connection between asymptotic behavior of wavelet transform coefficient and Hölder exponent provides the basis for wavelet-based fractal spectrum estimation methods.

A simple and intuitive way to estimate multifractal spectrum using the aforementioned expressions, can start with definition of structure function as follows

$$Z(q,s) = \int \left| W_\psi[x](s,t) \right|^q dt, \qquad q \in \mathbb{R}.$$

The problem of such definition is that a wavelet transform can assume very small absolute values and turn into 0, so when it is raised to negative power $q$, very large values can emerge. Thus, from calculation viewpoint, this way is very unstable. For solution of this problem two approaches are widely used: Wavelet Transform Modulus Maxima (WTMM) approach and wavelet leader based approach.

## *Spectrum estimation using maximum lines. WTMM*

The approach to multifractal spectrum estimation has been actively developed starting from [Mallat, 1992]. It turns out that expressions from the previous subsection hold, if instead of the sequence of wavelet transforms $W_\psi[x](s,t)$ under constant $t_0$ and decreasing scale $s$, we consider the curve of local maximums of absolute value $W_\psi[x](s,t)$. Let us start with the necessary definitions.

A modulus maximum is a point $(s_0, t_0)$ such that $W_\psi[x](s_0, t) < W_\psi[x](s_0, t_0)$, where $t$ is either left or right neighbor of $t_0$, while the strict inequality $W_\psi[x](s_0, t) < W_\psi[x](s_0, t_0)$ has to hold for at least one neighbor of $t_0$ (left or right).

A maxima line is a connected curve in the scale space $(s, t)$, along which all points are modulus maxima. The set of maxima lines obtained through wavelet transform of the function is called the skeleton.

Fig. 3.35 shows wavelet transform coefficients of wavelet transform for series T, C, and H with outlined skeletons.

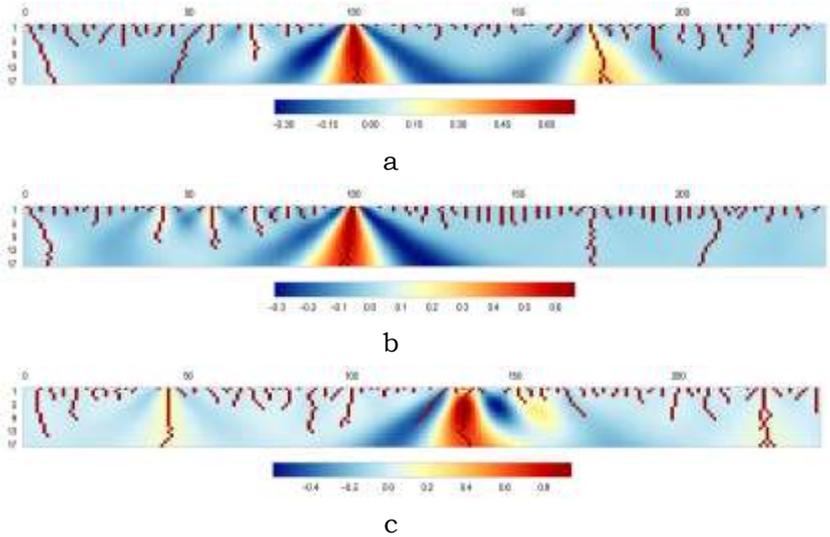

Fig. 3.35 – Wavelet transform coefficients for series T (a), C (b), and H (c) with outlined skeletons.

Now let us get back to the definition of a structure function. Instead of integration across all wavelet transforms under fixed scale $s$, let us make a transition to the discrete sum across local maximums.

$$Z(q,s) = \sum_{l \in \mathcal{L}(s)} \left( \sup_{\substack{(\tau,s') \in l \\ s' < s}} \left| W_\psi[x](t,s) \right| \right)^q, \quad q \in \mathbb{R}.$$

Then, according to respective formulas we can get to multifractal spectrum.

Fig. 3.36 shows the scale function and multifractal spectrum for series T, C, and H. On Fig. 3.36a, in addition to scale funcions of the series under consideration, the scale function for Brownian motion is shown.

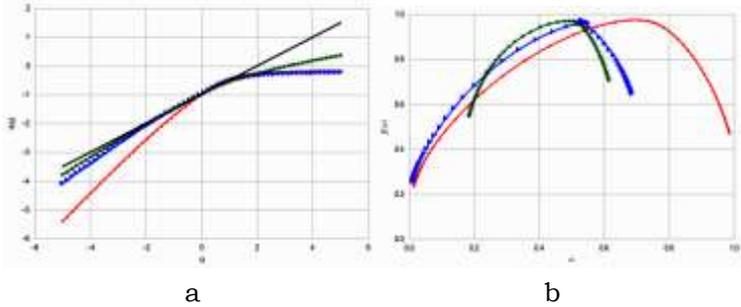

a                                    b

Fig. 3.36 – Scale functions (a) and multifractal spectrums (b) for series T, C, and H, obtained using WTMM method.

## *Wavelet leaders*

The concept of wavelet leaders and their application for calculation purposes was suggested by [Jaffard 2004]. The purpose of application of wavelet leaders is the same as the purpose of maximum lines – stabilizing the sum of wavelet coefficients when they are raised to negative power. Besides that, in contrast to maximum lines, the wavelet leader-based approach can be naturally generalized to the case of a $d$-dimensioned space [Jaffard 2006].

Wavelet leaders are defined through coefficients of discrete wavelet transform, so instead of arbitrary scale parameter $s$ only the values of $s = 2^{-j}$ are used.

Let $x$ be a limited function, then wavelet leaders of this function are defined according to formulas

$$d_j(t) = \max_{t' \in \{t-1, t, t+1\}} \left| W_\psi[x](t', 2^{-j}) \right|.$$

Having defined the wavelet leaders, we can introduce a formula for structure function

$$Z(q, j) = 2^{-j} \sum_t \left( d_j(t) \right)^q.$$

Just like in the case of WTMM we obtain the structure function, which is more stable from calculation viewpoint, than the structure function, defined directly through complete set of wavelet coefficients. Then estimation of scale function and multifractal spectrum is performed similarly to the initial formulas.

Fig. 3.37 shows the scale function and multifractal spectrum for series T, C, and H, estimated by wavelet leader method.

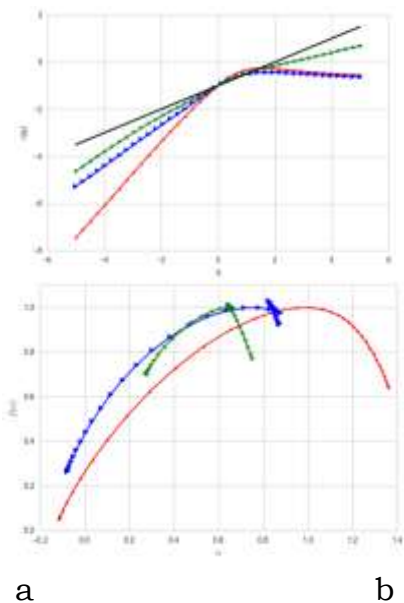

a      b

Fig. 3.37 – Scale functions (a) and multifractal spectrums (b) for series T, C, and H, obtained using wavelet leaders.

Different methods of multifractal spectrum estimation can produce different results. As an example, let us consider the methods' work for Brownian motion. It is known that Brownian motion is monofractal with Hurst indicator 0.5. That is why its scale function looks as $\tau(q) = 0.5q - 1$ and the spectrum is concentrated in one point $f(0.5) = 1$. Detailed comparison of methods MF-DFA and WTMM for different examples is presented in [Oswiecimka 2006]. Below we are going to provide the results of spectrum estimation using MF-DFA, WTMM, and wavelet leaders for Brownian motion.

Fig. 3.38 shows 50 generated Brownian motion trajectories of length 1000. For each process the scale function $\tau(q)$ was estimated using each of the three methods. On Fig. 3.39a, c, e the averaged $\tau(q)$ functions with standard deviation are displayed. Multifractal spectrums, which correspond to averaged scale functions, are shown on Fig. 3.39 b, d, f.

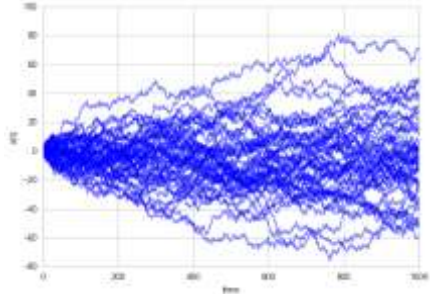

Fig. 3.38 – Generated Brownian motion trajectories

Multifractal spectrum can be used for comparison of time series. Multifractal spectrums of the same shape indicate similarity of the series, while different shape of spectrums indicates that there are significant differences in the nature of the respective series. This property is used in various studies. For instance, [Ivanov, 1999] analyzes the signals of physiological nature, such as heartbeat. It is shown that multifractal spectrums of time series, which correspond to healthy heartbeat, are much wider than in the opposite case. Thus, multifractal spectrum can serve as an indicator during analysis of heart rhythm dynamics. Similar approach to differentiation between healthy patients and those with balance problems is set forth in [Shimizu, 2002].

In [Sun, 2001] the daily index HangSeng at Hong Kong stock market is aalyzed. In this case the difference of multifractal spectrums also reflects the difference in the nature of economical processes. The results of research in economic area, based on comparisons of multifractal spectrums, are provided in [Zhou, 2009]. Using the Dow

Jones Industrial Average index, the economic factors influencing the spectrum shape are detected.

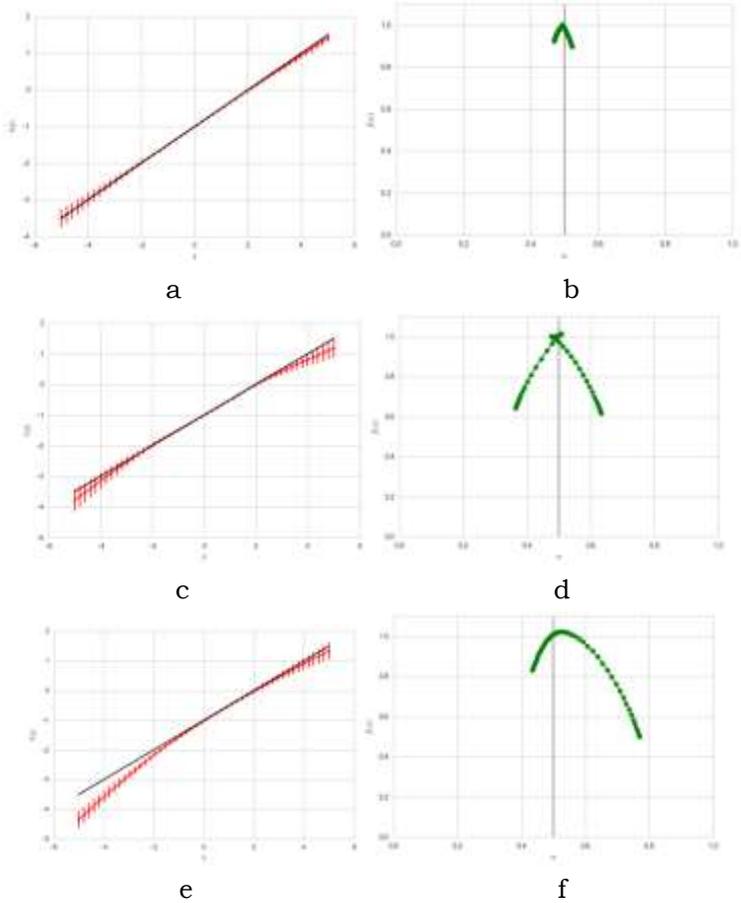

a            b

c            d

e            f

Fig.3.39 – Average scale function for generated Brownian processes, estimated using MF-DFA method (a), wavelet-leaders (c) and WTMM (e), and respective multifractal spectrums (b, d, f)

# 4. Analysis of network structures during information operation recognition

Lately a separate academic field started to emerge. We are talking about SNA or Social Networks Analysis, based on sociology and on Complex Networks at the same time [Newman, 2003]. As part of complex network theory, network properties are considered not only from topological viewpoint. Research subjects also include statistical phenomena, distribution of weights of particular edges and vertices (nodes), flow and conductivity effects. Although different types of networks are considered by the theory (including electric power grids, transport, information networks), studies of social networks provided the greatest contribution into the theory's development [Ландэ и др., 2009]

Three key directions are singled out within the complex network theory:

– Research of statistical properties that characterize the networks' behavior;
– Creation of network models;
– Forecasting of network behavior during changes of structural properties.

## 4.1. Network parameters

Most often applied researchers use such characteristics (common for network analysis) as network type, network density, centrality degree etc. In complex network analysis (just like in graph theory) the following parameters are researched:

– parameters of single nodes;
– parameters of a network as a whole;
– network sub-structures.

For separate nodes the following parameters are singled out:

– in-degree of the node, i.e. the number of graph edges that enter the node;
– out-degree of the node, i.e. the number of graph edges that exit the node;
– distance from a given node to each of other nodes;
– average distance from a given node to other nodes;
– eccentricity, i.e. the largest one among geodesic distances (minimum distances between nodes) from a given node to other nodes;
– betweenness, indicating how many shortest ways pass through a given node;
– centrality, i.e. the total number of connections of a given node in relation to other nodes.

To analyze the network as a whole the following parameters are used:

– number of nodes;
– number of edges;
– geodesic distance between nodes;
– average distance from a node to other nodes;
– density – ratio between the actual number of edges in the network and the maximum possible number of edges under given number of nodes;
– number of symmetric, transitive, and cyclic triads;
– network diameter – maximum geodesic distance in the network, etc

There are several topical problems of mathematical research of social networks, including the key ones.

An important characteristic of a network is the node degree distribution function $P(k)$, that is defined as the probability of the fact that the node number $i$ has the degree of $k_i = k$. Networks characterized by different $P(k)$, demonstrate different behavior, while in some cases $P(k)$ may be Poisson ($P(k) = e^{-m} m^k / k!$, where $m$ – is the mathematical expectation), exponential ($P(k) = e^{-k/m}$) or power ($P(k) \sim 1/k^\gamma$, $k \neq 0$, $\gamma > 0$) distribution.

Networks with power distribution of node degrees are called scale-free networks. It is the scale-free distributions that are most often witnessed in real social networks. Under power distribution, nodes with extremely high degrees can exist, while under Poisson's distributions we can almost never witness such a thing.

The distance between nodes is defined as the number of steps that need to be made in order to get from one node to the other using only the existing edges. Naturally, the nodes can be connected both directly and through intermediary links. The shortest way from $i$ to $j$ is the minimum distance between them. For the whole network the concept of an average way can be introduced as the average shortest distance across all node pairs:

$$l = \frac{2}{n(n+1)} \sum_{i \geq j} d_{ij},$$

where $n$ us the number of nodes, $d_{ij}$ – is the shortest distance between nodes $i$ and $j$.

Hungarian mathematicians P. Erdős and A. Rényi showed that the average distance between two nodes in a random graph increases proportionally to the logarithm of the number of nodes in it [Erdős, 1960].

Some networks may turn out to be disconnected, i.e. there will be nodes, distance between which is infinite. Consequently, the average way length will also equal infinity. In order to take such cases into consideration, the concept of average inverse way between the nodes is introduced (sometimes it is also called "network efficiency"). It is calculated as follows:

$$il = \frac{2}{n(n-1)} \sum_{i > j} \frac{1}{d_{ij}}.$$

Networks are also characterized by such parameters as diameter or maximum shortest length equaling maximum value among all $d_{ij}$.

In 1998 D.Watts and S.Strogatz introduced such network parameter as clustering coefficient [Watts, 1998], characterizing the level of connectivity of nodes in the network, the tendency towards formation of groups of interconnected nodes, or so-called cliques. Besides, for a specific node, clustering coefficient indicates, how many neighbors of this node are also immediate neighbors of each other.

Clustering coefficient can be calculated for both a single node and the whole network. Clustering coefficient of a network is defined as the sum of respective coefficients of separate nodes, normalized by the number of nodes.

Clustering coefficient of a single node is calculated as follows. Suppose, $k$ edges come out of a given node, connecting it to $k$ other nodes, its immediate neighbors. If we assume that all immediate neighbors of the node are directly connected to each other, then we will get $1/2 \cdot k(k-1)$ edges between them. That is, this is the maximum number of edges that could possibly connect immediate neighbors of the given node. The ratio between the actual number of edges connecting immediate neighbors of a given node and the maximum possible number (i.e. the number of edges under which all immediate neighbors of the node are directly linked to each other) is called the clustering coefficient of the node $i - C(i)$. Naturally, the value never exceeds one.

Betweenness is a parameter, indicating how many shortest ways go through a node. This characteristic reflects the role of a given node in establishment of connections within the network. Nodes with maximum betweennes play the most important role in connecting other nodes of the network. Betweennes $b_m$ of the node $m$ is calculated according to the formula:

$$b_m = \sum_{i \neq j} \frac{B(i,m,j)}{B(i,j)} \ ,$$

where $B(i,j)$ is the general number of shortest ways between nodes $i$ and $j$, $B(i,m,j)$ is the number of shortest ways between nodes $i$ and $j$, going through node $m$.

We can consider "community structure" when we have node groups with high density of edges within them, while the density of edges connecting separate groups is low. Cluster analysis is a traditional method for detection of community structures. There are dozens of applicable methods for solving this problem. They are based on different measures of distance between nodes, weighted way indices etc. Particularly, for large social networks, the presence of community structures turned out to be an essential property.

Properties of real social networks also include so-called "weak" connections. Relations with distant acquaintances and colleagues can provide analogue of weak social connections. In some cases these connections turn out to be more efficient than "strong" ones. For instance, a group of researchers from the UK, USA, and Hungary, drew a conceptual conclusion in the area of mobile communication: "weak" social connections between individuals turned out to be the most important ones for the existence of a social network [Bjorneborn, 2004].

For research purposes the calls made by 4.6 million subscribers (i.e. 20% of population of a certain European country) were analyzed. That was the first case in the world practice when researchers managed to obtain and analyze such a large data set in the interpersonal communication sphere.

7 million social connections were detected in the social network of 4.6 nodes (i.e. mutual calls from one user to another and back, providing return calls were made within 18 weeks' period). Frequency and duration of the calls were used to define the strength of each social connection.

It was detected that weak social connections (one-two return calls during 18 weeks) were the ones holding the large social network together. Ignore these connections, and the network will fall apart into separate fragments. If you neglect strong connections, then the connectivity of the network won't be violated (Fig. 4.1). So, it turns out that weak connections are the phenomenon that connects the network into a unified whole.

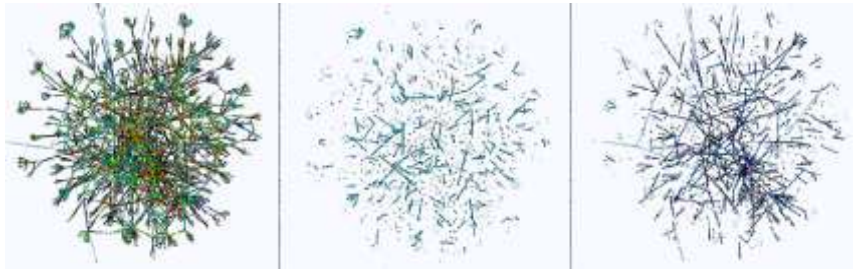

1                       2                       3
Fig. 4.1 – Network structure:
1) complete social communication map;
2) a social network, from which weak connections are withdrawn;
3) a network, from which strong connections are withdrawn: the structures maintains connectivity

In spite of vast sizes of some social networks, in many of them distances between any two nodes (geodesic distances) are relatively small. In 1967, as a result of large-scale experiments, psychologist S.Milgram calculated that an acquaintance chain of six links (on the average) connects two randomly chosen US citizens [Milgram, 1967].

D.Watts and S.Strogatz detected a phenomenon, common for many real networks, called Small Worlds effect [Watts, 1998]. While researching this phenomenon they suggested a procedure for construction of a visual model of a network, in which the phenomenon was witnessed. The three conditions of the network are presented on Fig. 4.2:

regular network – every node in it is connected to four neighboring nodes, same network where "close" connections are randomly replaced by "distant" ones (it is in this particular case that the small worlds effect emerges), and random network, where the number of such replacements exceeds some threshold value.

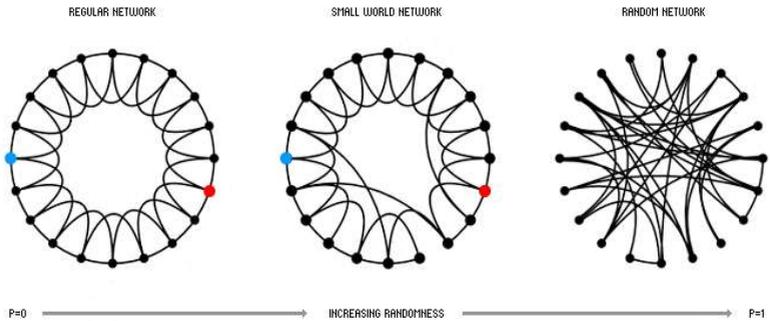

Fig. 4.2 – Watts-Strogatz model

In reality it turned out that the networks, where nodes simultaneously had certain numbers of local and random "distant" connections, displayed small world effect and high clustering degree at the same time.

Fig. 4.3 shows the change of average way length and clustering coefficient of an artificial network of D.Watts and S.Strogatz, depending on probability of establishment of "distant" connections (in semi-logarithmic scale).

For example, WWW is a network, for which the small world effect is confirmed. Web topology analysis, conducted by S.Zhou and R.J.Mondragon from London University, indicated that nodes with large numbers of outbound hyperlinks have more connections among themselves than with low-degree nodes, while the latter have more connections with larger-degree nodes than among themselves. This property was called the "rich-club phenomenon". The study indicates that 27% of all connections take place between just 5% of the largest nodes, while 60% are connections of other 95% of nodes with these largest 5%, and only 13% are connections between nodes, that do not belong to the leading 5%.

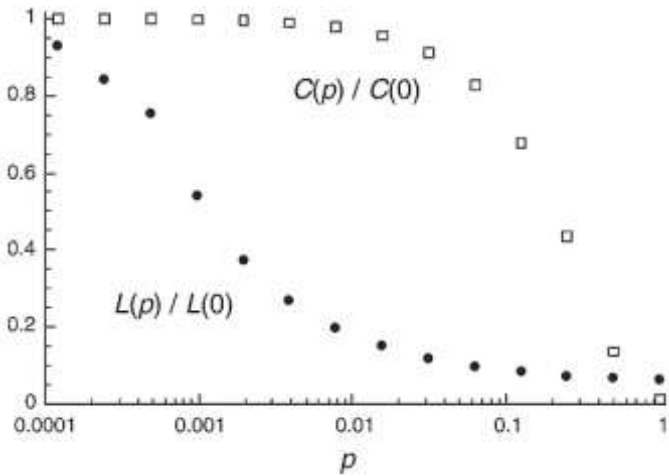

Fig. 4.3 – Change dynamics of way length and clustering coefficient in the Watts-Strogatz model in semi-logarithmic scale (axis 0X is the probability of replacement of close connections with remote ones)

The research provides grounds for assuming that dependence of WWW on large nodes is much more significant than expected, making the network even more vulnerable to deliberate attacks. The "small world" concept brings on a practical approach, called "network mobilization", implemented upon "small world" structure. Particularly, as a result of the small world effect, information dissemination speed in real networks increases, in comparison to random networks, as most pairs of nodes in real networks are linked by short ways.

It is empirically confirmed [Rothenberg, 2002], that terrorist networks are, most often, not just scale-free; they also possess the small world properties, so that presence of closely connected clusters (groups of strongly linked nodes) is ensured by local connections even in case of successful attacks, when concentrators (largest intermediaries) are disabled.

In the process of studies of "small worlds" an interesting approach emerged, associated with percolation concept [Broadbent, 1957], [Снарский, 2007]. As it happens, many issues, emerging during analysis of Internet security, are directly related to this theory. The simplest formulation of percolation theory (cleansed from all layers of physical and mathematical meanings) is as follows: "There is a network, where a random number of edges conduct the signal, while the remaining edges don't. The key question is: what is the minimum concentration of conducting connections, under which a way through the whole network exists?" Problems, solved within the boundaries of percolation theory and network analysis, include definition of limit conductivity level, change of way length and its trajectory when conductivity gets close to limit level, calculation of the number of nodes that should be disabled to breach the network connectivity.

Lately security experts started associating the "small world" effect with terrorist organization networks, the so-called overlay networks, built on top of the Internet.

By analyzing connections in a network we can learn about its important properties, for example, detect clusters, determine their constitution, connectivity differences within and between clusters, and identify key elements, connecting clusters with each other etc. At the same time, incomplete information on connections between separate nodes poses a serious obstacle for analysis.

Recently a group of researchers from Santa Fe Institute presented an algorithm, allowing us to automatically get information on the hierarchic structure of such networks [Clauset, 2008]. The new method of connection reconstruction can become a handy tool for both special agencies and competitive intelligence units of companies. For example, knowledge of just a half of connections between terrorists might allow you to reconstruct, with high probability, the missing links of the whole chain.

Even without complete description of a system one can get a representative set of connections, and try to reconstruct the whole network based on it. Analysis of the graph you get in the process, allows you to detect potentially important connections, which were impossible to detect in a real network. For example, once you have information on just half of contacts among network participants, with the probability of 0.8 you can predict connections, which you were unaware of initially. Obviously, the approach can be very useful for detection of hidden network groupings, and, thus, allow users to bring the task of ensuring state and business security to a conceptually new level.

For analysis of complex networks of terms and concepts, mentioned in specific documents from information streams, deep text analysis (content-monitoring, to be exact) methods, and extraction of such concepts as persons, companies, toponyms (geographic names) etc, can be used.

One of the directions of social network analysis is visualization, which means a lot, because often it allows users to draw important conclusions regarding the nature of interaction between subjects-nodes without resorting to rigorous analysis methods. While displaying the network model it might make sense to:

- place network nodes in two dimensions;
- order objects in space in one dimension in accordance to certain qualitative properties;
- use the methods, which are common for all network diagrams, for displaying qualitative and quantitative properties of objects and relationships.

## 4.2. Network-specific features of information operations

As an extension of the above-mentioned multi-agent information dissemination model we can consider a model, in which the structure of the network being formed is taken into consideration [Пугачев, 2015]. Within this model every agent (information source), instead of "potential" has a certain rating (which corresponds to the size of the respective node on the figures). Connections in such a network correspond to facts of reprinting or "retelling".

The model is based upon the assumption that during information operations the top-rated sources reprint information from low-rated ones, or clusters of low-rated publishers reprinting one and the same piece of news are formed.

Fig. 4.4 displays an example of typical information operations, recognized within the described model.

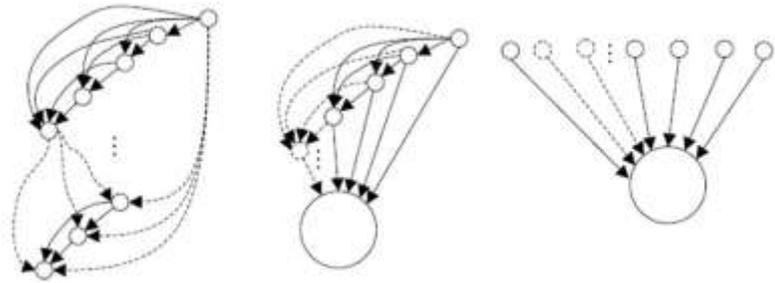

Fig. 4.4 – Examples of information dissemination networks that have the features of information operations

As part of formalization of this same model several dozens of information dissemination network topology parameters are selected, including diameter, density, clustering, betweenness etc, which are compared to certain benchmark values.

The advantages of this model include formal strictness and correspondence to the rapidly developing modern concept of Complex Networks. Thus, we should anticipate further development of the Complex Networks field in the nearest time. The disadvantages of the model, probably, include low correlation with the content aspect of information operations under consideration, as well as certain computational difficulty during detection of fuzzy information duplicates of documents.

## *Studying mutual impact of information sources: technological phases of the process*

For efficient study of mutual impact of information sources from the Internet (web-resources, social media) we suggest a sequence of steps, i.e. information processing stages, every one of which allows us to obtain an analytical product. The combination of these stages, based on usage of necessary and available tools and special techniques, can be considered a procedure for conduction of activities, targeted at obtaining of analytical materials, including construction and analysis of a network of their mutual impact.

During such content monitoring-based information and analytical research the topical problems might include the following ones:

– Finding relevant publications on the given topics.
– Detection of mutual contextual links and reprints in the documents, provided by different information sources.
– Construction of impact network, analysis and visualization of mutual connections of information sources, including ranking of nodes of the constructed network according to their impact degrees.
– Detection of potential information operations and building scenarios for counteracting information operations in the network environment.

### *Obtaining of a representative array of publications*

In order to get a representative array of publications on a selected topic we should select the content-monitoring system that provides a stream of information communications on a certain topical scope. The topical scope may be described by a query on a language of an information search system.

The authors chose InfoStream content-monitoring system that, as of now, covers about 10 thousand Ukrainian and Russian-language information sources. Every day more than 100 thousand documents are added to the system's data bases. InfoStream system performs search and allows users to view the lists and full texts of relevant documents.

The example displayed on Fig. 4.5 shows a fragment of the system's interface, which processes the query regarding the discussion of resignation of the Prime-Minister of Ukraine A.Yatesenyuk in January 2016.

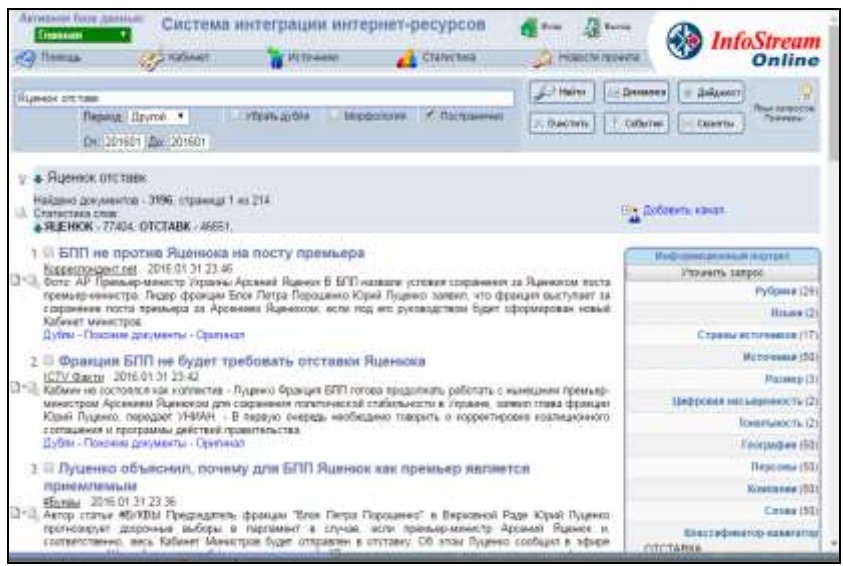

Fig. 4.5 – A fragment of content-monitoring system interface

As a result, a thematic information array was formed. It covered 3196 documents.

## *Definition of context links*

The basis for information source impact network is provided by contextual links and reprints within the topical information stream. Contextual links are recognized through identification of templates (Table 1) within the documents of the selected information array and signs of exact reprints, identified by plagiarizing detection methods. The templates themselves, in their turn, are regularly redefined and appended by the experts in an automated mode through contextual analysis of document stream of the content-monitoring system based on Text Mining methods.

## *Construction of information source impact network*

Contextual links and reprints, found in the texts, allow us to build a citation matrix. By transposing it in accordance to the above-listed hypothesis, we can build an impact matrix. Source impact network corresponds to this matrix. An example of network visualization for the above-mentioned topical information array using the Gephi system is provided on Fig. 4.6.

## *Research of the information source impact network*

The newly-built information source impact network can be researched using the generally-acknowledged tools (for example, Gephi system allowed us to obtain the following parameters of the network: number of nodes – 141, number of edges – 196, graph density – 0.01, average clustering coefficient – 0.026, average way length – 1.26 etc).

Table 1. Templates of information resource names (a fragment)

| # | Source code | Template 1 | Template 2 |
|---|---|---|---|
| 1 | srd06193 | Depo | "Депо" |
| 2 | srd03176 | Українські національні новини | УНН |
| 3 | srd00045 | Сегодня.ua | "Сегодня" |
| 4 | srd03076 | ТСН.ua | "ТСН" |
| 5 | srd07509 | 112.ua | "112" |
| 6 | srd02348 | Gazeta.ua | |
| 7 | srd00069 | Корреспондент.net | "Корреспондент" |
| 8 | srd07487 | Еспресо TV | "Еспресо ТВ" |
| 9 | srd02535 | Телеканал новин "24" | "24" |
| 10 | srd06453 | Телеграф.com.ua | "Телеграф" |
| 11 | srd01351 | ЗІК | "ЗІК" |
| 12 | srd02514 | РБК-Україна | РБК-Украина |
| 13 | srd04508 | Українські Новини | |
| 14 | srd07686 | "Антикор" | |
| 15 | srd00253 | "Обозреватель" | |
| 16 | srd00057 | Интерфакс | Інтерфакс |
| 17 | srd00404 | ICTV Факти | ICTV |
| 18 | srd04125 | РИА Новости Украина | |
| 19 | srd02732 | УКРІНФОРМ | УКРИНФОРМ |
| 20 | srd00095 | УНІАН | УНИАН |
| 21 | srd00094 | Українська правда | |
| 22 | srd01408 | Цензор.Нет | |
| 23 | srd00064 | ЛІГАБізнесІнформ | |
| 24 | srd00039 | Газета День | |
| 25 | srd07038 | Вести.ua | |

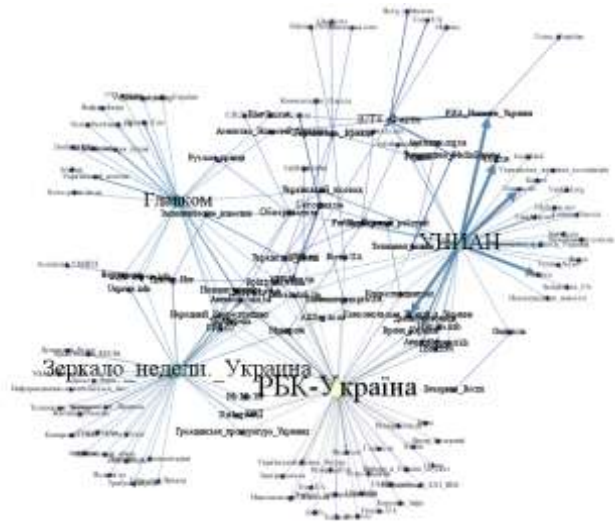

Fig. 4.6 – A fragment of a network of connections between sources within a selected thematic scope

Weights of network nodes play an important role for content analysis; the list of important nodes (according to output capacity) is provided in Table 2.

Table 2. The most influential nodes according to numbers of citations

| # | Web-resource | Output degree |
|---|---|---|
| 1 | РБК-Україна | 50 |
| 2 | Зеркало недели | 38 |
| 3 | УНИАН | 35 |
| 4 | Главком | 28 |
| 5 | ICTV-Факты | 10 |
| 6 | Сегодня.ua | 7 |
| 7 | Українська правда | 7 |
| 8 | Обозреватель | 6 |
| 9 | Forbes-Украина | 4 |
| 10 | Цензор.Нет | 3 |

A promising approach to ranking of sources in accordance to impact degrees is the HITS algorithm, suggested by J.Kleinberg in [Kleinberg, 2006].

HITS algorithm ensures selection of the best "authors" from the network (the nodes the links lead to) and "mediators" (nodes, from which inclusive links outgo).

According to HITS algorithm, for each network node its significance as an author $a(v_j)$ and mediator (hub) is calculated $h(v_j)$ according to formulas:

$$a(v_j) = \sum_{i \to i} h(v_i);$$
$$h(v_j) = \sum_{i \to i} a(v_i).$$

In these formulas summation is done across all nodes that refer to the given node (or across nodes that the given node refers to).

If we reformulate the denominations offered by [Kleinberg, 2006], particularly, replace "authorship" with "proneness to impact" and "betweenness" with "impact significance", we can calculate the respective characteristics of impact network nodes with relatively small computational efforts.

Besides that, a significant role in detection of information impacts belongs to identification of "latent" connections. The methodology for detection of latent (hidden) connections and impacts is set forth in [Snarskii, 2016].

## *Recognition of potential information operations*

Information source impact network allows us to efficiently identify potential information operations according to approaches suggested in [Потемкин, 2015]. We assume that the probability of an information operation happening is low, if the information on some event is initially generated by an influential information source, and then reprinted (with respective references or without them) by less influential sources (Fig. 4.7). Opposite scenarios, when more influential publications reprint information from less influential (although sometimes, numerous) ones, may be a signal of an information operation or attack (Fig. 4.8). It is these particular situations that were witnessed during network analysis of real topical information streams (Fig. 4.9).

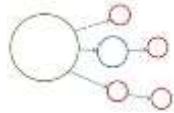

Fig. 4.7 – A typical information dissemination scenario

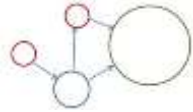

Fig. 4.8 – Information dissemination scenario, common for an information operation

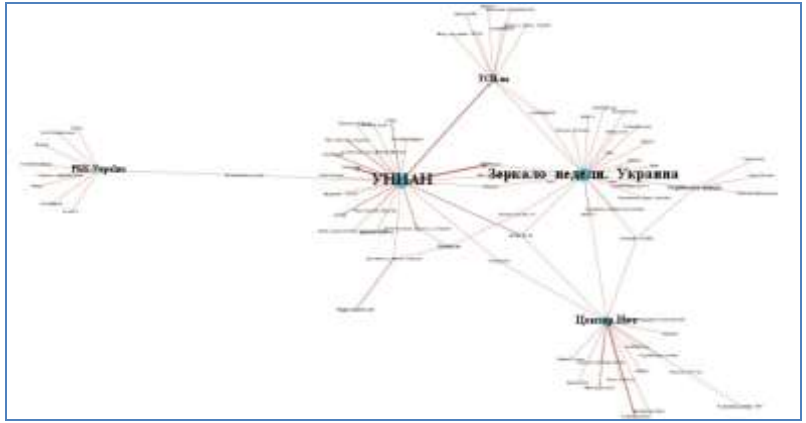

Fig. 4.9 – A fragment of a network of connections between information sources within specific topical scope

# 5. Decision-making support during information operation detection

Planning of activities on strengthening of information security and prevention of negative/enemy information impacts and information operations (as well as planning of successful operations in the process of information struggle) calls for clear understanding and knowledge of the subject domain. However, the information security sphere is a weakly-structured subject domain which is difficult to describe, particularly, in formal quantitative terms. Papers [Андрейчук, 2016; Andriichuk, 2017; Каденко, 2016; Kadenko, 2016] show that information operations (IO), indeed, belong to weakly-structured subject domains. Expert data-based decision support technologies are a powerful tool for problem solving in these domains [Тоценко, 2002]. That is why, the aspects of usage of expert decision support technologies in the information security sphere and specific context of usage of particular methods deserve to be addressed separately.

## *Information security as a weakly-structured subject domain*

In [Таран, 2006] the following properties of weakly-structured subject domains are listed (fig. 5.1): absence of functioning goal which could be formalized, absence of optimality criterion, uniqueness, dynamics, incomplete description, presence of subjective human factor, impossibility of analytical model building, lack of benchmarks, high dimensionalities.

In Chapter 1 it is stated that informational operations are influenced by numerous solely qualitative criteria, factors, and parameters (including socio-psychological ones). It is problematic to provide a formal mathematical (analytic) description of these factors.

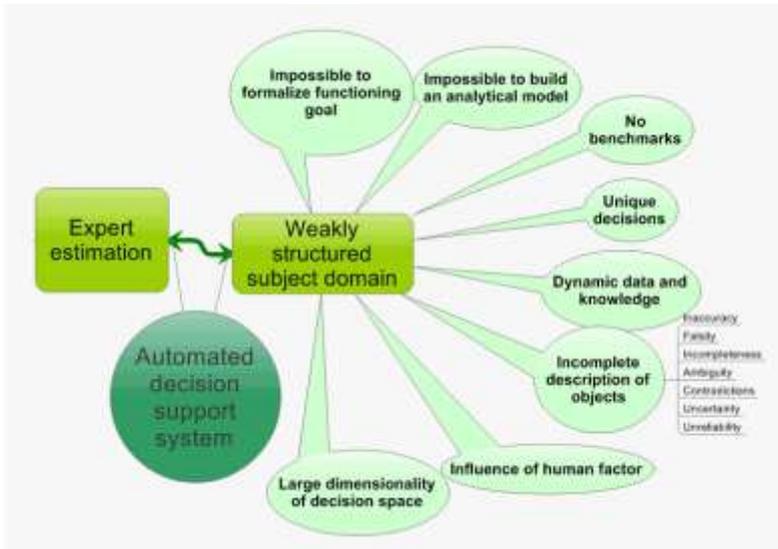

Fig. 5.1 – Properties of weakly-structured subject domains

Authors of [Горбулін, 2009] point out the impossibility of development and practical application of some universal methodology for modeling of informational operations, first and foremost, due to weak formalization of concepts and factors. These authors stress, that in each particular case one should consult the analysts (i. e. information operation experts) and rely on their competency. The analysts are sometimes able to build accurate forecasts of certain dependencies, which are later confirmed by practice. Analysts (experts) should be consulted to provide description of subjective factors. When it comes to objective factors, their description and analysis can be performed using well-known methods, which operate determined data, including mathematical statistics and analysis of time series. However, these methods are targeted only at description of formal aspects, and do not touch upon content-related aspects. As a result, authors of [Горбулін, 2009] stress the necessity of extension of technological arsenal, which can be used for analysis and modeling of informational operations.

As we can see, informational operations (like all other operations, requiring human participation) represent a vivid example of a weakly-structured subject domain. We feel that expert data-based decision support technologies should become one of the technical tools for analysis and modeling of informational operations. The relevance of expert data usage in weakly structured subject domains is also corroborated by the research, conducted by Delphi Group, listing the sources of knowledge, possessed by organizations [Тузовский, 2005]: electronic documents - 20%, paper documents - 26%, electronic knowledge bases - 12%, minds of specialists - 42%.

The research indicated that a considerable share of knowledge was stored not in the databases, or on paper and digital mediums, but rather in the minds of experts (analysts, specialists). Consequently, in the context of description and analysis of informational operations, expert knowledge should, definitely, be involved, especially when it comes to subjective qualitative factors.

## 5.1. Hierarchic decomposition and complex target-oriented dynamic evaluation of alternatives

As it is stated in [Горбулін, 2009], an informational operation is an inter-disciplinary set of methods and technologies that encompasses multiple spheres, from military science to sociology. At the same time, there is no universal standard information operation technology (which could serve both the military and the management of large governmental or business agencies). So, according to Gorbulin et al, development of a scientific basis of informational operations remains an extremely relevant issue.

With the inter-disciplinary nature of informational operations in mind, we feel that expert knowledge-based hierarchic decomposition should become a handy informational operation description tool. Particularly, a hierarchic approach provides the basis for the method of complex target-oriented dynamic evaluation of alternatives (MCTDEA), suggested by [Тоценко, 2002] and further improved by V.Tsyganok [Цыганок, 2013]. The method allows a decision-maker (DM) to aggregate a large quantity of criteria of diverse nature (i.e. belonging to different spheres), that influence a specified main goal, into a unified hierarchy.

Depending on the type of a specific informational operation (offensive or defensive [Горбулин, 2009]), the analyst (expert) or the DM himself can formulate the main goal. Any information operation is decomposed into certain stages (or steps), which are listed in [Горбулин, 2009]. The contents of these steps can vary, depending, again, on the type and context of the operation. For example, during modeling and decomposition of an informational attack against the Academy of sciences (using MCTDEA), a goal formulated as "Discredit an academic institution in the media" can be decomposed into the following lower-level sub-goals: "Discredit academic papers and achievements" and "Discredit academic researchers".

In MCTDEA decomposition is performed down to the level of "atomic", elementary sub-goals (factors, criteria) that can be directly influenced by the DM. These goals are called projects, and, generally, can be characterized by a certain value (absolute/relative numeric, Boolean, or threshold-type).

The general conceptual task of decision support methods, involving hierarchic decomposition of a problem, particularly, MCTDEA and analytic hierarchy/network process (AHP/ANP) [Saaty, 2008], is to build a rating or ranking of objects (alternatives, projects). Based on such rating, a DM can make an informed choice of the best alternative (decision variant) from a given set, or set priorities in his/her activity (i.e., determine, which factors or actions are most important for achievement of a given main goal). In order to build such a rating, one should define the relative importance of all goals in the hierarchy graph, built by experts (or by knowledge engineers through dialogues with experts). In order to define the relative importance of sub-goals of a given goal (its "descendants" in the hierarchy graph), experts should compare them pair-wise (unless they are able to provide direct estimates). Evaluation of impacts (weights) can be conducted by experts in different pair-wise comparison scales. Recent research of V.Tsyganok [Цыганок, 2011; Цыганок, 2012] has shown that an expert should be given an opportunity to input every single pair-wise comparison value in the scale, which is most convenient for him or her (i.e. which reflects his/her knowledge of the subject domain most accurately). When the experts have evaluated all relative weights (priorities) in the hierarchy graph, relative impacts of all the projects upon the main goal (their relative efficiencies) can be calculated as shown in [Тоценко, 2002].

If evaluation is done by several experts, then a few important aspects should be taken into consideration. The first aspect is expert competence. If it is known that the experts, who evaluate criteria, projects, or alternatives, have different competencies, their relative competence levels should be calculated based on three components: self-assessment, mutual assessment, and objective data (as shown in [Тоценко, 2002]). Differences in relative competence levels of the experts can be neglected only when the size of the expert group is large enough [Tsyganok, 2012]. The second important aspect is consistency of estimates, provided by different expert group members. Expert estimates' consistency should be checked (as recommendations to the DM, based on inconsistent or incompatible expert data will provoke distrust). In order to evaluate consistency level of expert estimates, in our view, it is appropriate to use so-called spectral methods, described, for instance, in [Zgurovsky, 2004] and [Olenko, 2016].

The advantage of spectral methods over other approaches to consistency evaluation (for example, those suggested by Saaty and colleagues [Forman, 1998]) is as follows. If necessary (i.e., if consistency level of expert estimates in the group is low), spectral methods allow the knowledge engineer to organize step-by-step feedback with experts. Experts are requested to change the respective outlying estimates so as to make overall consistency level reach the required threshold. When estimates, provided by different members of the expert group, achieve the level of consistency, that is high enough, they can be aggregated (used for calculation of generalized group estimates). For aggregation of expert estimates, we suggest using the combinatory aggregation method [Циганок, 2000]. The most important of the method's advantages over other aggregation methods are as follows. First, it can be used for aggregation of incomplete pair-wise comparison matrices. Second, it utilizes the redundancy of information most thoroughly.

In the context of MCTDEA the "weighted" hierarchy of criteria (goals) is called the knowledge base (KB) of the subject domain. This paper is focused on subject domains, related to information security, particularly, information operations. In terms of content, such KB represents one of the types of subject domain models. The KB is built by experts (or by knowledge engineers through dialogues with experts) using special software tools – automated decision support systems (DSS).

We should note that MCTDEA does not require all the data, input into DSS KB, to be represented by expert estimates only. For instance, when it comes to comparison of several alternatives according to some criterion, estimate values should not necessarily be expressed by grades of pair-wise comparison scales. Sometimes, the values can be absolute ones, available from open information sources. For example, in order to analyze the information policy or campaign of some organization absolute values are often used, such as "number of publications with negative flavor per month". An indicator like this "has all the rights" to be included into a hierarchy of criteria, describing the information policy of an organization.

## 5.2. Peculiarities of working with experts

Research indicated that a considerable part of knowledge was located not in the knowledge bases or on paper and electronic mediums, but rather in the minds of the experts (analysts, specialists). An expert (analyst), usually, comes from a narrow-focused subject domain and, in the general case, is not familiar with decision support methods and technologies. Consequently, it is extremely important to make the process of expert data input into the DSS as comfortable for experts as possible. Formal side of the process (expert data formalization) should be delegated to the knowledge engineer, and mathematical calculations – to the automated DSS. In order to achieve this objective during expert examination it is appropriate to keep several important issues in mind.

First, if an expert is completely new to decision support technologies, it makes sense to familiarize him/her at least with the general agenda of the examination. In the ideal case examination participants should be given a thorough explanation of the whole technology, used to process their estimates and to form recommendations for the DM as to decision variant selection. Such an explanation will increase the level of trust the experts have towards the process and allow them to input data into the DSS in the most convenient format. So, before starting to collect information from the experts, it is reasonable to hold coaching sessions with them.

Second, criteria (goal) hierarchy (fig. 5.3) should be well-balanced. It is preferable to locate all the projects on one level (because, as it is shown in [Saaty, 2008], their weights should belong to one order of magnitude). With human psycho-physiological constraints taken into consideration, any goal in the hierarchy graph should not have more than 7±2 descendants [Miller, 1956]. In the process of hierarchy building, it is preferable not to "bother" the expert with multiple similar questions (particularly, concerning positive or negative character of impact of goals upon their common "ancestor" in the graph, or pair-wise compatibility of "descendant" goals).

Third, when evaluating the impacts, verbal values should be used rather than their numeric equivalents (for instance "1" – equality, "2" – weak preference, ... , "5" – extremely strong preference).

On the whole, interface of an automated DSS should be user-friendly, intuitively understandable, and easy to master.

More detailed analysis of peculiar features of working with experts is provided in [Каденко, 2016]. Let us now address the problems, emerging at different phases of dialogue with an expert.

## Coaching sessions, preceding the examination

As we have already mentioned, the first thing an expert examination organizer should remember: in the general case the DM and experts are not familiar with the mathematical methods that are used to input and process expert data. If the DM and experts obtain some understanding of the overall process of the examination, it will be easier for them to input data in the required format. So, before the examination begins, it makes sense to hold the respective coaching sessions. The author of the AHP, Tom Saaty [Saaty, 1996] and his followers pay considerable attention to such coaching sessions.

## Peculiarities of working with experts at the hierarchy building phase

At the phase of hierarchy building we should take several peculiar features of working with experts into account.

1) We should keep in mind the "balance" of the hierarchy: projects (atomic, non-dividable sub-goals) are to be located at the same level. Otherwise at the phase of priority calculation a project, located at a higher level will get much greater weight than projects located at lower levels. This weight will be even larger if this project influences several goals, because its weight (i.e. impact upon the main goal) is calculated as a sum of products of relative impacts across all ways leading from the project to the main goal in the hierarchy graph [Тоценко, 2002; Каденко, 2008]. Adding of a level to a hierarchy of criteria decreases the magnitude order of weights of goals (projects), located at lower levels.

The problem of optimal hierarchy structure definition was addressed, for example, by Stan Lipovetsky [Lipovetsky, 2006; Lipovetsky, 2014].

2) Experts should be focused on the decomposition of the main goal, leaving other, more general concepts, outside the scope of their attention.

3) In order to prevent ambiguity, emergence of sub-goals and projects with similar formulations in the hierarchy should be avoided [Андрійчук, 2014].

4) The process of input of links (edges) and impacts into the hierarchy should be clear and consecutive; from this perspective, it is more convenient for an expert to select sub-goals and projects from the list of already input formulations, or input new formulations into an already available list.

5) During dialogue with an expert multiple repetitions of similar procedures (such as determination of impact nature (positive or negative) or pair-wise verification of compatibility of goals) should be avoided. For instance, in order to verify compatibility of $m$ goals, an expert has to answer $m\ (m\text{ - }1)\ /\ 2$ similar questions.

6) Special attention should be given to projects; their atomic nature should be stressed (for this purpose it makes sense to ask an expert a clarifying question, such as: "can this sub-goal be considered a project? Or can it be further decomposed?").

7) We should remember about psycho-physiological limitations of a human; so during hierarchy building the allowed number of immediate sub-goals of every goal should not exceed $7 \pm 2$ [Miller, 1956].

Table 5.1. Saaty's fundamental scale grades and their verbal equivalents

| Numerical values | Verbal term | Explanation |
|---|---|---|
| 1 | Equally important | Two elements have equal importance |
| 3 | Moderately more important | Experience or judgment slightly favors one element |
| 5 | Strongly more important | Experience or judgment strongly favors one element |
| 7 | Very strongly more important | Dominance of one element proved in practice |
| 9 | Extremely more important | The highest order dominance of one element over another |
| 2,4,6,8 | Important intermediate values | Compromise is needed |

## *Avoiding quantitative values at the estimation stage*

At criterion (factor) impact estimation stage it is desirable to avoid quantitative values and offer an expert to use their verbal equivalents instead ("the best", "the worst", "better", "worse", "select the best alternative", "is superior to", "is inferior to"; "rank the alternatives" instead of "rank 1" or "rank 5"; "strongly dominates", "is extremely weaker" instead of "5 times better", "9 times worse").

Ranks and cardinal values (including pair-wise comparisons) should be brought into compliance with verbal equivalents automatically (as shown in table 5.1).

As an example, let us consider the correspondence between Likert agreement scale grades and their verbal equivalents [Likert, 1932]: 1 – "totally disagree", 2 – "disagree", 3 – "unable to make a choice", 4 – "agree", 5 – "totally agree". Such a scale is used in multiple contemporary surveying methodologies, for example in Space Shaper methodology of location efficiency evaluation [St. James Park, 2007]. Respondents do not have to deal with numbers at all. Their answers of "agree or disagree" kind are adequately processed, then general satisfaction indices are calculated, and on their basis illustrative spider-web or bar diagrams are built (fig. 5.2).

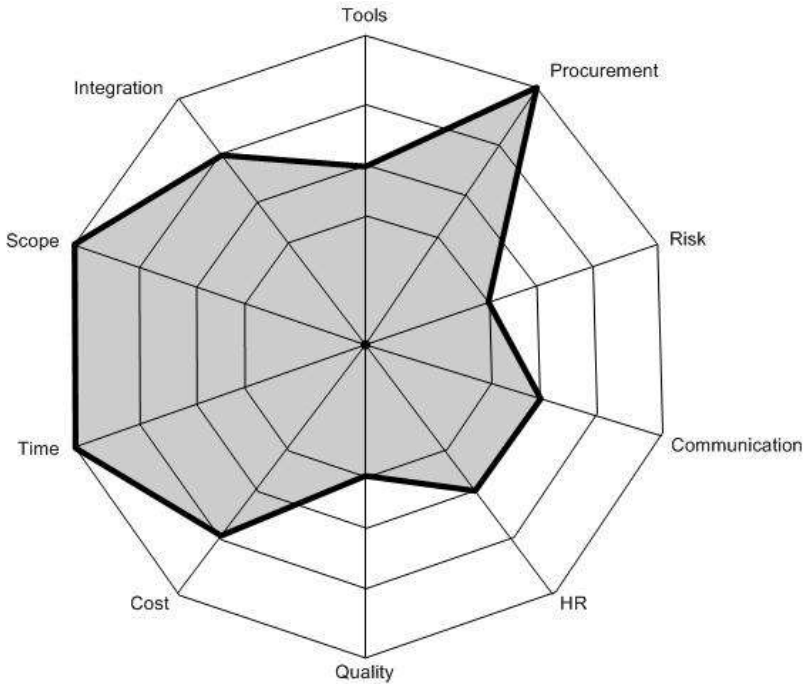

Fig. 5.2 – Spider-web diagram example

Contemporary expert information collection and processing methodologies, as well as automated DSS interfaces should be developed with the listed peculiarities of dialogue with experts in mind. In the context of working with experts, the key concerns of methodology and DSS developers should be to create comfortable working environment for the experts and to delegate as many of the required logical and mathematical operations to the automated DSS as possible. Some of the modern DSS methods (such as combinatorial approach, described, for instance, in [Цыганок, 2000]) allow us to use expert information most thoroughly. (Under conditions of connectivity of preference graph on a set of alternatives, thanks to redundancy of expert information (even if it is provided in the form of an incomplete pair-wise comparison matrix, input by a single expert in verbal scale), the combinatorial approach always allows the user to calculate alternative weights).

As we can see, although procedures of obtaining data from experts and inputting these data into the DSS may seem trivial, they are characterized by certain important properties. The key phases of expert examination (particularly, those involving experts) have been analyzed. A set of recommendations for knowledge engineers, expert examination organizers, and DSS software developers has been suggested. These recommendations allow us to make the process of obtaining of expert data and their input into DSS more constructive, and, thus, increase the level of trust of the DM towards the expert examination results.

## 5.3. A methodology of DSS application during information operation detection

Information sources exert considerable influence upon people. During the last few years it became evident that mass media could be efficiently used for spreading disinformation. Besides that, social experiments indicate that many people believe in unconfirmed news and keep spreading this news themselves. For instance, a paper by [Lewandowsky, 2012] presents an overview of known false statements and disinformation in American society. [Berinsky, 2017] describes actual social experiments, in which people's proneness to believing political rumors related to healthcare reform launched by the Congress in 2010, was studied. Depending on the way of information presentation, 17-20% of experiment participants believed in rumors, while 24-35% of participants did not have any specific opinion, and 45-58% of the respondents denied the rumors.

According to [Горбулін, 2009; Додонов, 2016; Ландэ, 2016b] an information operation (IO) is a complex of information activities (news articles published online and in papers, news on TV, comments in social networks, forums etc), aimed to change public opinion about a certain object (person, organization, institution, country, etc). For example, by spreading rumors about problems in a bank, one can make deposit holders withdraw their deposits, and thus, push the bank towards bankruptcy. In general, these are the disinformation activities.

[Горбулін, 2009; Додонов, 2016; Ландэ, 2016b] describe IO recognition methods, relying on analysis of time series, built based on content-monitoring of thematic information flow. Let us outline a set of problematic situations which can emerge during IO identification due to shortcomings of the respective methods and techniques:

1) Against the background of a fairly large number of publications about the IO object the number of publications (planted information precedents) regarding its separate component can be very insignificant, and, as a result, the respective systematic violations of information scenario dynamics (such as "Mexican hat" and Morlet wavelets on the scalogram) will remain undetected. Some IO can be complex ones, and respective information causes can be planted in a stepwise manner and concern different components of IO object at different moments of time. If their number is blurred against the general number of publications regarding the IO object ("information noise"), and respective information attacks remain undetected, we can miss the beginning of an information campaign on discrediting of IO object, and a certain informational damage to its image will not be taken into account.

2) Content-monitoring means working on queries, consisting of keywords; as a result the respective publications are found. Keywords are formulated based on IO object name. However, a complex IO object can have a considerable number of components with respective names that are not considered within the queries, and, as a result, not all publications on the given topic are found.

3) Queries concerning IO object have different degrees of importance in relation to IO components they touch upon. Lack of information on values of these importance degrees (i.e. their equivalence) leads to decrease of adequacy of IO model.

In order to overcome the listed drawbacks we suggest using the following methodology of DSS application during IO detection.

## *Methodology contents*

The contents of the suggested methodology of expert decision-making support application to IO identification is as follows:

1) Preliminary study of the IO object is performed; its target parameters (indicators) are selected. Then we assume that before (in retrospective) IO against the object already took place, and its state (i.e. respective target indicators) worsened as a result of IO.

2) Group expert examination on decomposition of IO goals and estimation of impact degrees is performed. The IO object is decomposed as a complex weakly structured system. For this purpose distributed expert information collection and processing tools (DEICPT) are used. To obtain thorough and undistorted expert information, expert estimation system (EES) is used.

3) The respective knowledge base (KB) is built using the DSS tools, based on the results of group expert examination conducted by DEICPT and available objective information.

4) Dynamics of thematic information flow is analyzed using content-monitoring system (CMS) tools. DSS KB is appended.

5) Based on the constructed KB, using DSS tools, recommendations for the decision-maker are calculated. For this purpose the degrees of IO goals are calculated in retrospective and compared to the respective changes of the object's state. Average values of IO goal achievement degrees, under which the target parameters of the IO object significantly worsened, are calculated. Thus, through monitoring of the IO object's state during the current period of time, we can predict worsening of target IO object parameter values based on comparisons of calculated current IO goal achievement degrees with the above-mentioned average value. Under statistically sufficient set size and sufficient correlation between IO goal achievement degrees and deterioration of target IO object parameters, we can forecast quantitative values of target IO object parameters for the current period of time.

The advantages of the suggested methodology are as follows:

1) High degree of detail of the model – against large number of publications about the IO object in general, changes of dynamics of the number of publications, induced by planted information about one of IO object components, will be insignificant, and, consequently, they might remain undetected.

2) The volume of identified topical publication increases, as there are more queries and keywords.

3) Weights of IO components allow us to prevent a situation when all components are equally important. IO model, built in such a way, is going to be more adequate.

4) Once built, the KB can be used in future during a significant period of time, without the need to perform new expert examinations.

5) Usage of DEICPT allows experts to work in the global network, and, thus, ensure savings of time and resources.

The disadvantages of the methodology are as follows:

1) Application of expert technologies calls for time and financial expenditures, which are necessary to perform group expert examination. Besides that, we should actualize the KB in time, in order to be able to use it in future.

2) Complexity and, sometimes, ambiguity of representation of certain rather sophisticated formulations of IO components as queries for content monitoring system.

## *Methodology application example*

Let us demonstrate the methodology, suggested in the previous section, on an example of an information operation against the National academy of sciences (NAS) of Ukraine. As we know, the NAS of Ukraine has seen better times. During recent years financial situation has been constantly deteriorating: budget of the NAS of Ukraine has been decreasing, and its share within the national budget has been decreasing as well. This process can be tracked if we take a look at data on the distribution of expenditures of the State budget of Ukraine, say, in 2014 – 2016 [Закон України, 2014; Закон України, 2015; Закон України, 2016]. Let us assume that such decrease of funding volumes results from an information operation against the NAS of Ukraine.

Group expert examination was performed using DEICPT "Consensus-2" [Циганок, 2017]. After that the KB was built using the DSS "Solon-3" [Тоценко, 2003]. For expert estimation the EES "Riven" ("Level") was used [Циганок, 2012b].

As a result of group expert examination 15 expert formulations of anti-NAS IO components were obtained:

1) Bureaucracy in the NAS of Ukraine;

2) Inefficient staff policy of the NASU;

3) Corruption in the NAS of Ukraine;

4) Underestimation of the level of scientific results obtained by the NAS of Ukraine;

5) Lack of applications of scientific research results in production;

6) Underestimation of international cooperation level;

7) Improper and inefficient usage of the NAS property;

8) Improper and inefficient usage of the NAS land resources;

9) Discrediting the president of the NAS of Ukraine;

10) Discrediting the chief executive officer of the NAS of Ukraine;

11) Discrediting other well-known persons from the NAS of Ukraine;

12) Contraposition of scientific results of the Ministry of education and science and the NAS;

13) Contraposition of scientific results of the NAS of Ukraine and other academic organizations;

14) Contraposition of achievements of Ukrainian companies and the NAS of Ukraine;

15) Contraposition of scientific results of foreign organizations and the NAS of Ukraine.

Using the tools of CMS InfoStream [Григорьев, 2007], dynamics of topical information stream was analyzed. For this purpose, based on each of the above-listed IO components, queries were formulated on a specialized language. According to these queries, further analysis of publication dynamics on target topics was performed.

Express-analysis [Ландэ, 2016b] of topical information stream, related to the IO object – NAS of Ukraine – has been performed. As a result of express-analysis, using the CMS InfoStream, the respective topical information stream from Ukrainian web-space segment has been obtained. In order to detect planted information using the available analytical tools, dynamics of publications on the target topic was analyzed.

In order to detect the degree of similarity of the respective time series and the IO diagram under different scales, "wavelet-analysis" was applied. Wavelet coefficients indicate, to which extent the behavior of the process in a certain point is similar to the wavelet of a certain scale. At the respective wavelet-scalogram we can see all characteristic features of the initial series: scale and intensity of periodic changes, direction and significance of trends, presence, location, and duration of local peculiarities.

IO dynamics is most precisely reflected by "Mexican hat" and Morlet wavelets [Додонов, 2015b]. That is why we are analyzing the time series, corresponding to each of the 15 IO components during the 4 periods (01.01.2013 –

31.12.2013, 01.01.2014 – 31.12.2014, 01.01.2015 – 31.12.2015, and 01.01.2016 – 15.12.2016) and detecting the above-listed wavelets.

Based on detected planted information precedents (attacks) and their parameters (location and duration) the knowledge engineer appended the KB of "Solon-3" DSS. Particularly, planted information precedent on the IO component "Underestimation of scientific results of the NAS of Ukraine", created on 30.11.2015, which lasted for 14 days, was detected. Consequently, an additional parameter of the project "Underestimation of scientific results of the NAS of Ukraine" was introduced: "project implementation duration" of 14 days; besides that, the characteristic parameter of 10-months' delay of project impact was introduced as an additional parameter of impact of the project "Underestimation of scientific results of the NAS of Ukraine" upon the goal "Discrediting the scientific results of the NAS of Ukraine". For other detected planted information precedents, project and impact characteristics were input in a similar way.

So, for the period 01.01.2015 – 31.12.2015 the KB, shown on fig. 5.3, fig. 5.4, and fig. 5.5, respectively, has been appended.

We should note that for certain IO components, namely, "Corruption in the NAS of Ukraine", "Bureaucracy in the NAS of Ukraine", "Inefficient staff policy of the NASU", "Improper and inefficient usage of the NAS land resources", and "Improper and inefficient usage of the NAS property", two information precedents planted for each IO component during 2015, were detected; consequently, for calculation purposes, the respective projects were input into the KB 2 times. For example, the IO component "Bureaucracy in the NAS of Ukraine" corresponds to two projects: "Bureaucracy in the NAS of Ukraine 1" and "Bureaucracy in the NAS of Ukraine 2". Each of them has different characteristics of duration (9 and 15 days), and respective impacts have different delays (9 and 11 months).

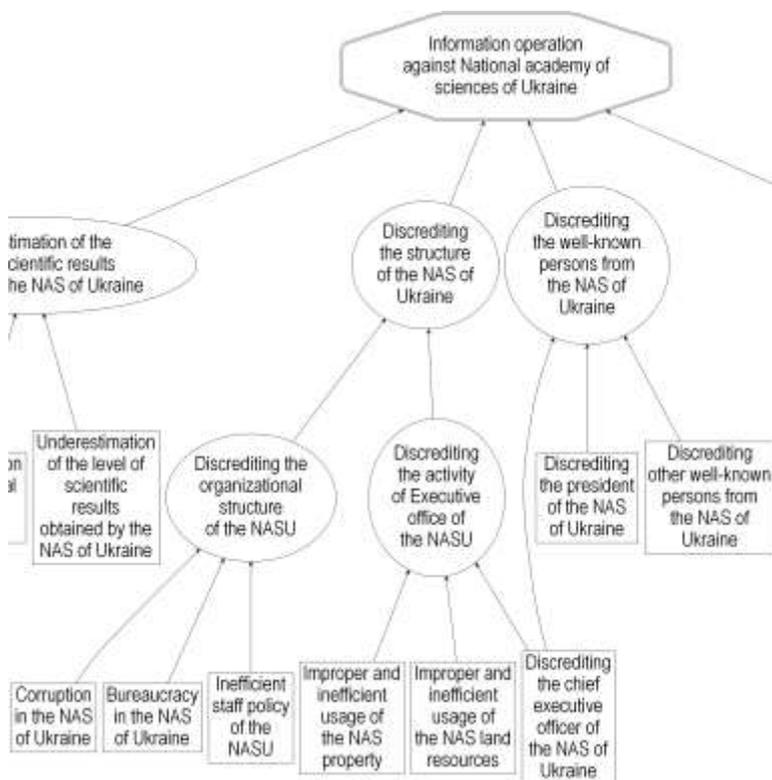

Fig. 5.3 – A fragment of goal decomposition network:
"Discrediting the structure of the NAS of Ukraine" and
"Discrediting the well-known persons from the NAS of
Ukraine"

Then, using "Solon-3" DSS, the degrees of project
implementation are input. If for some IO components no
planted information precedents were detected (as in the
case of "Contraposition of achievements of Ukrainian
companies and the NAS of Ukraine" and "Discrediting the
executive office of the NAS of Ukraine") the respective
projects' implementation degrees are set to 0%. For all
other projects the implementation degrees equal 100%.

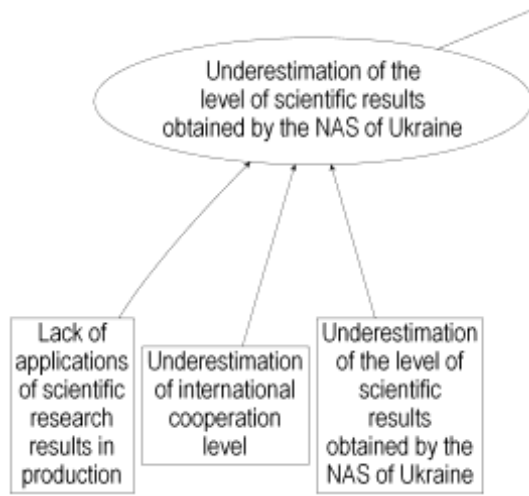

Fig. 5.4 – Decomposition of goals "Underestimation of the level of scientific results obtained by the NAS of Ukraine" and "Discrediting the organizational structure of the NASU"

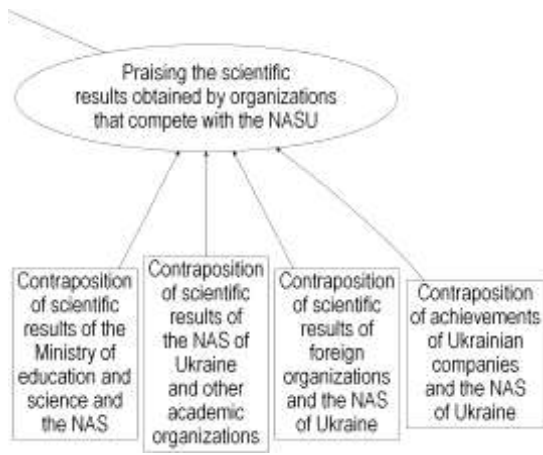

Fig. 5.5 – Decomposition of the goal "Praising the scientific results obtained by organizations that compete with the NASU"

Then the recommendations are calculated, particularly: the degree of the main IO goal achievement (fig. 5.6) and project efficiencies (relative contributions to achievement of the main goal).

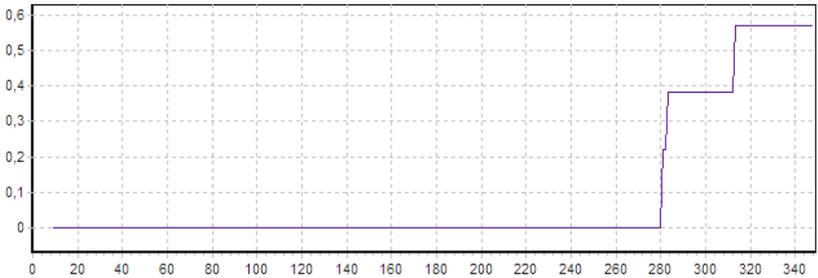

Fig. 5.6 – Dynamics of the degree of the main IO goal achievement

During the periods of 01.01.2013 – 31.12.2013, 01.01.2014 – 31.12.2014, 01.01.2015 – 31.12.2015, and 01.01.2016 – 15.12.2016 the degrees of main goal achievement are: 0.380492, 0.404188, 0.570779, and 0.438703, respectively.

In retrospect, the average value of the main goal achievement degree is: (0.380492 + 0.404188 + 0.570779) / 3.0 ≈ 0.45182.

As the retrospective average and current values of the main IO goal achievement are rather close (the difference is less than 3%), we have concluded that during the period under consideration, the IO could, possibly, result in worsening of target parameters of the object.

## 5.4. A methodology of DSS knowledge base building during detection of information operations

During detection of IO we should consider the impact of different publication topics upon formation of information space.

Application of the approach, described in [Андрейчук, 2016; Andriichuk, 2017], and the respective methodology calls for availability of a group of experts. Experts' work costs quite a lot and requires considerable time. That is why reduction of expert information usage extent in the process of DSS KB building during IO detection remains a topical problem.

The purpose of the current research is to develop a methodology for DSS KB building during IO detection, which would allow us to determine the information impact rating of publication topics, and, at the same time, minimize expert information usage level.

The essence of the suggested methodology of DSS KB building during IO detection is as follows:

1) Group expert examination on definition and decomposition of IO goals is performed. The IO is decomposed as a complex weakly structured system. For this purpose the means of DEICPT are used.

2) The respective knowledge base (KB) is built using the DSS tools, based on the results of group expert examination conducted by DEICPT and available objective information. In order to adjust the queries to the CMS, and to append the DSS KB with missing objects and links, we are using a network of keywords from the subject domain of the respective IO.

3) Dynamics of thematic information flow is analyzed using CMS tools. DSS KB is appended with partial impact coefficients.

4) Based on the constructed KB, using DSS tools, recommendations for the decision-maker are calculated.

## *Methodology application example*

The methodology is demonstrated on the example of Brexit topic, which was researched by the academic community [Bachmann, 2016].

Based on the interpretation of the database, formed using "Consensus-2" DEICPT, the knowledge engineer built the respective KB in "Solon-3" DSS. The goal hierarchy structure of this KB is presented on fig. 5.7.

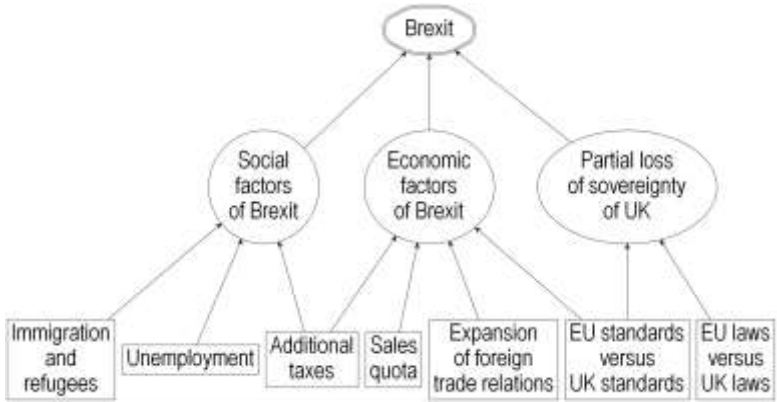

Fig. 5.7 – KB hierarchy structure

On further phases, analysis of thematic information flow dynamics was performed, using InfoStream CMS. For this purpose, based on each of the IO components, queries were formulated on a specialized language. According to these queries, further analysis of publication dynamics on target topics was performed. During query formulation, in accordance to the specificity of the goal hierarchy structure (fig. 5.7), the following rules were used:

- On the way from top to bottom, the respective queries of lower-level IO components were appended by higher-level components (after the "&" sign) for clarification;

- In case of abstract, vague character of IO components, the hierarchy was analyzed from bottom to top, while the respective queries were appended by queries concerning lower-level IO components (using the "|" sign);

- In case of high specificity of IO components, the query was made unique.

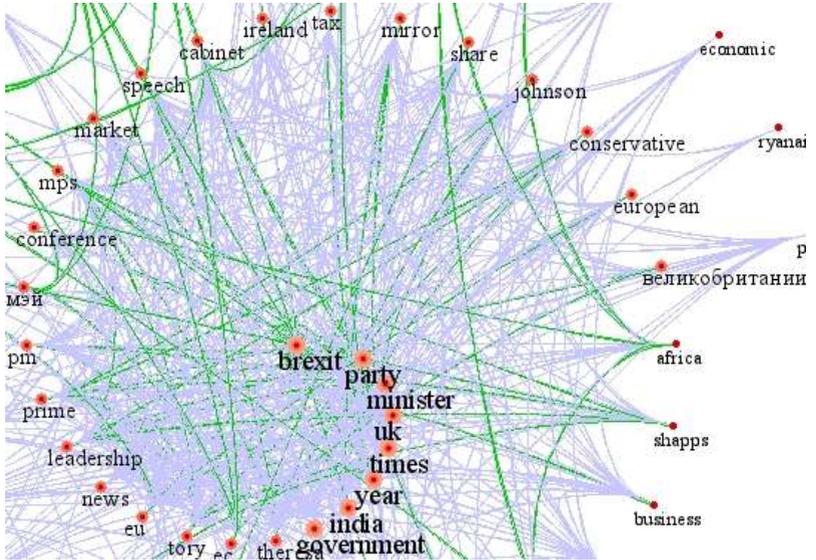

Fig. 5.8 – A network of keywords for Brexit topic

In order to clarify queries to the CMS and append the DSS KB with the missing objects and links, the network of keywords for Brexit topic (shown on fig. 5.8) was used.

Based on the results of fulfillment of queries in InfoStream system, particularly, on the number of documents found for each query, the respective partial impact coefficients (PIC) were calculated. During calculation of PIC values the following assumption was applied: the degree of impact of an IO component is proportional to the number of the respective documents found.

The obtained PIC values were input into the KB. Thus, we managed to refrain from addressing the experts for estimation of impact degrees of IO components.

Based on the KB (constructed and appended as described above) "Solon-3" DSS calculated the recommendations. The following rating of information impacts of publication topics was obtained: "Immigration and refugees" – 0.364, "EU standards versus UK standards" – 0.177, "EU laws versus UK laws" – 0.143,

"Expansion of foreign trade relations" – 0.109, "Additional taxes" – 0.084, "Sales quota" – 0.08, "Unemployment" – 0.043.

## 5.5. Considering Importance of Information Sources during Aggregation of Alternative Rankings

In the process of informational and analytic research a common problem, often faced by analysts is compiling a ranking of a set of objects or alternatives (products, electoral candidates, political parties etc) according to some criterion. This criterion may be explicitly formulated by a decision-maker, or presented in the form of a particular topical query (such as "top comic actors", "best online footwear stores", "downtown restaurants", etc). Even these randomly selected formulations demonstrate that it is problematic to select these alternatives, satisfying the respective queries based on solely numeric data, as there is no quantitative criterion, according to which the alternatives could be measured. A common approach, summarized in the most explicit way, perhaps, by Tom Saaty ([Saaty, 1994]), is as follows: if you cannot measure the alternatives, the best way to numerically describe them, is to compare them among themselves. Conceptually, comparisons of alternatives according to some pre-defined criterion can be classified as either cardinal or ordinal estimates. Cardinal pair-wise comparisons of alternatives bear information about numeric ratios between them according to the given criterion, while ordinal comparisons indicate only the ordering of these alternatives. Cardinal pair-wise comparisons are good for relatively small numbers of alternatives, lying within the same order of magnitude. Ordinal pair-wise comparisons are easier to obtain and processed, and they can be used to compile rankings of large numbers of alternatives.

Amount of data, used in informational and analytical research is, usually, large, and that is why ordering of alternatives is more frequently used. For example, this is one of the possible reasons why web search engines operate mostly with rankings (orderings) of references and not with their numerically expressed ratings of any kind.

In order to compile a ranking of alternatives based on all available information, rankings, coming from several information sources (IS), need to be taken into consideration and aggregated in some way. In order to provide at least some analytical description of a weakly structured domain, data, coming from any available sources, needs to be taken into consideration.  In this section we will try to address the most general case, when information sources can include paper and online documentation, web search results, and expert judgments.

The problem with aggregation of data, coming from different information sources is that, in the general case, these sources have different weights, depending, again, on several factors. For example, it is reasonable to assume, that more credible information source, providing larger amount of information, should be assigned greater weight prior to data aggregation. While there is, roughly, a dozen of methods of ranking aggregation (proposed by Borda, Condorcet, Kemeny, and others), that are commonly known and described in a multitude of publications, the problem of defining the weights of data sources (voters, judges, experts, search engines etc.) during data aggregation is still relevant and open to discussion.

So, in further subsections, we are going to set forth a method, allowing to aggregate data, presented in the form of alternative rankings, coming from different information sources, particularly focusing on different aspects of information source weight definition problem.

### *Problem description*

The problem of rank aggregation became relevant back in the ancient times, when democratic societies and voting systems started to emerge. An extensive analysis of voting rules and their evolution is provided, for example, in [Rzążewski, 2014].

The first rank aggregation methods, that emerged in the process of democratic voting evolution, which are still relevant today, were suggested in the late 18th century by Borda ([Borda, 1781]) and Condorcet ([Condorcet, 1785]). Since then multiple approaches to aggregation of individual preferences using different social welfare functions were devised. We find it most appropriate to mention the period of 1950-s and 1960-s – the time when Arrow's impossibility theorem was formulated (see [Arrow, 1970]) and attempts were made to bypass its constraints (particularly, by [Kemeny, 1959; Copeland, 1951]). F.Aleskerov summarizes Arrovian aggregation rules in his book [Aleskerov, 1999].

In the 2000s, with growing popularity of online information search engines, the problem of rank aggregation became even more relevant, as it became evident, that beside data obtained from voters, experts, and analysts, it was necessary to take online information sources into account (sometimes, with minimum human involvement). In this context, we should mention several papers from the 2000s, particularly [Dwork, 2001; Renda, 2003; Аборок, 2005]. In these publications the authors suggest and compare several approaches, targeted at aggregation of data from online information sources.

Definition of weights of information sources and weighted aggregation of rankings represents a separate matter. Methods of Borda and Condorcet can be easily extrapolated to the case when information sources have different weights, as shown by Totsenko in [Totsenko, 2005]. Kemeny's median is a more problematic case when information sources have different weights. Ordinal factorial analysis methods, described in [Kadenko, 2008; Kadenko, 2013] allow analysts to calculate weights based on available sets of alternative rankings, previously provided by experts. This approach can be extrapolated to calculation of weights of online information sources as well, if evaluators provide some global alternative ranking a priori.

Dwork et al in [Dwork, 2001] stress the importance of involvement of human evaluators in the process of preliminary information source weight definition. However, their paper focuses mostly on comparative analysis of rank aggregation methods and does not address the problem of IS weight calculation directly.

As for IS weights, we can, again, mention Tom Saaty ([Saaty, 1994]) and his academic school (including Ramanathan & Ganesh [Ramanathan, 1994], Forman & Peniwati [Forman, 1998], as well as Yang et al [Yang, 2017]). However, their methods are primarily targeted at aggregation of expert estimates, represented in the form of pair-wise comparison matrices (PCM), provided in some ratio scales (and not rankings).

In our current research we are going to suggest information source weight calculation methods based on both preliminary human evaluation of information source credibility and amount and quality of information, provided by a given information source.

## Problem statement

**What is given**: $n$ information sources ($IS_1 - IS_n$). Each of IS provides a ranking $R_i$, that includes $m_i$ alternatives ($i=1..n$). Information sources can be experts, search engines, analysts who monitor online content, paper or online documents, etc. We should stress, that the number of alternatives, provided by different information sources is, in the general case, different ($m_i \neq m_j$; $i, j = 1...n$).

**We should find**: an aggregate (global) ranking of alternatives.

## Solution ideas for the case of equal weights of information sources

We suggest building a solution algorithm based on available methods of aggregation of individual rankings (ordinal estimates). The most common ranking aggregation methods were proposed by Borda, Condorcet and, perhaps, Kemeny. We should stress that if we are dealing with experts, the number of alternatives an expert can analyze "in one go" is no more than 7±2. However, search engines can return rankings of hundreds of links. That is why, usage of domination matrix – based methods is not recommended for such dimensionalities. For example, if we are dealing with 10 IS, providing rankings of 1000 alternatives each, then we have to analyze and process 10 matrices with 1000×1000 cells. In such cases Borda method seems to be the most appropriate option, as it operates with ranking vectors and not domination matrices. Besides, it is easier to extrapolate Borda method to the case when IS have different weights (while, for instance, extrapolation of Markov chain-based methods looks more problematic). If all IS have the same weight, we can use the following algorithm.

1) Find the length of a ranking with all unique alternatives, featured in individual rankings
$$M = card\left(\bigcup_{i=1}^{n}\{A_j^{(i)}; \ j=1...m_i\}\right).$$

2) Form a unified set of alternatives. For this purpose we should add to each ranking $R_i$ of $m_i$ alternatives all the remaining alternatives $\bigcup_{k=1}^{n}\{A_j^{(k)}; \ j=1...m_k\}/\{A_j^{(i)}; \ j=1...m_i\}$ with rank $m_i+1$. During unification of alternatives, provided by different IS, we should check whether alternatives are conceptually the same. If IS are experts, then for this purpose we can use the methods of semantic similarity determination, suggested in [Андрійчук, 2014].

3) As a result, we will get $M$ alternatives in each ranking. Each ranking $R_i$ will include $m_i$ alternatives with different ranks and $(M - m_i)$ alternatives with the same rank $(m_i+1)$. (A similar approach was mentioned by Dwork et al in [Dwork, 2001]).

4) Add the ranks of each alternative across all IS.

5) Sort (order) the alternatives in the order of increment of individual rank sums. This sorting will result in the rank order that we should find.

$$\sum_{j=1}^{n} r_{kj} > \sum_{j=1}^{n} r_{lj} \Rightarrow r_k < r_l, k,l = 1..M, \qquad (5.1)$$

where $r_{kj}$ and $r_{lj}$ – are the ranks of alternatives $A_k$ and $A_l$ in the ranking, provided by IS number $j$.

## *Usage of different methods under smaller cardinality of the set of alternatives*

If the number of alternatives amounts to one or two dozens, or if the decision-maker (analyst) is interested only in the rank order of the first few alternatives, provided by IS, then it makes sense to use other ranking methods (beside or instead of Borda) in order to aggregate the individual ranking results.

For example, let us assume, we have 10-15 IS, and each of them provides a ranking of alternatives, from which we are particularly interested in the top 10-20 items. If all alternatives (items) are different, then we have 300 ($15 \times 20$) different items at most. In fact (especially if we are talking about search engines), many items, featured in individual rankings across different IS, are the same. So, if we have 15 IS and each of them provides a ranking of 20 alternatives, then the cardinality of the set of unique items will amount to approximately 100 alternatives.

Consequently, if we use aggregation methods operating with ordinal pair-wise comparison matrices (PCM) (such as Condorcet rule or Kemeny's median) and not with ranking vectors (like Borda or Markov chain based methods), we will have to analyze 15 square matrices containing $100 \times 100$ cells. As strict rank order relationship is reciprocally symmetrical (if alternative $A_1$ ranks higher than $A_2$, then $A_2$ ranks lower than $A_1$), during aggregation of PCM we will have to analyze just matrix elements above the principal diagonal $\{d_{ik}, i<k\}$. That is why, even if the aggregate ranking features 100 alternatives, then we will have to analyze ($100 \times 99$)/2 or 4950 cells in each PCM. These calculations demonstrate, that under smaller cardinality of alternative set, we can use Condorcet rule and Kemeny's median for aggregation of individual ranking results.

In our view, the disadvantage of Borda method is that during aggregation of individual rankings we are witnessing an explicit heuristic transition from ordinal preference scale to ratio scale. However, this transition is inevitable, because representation of alternative estimates in ratio scale is the necessary and sufficient condition of existence of linear global criterion, allowing us to aggregate these estimates (as proven by Litvak in [Литвак, 1982]). For example, an alternative with rank 2 does not necessarily have to be exactly 2 times better than alternative with rank 4 (we should remind, that dominating alternatives are assigned smaller ranks).

Condorcet rule, Markov chains, and Kemeny's method allow us to avoid such explicit transitions from ranks to ratios.

### *Existing methods: a brief overview*

**Condorcet rule**. The key idea of Condorcet method is as follows. If alternative $A_i$ dominates over alternative $A_k$ in the majority of individual rankings, then this preference relationship should be maintained in the global (aggregate) ranking.

In order to get an aggregate ranking we should first build individual ordinal PCMs. If in some individual ranking $R_j$ alternative $A_i$ dominates over $A_k$, then the respective element of ordinal PCM equals 1, if $A_i$ is dominated by $A_k$ – (-1), if the alternatives are equal – 0:

$$d_{ik}^{(j)} = \begin{cases} -1, & A_i \prec A_k \\ 0, & A_i \equiv A_k \\ 1, & A_i \succ A_k \end{cases} \qquad (5.2)$$

In order to aggregate rankings, provided by several information sources according to Condorcet's rule, we should limit the number of alternatives to be featured in the global ranking by a fixed number $M$ and build ordinal PCM based on rankings, provided by all IS $\left\{ D^{(j)} = \left\{ d_{ik}^{(j)}; \ i,k = 1..M \right\}; \ j = 1...n \right\}$. After that we should calculate the sums of all respective elements of individual PCM and fill ordinal PCM, corresponding to the global rank ordering relationship (tournament).

$$D = \{d_{ik}; \ i, k = 1...M\},$$

$$where \ \ d_{ik} = \begin{cases} 0, \ \sum_{j=1}^{n} d_{ik}^{(j)} = 0 \\ 1, \ \sum_{j=1}^{n} d_{ik}^{(j)} > 0 \\ -1, \ \sum_{j=1}^{n} d_{ik}^{(j)} < 0 \end{cases}.$$

(5.3)

After that we should add the elements in each line of the global rank ordering relationship matrix $D$ and rank the obtained line sums.

$$\sum_{k=1}^{M} d_{ik} > \sum_{l=1}^{M} d_{lk} \Rightarrow r_i < r_l$$

(5.4),

where $r_i$ and $r_l$ are ranks of alternatives $A_i$ and $A_l$ in the global ranking $R$.

The disadvantage of Condorcet's rule (beside dimensionality limitation) is that equal alternative ranks may emerge and transitivity of the global preference relationship may be violated as a result of the so-called "Condorcet paradox". The paradox is one of the specific cases of violation of Arrow's requirements to social choice functions [Arrow, 1970]. If feedback is acceptable, then transitivity of the global preference relationship can be achieved if we use algorithms, described in [Tsyganok, 2010b; Kadenko, 2012].

In the context of aggregation of results of online data search, provided by several engines, the advantage of Condorcet's rule (mentioned by Dwork et al in [Dwork, 2001]) is its ability to filter spam content.

A more "just" aggregation rule in view of Arrow's impossibility theorem is Kemeny's rule (sometimes called Kemeny's median).

***Kemeny's median*** can be considered the analogue of average mean for ordinal (non-cardinal) estimates.

As ordinal estimates (or ranks) do not bear information on quantitative relation between alternatives, such concepts as Euclidian distance metric or center of mass (center of gravity) do not apply to rank order vectors.

Distance between two rankings of a given set of alternatives depends upon the number of elementary permutations that is needed to obtain one ranking from the other. Kemeny's distance between two rankings is based on Haming metric. If we have two rankings $R_1$, $R_2$ of a set of alternatives $A=\{A_1..A_m\}$, and respective domination matrices $D_1$ and $D_2$ are built based on these rankings according to formula (5.2), then Kemeny's distance $K$ is calculated as follows.

$$K( R_1, R_2 ) = \sum_{i=1}^{m} \sum_{k=1}^{m} \left| d_{ik}^{(1)} - d_{ik}^{(2)} \right| \qquad (5.5),$$

If we have a set of $n$ rankings $\{R_1..R_n\}$ of alternatives $A=\{A_1..A_m\}$, then, by definition ([Kemeny, 1959]), Kemeny's median of this set of rankings is a ranking $R$:

$$R = \arg\min_{R \in T} \sum_{j=1}^{n} K(R, R_j) \qquad (5.6),$$

where $T$ is a set of all possible rankings of alternatives. Kemeny's median calculation procedure is very labor-intensive. Some estimates of complexity of this problem are provided in [Dwork, 2001]. One of the "simplest" algorithms of Kemeny's median calculation is provided by Litvak in [Литвак, 1982].

## Generalization of suggested approaches to the case of different weights of information sources

Let us now assume that IS weights $\{w_1...w_n\}$ (reflecting their credibility) have been provided by experts, calculated based on previous estimation experience (as described in [Kadenko, 2008; Kadenko, 2013]), or obtained in some other way. In this case, respective formulas for ranking aggregation will change, as all rankings, provided by individual IS, will be taken into consideration together with their respective weights. Formula (5.1) (corresponding to Borda aggregation method) will look as follows.

$$\sum_{j=1}^{n} w_j r_{kj} > \sum_{j=1}^{n} w_j r_{lj} \Rightarrow r_k < r_l, \; k, \; l = 1...M \qquad (5.7)$$

Formula (5.3) (corresponding to Condorcet method) will look as follows.

$$D = \left\{ d_{ik}; \; i,k = 1..M \right\}$$

$$where \quad d_{ik} = \begin{cases} 0, \; \sum_{j=1}^{n} d_{ik}^{(j)} = 0 \\ 1, \; \sum_{j=1}^{n} d_{ik}^{(j)} > 0 \\ -1, \; \sum_{j=1}^{n} d_{ik}^{(j)} < 0 \end{cases} . \qquad (5.8)$$

In formula (5.6) weight coefficients will become multipliers for respective Kemeny distance values.

$$R = \arg\min_{R \in T} \sum_{j=1}^{n} w_j K(R, R_j) \qquad (5.9).$$

Extrapolation of Markov chain based methods, described in [Dwork, 2001], represents a problem that should be addressed in a separate research.

Experience and numerous publications indicate that when a group of experts participates in decision-making process, individual expert competence should be taken into consideration (providing, the expert group is relatively small) (for details – see [Циганок, 2011], [Tsyganok, 2011], [Tsyganok, 2012]). Similarly, if several IS are used to build a ranking of alternatives according to their relevance in terms of a particular information query or in the process of some analytical research, we should take their weights into account as well.

Now let us address the issue of calculation of IS weights in greater detail. So far we have come up with several conceptual approaches to IS weight calculation. These approaches are outlined in the next subsection.

## *Approaches to calculation of the relative weights of information sources*

***1) Experience-based approach*** was set forth in [Kadenko, 2008; Kadenko, 2013]. In essence, experts or evaluators provide their ranking of alternatives and then, based on their ranking and rankings, provided by individual IS, the weights of IS are calculated (under assumption that rankings are aggregated using Borda or Condorcet rule).

***2) Definition of relative IS weights based on quantity and quality of provided information.*** It is reasonable to assume, that the relative weight of an IS should depend on quantity and quality of information, provided by the source in terms of every information query (in the given subject domain). Criteria, representing quality and quantity of information search, following a given query, can be formulated as follows.

– Number of alternatives (references or links in case of online information search), provided by the IS in response to a particular query;

– Relevance of these alternatives (references, links), i.e. their correspondence to the specific query.

Relevance of the results of functioning of an IS can be defined based on the estimates of users (evaluators, experts) from the given subject domain. These estimates of IS can be based on previous experience of information search queries.

Relevance of results of IS work can be considered in terms of search queries of some particular type (search for information in some given language, formulas, images etc.). If it is possible to outline some query type, expert estimate will depend upon IS capability of processing this particular type of queries. Otherwise (when it is problematic to outline specific query types) it makes sense to use an average value of expert estimate (across different query types).

In the context of this subsection we suggest calculating IS weight using the approach, similar to methods, used by Totsenko for calculation of expert competence in [Тоценко, 2002], [Totsenko, 2002]. In accordance to this approach, expert competence should depend on self-estimate, objective estimate, and mutual estimate.

$$k = s(x_1 b + x_2 v) , \qquad (5.10)$$

where $k$ is the relative competence of the expert, $s$ is self-estimate, $b$ is an objective component, $v$ is mutual estimate of expert group members, while $x_1$, $x_2$ represent the coefficients of relative importance of objective and mutual estimates' respectively.

When the weights of "generalized" IS (i.e. experts, search engines, documents, etc.) are calculated, *mutual estimate* can be replaced by the ratio between the number of alternatives, provided by the *i*-th IS and the total number of alternatives, provided by all IS - $V_i$ :

$$V_i = \frac{m_i}{\sum_{k=1}^{n} m_k} , \qquad (5.11)$$

where $m_i$ is the number of alternatives, provided by *i*-th IS, $n$ is the total number of information sources.

*Self-estimate* in formula (5.10) can be replaced by the value of expert estimate $E_i$ . This estimate depends on previous experience of IS usage. For each particular query type (for instance, Ukrainian, English, image, formula, other) we should apply the respective average value of $E_i$ . Definition of query types is not the subject of this paper and should be addressed separately.

*Objective estimate* in formula (5.10) can be replaced by the ratio between the number of alternatives, provided by *i*-th IS and the total number of unique alternatives, provided by all IS $O_i$ (thus, it will reflect the ability of IS to provide unique new information).

$$O_i = \frac{m_i}{P} , \qquad (5.12)$$

where $P$ is the number of unique alternatives, provided by all IS, that is $P = card\left(\bigcup_{i=1}^{n}\left\{A_j^{(i)}; \ j = 1...m_i\right\}\right)$. If IS are experts, semantic similarity methods, suggested in [Андрійчук, 2014] can be used to define whether the alternatives are relevant in the context of the given query. Similarly to expert competence, weight of IS will be calculated as follows.

$$w_i^* = E_i(x_1 O_i + x_2 V_i), \tag{5.13}$$

where $w_i^*$ is the non-normalized IS weight; $x_1$, $x_2$ are the respective coefficients of components' relative importance, and $E_i$ is the non-normalized expert estimate value. Normalized values are calculated as follows.

$$w_i = \frac{w_i^*}{\sum_{i=1}^{n} w_i^*}, \tag{5.14}$$

where $w_i$ is the normalized relative IS weight.

We propose to follow the assumption that "objective" and "mutual" IS estimates should form a convex combination ($x_1 + x_2 = 1$), so $x_1$ and $x_2$ should lie within the $[0;1]$ range and vary depending on particular information query.

As a result of each information search, every IS provides several alternatives (in the general case they are different, as mentioned above). Figure 5.9 represents an example, where 5 IS provide 9 unique alternatives. 1st and 6th alternatives were provided by 2 IS, 2nd and 5th – by 3 IS, 3rd alternative – by 5 IS, while other alternatives were provided by 1 IS each.

In order to illustrate the behavior of dependence of $x_1$ and $x_2$ on information search results, let us consider several extreme cases.

***Extreme case 1*** Let us assume that all IS provided completely different alternative sets. Generalized information search results are displayed on Fig. 10: 5 IS provide 9 different alternatives and each of the alternatives is provided by a single IS (for instance, IS1: a1, a2; IS2: a3, a4; IS3: a5; IS4: a6, a7, a8; IS5: a9).

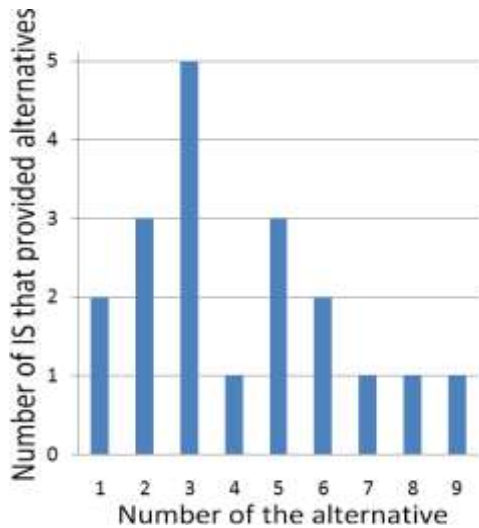

Fig. 5.9 – Example of a unified set of alternatives, provided by several IS

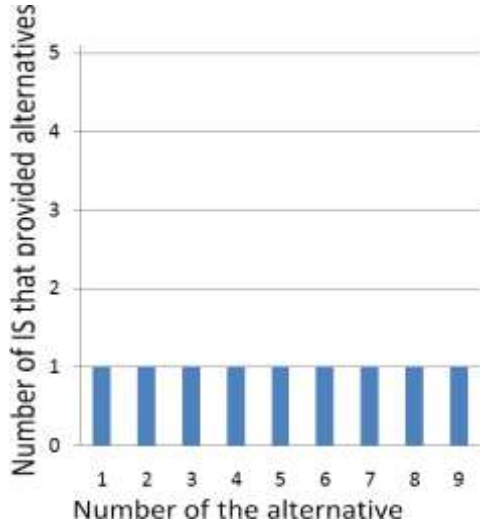

Fig. 5.10 – Example of a unified set of alternatives, where
each alternative is provided by a single IS

If all IS provide different alternatives in response to
some query, the relevance of these alternatives seems
questionable. In such cases we suggest using expert
estimate (based on previous experience of IS usage) as the
key component of IS weight, as it reflects the experience of
previous information search sessions and the ability of IS
to find new information.

*Extreme case 2*. Let us assume that all IS provided the
same alternatives. An example of a hypothetical
information search results are presented on fig. 5.11: 9
alternatives are provided by all 5 IS (although the order, in
which alternatives are ranked by different sources, may be
different).

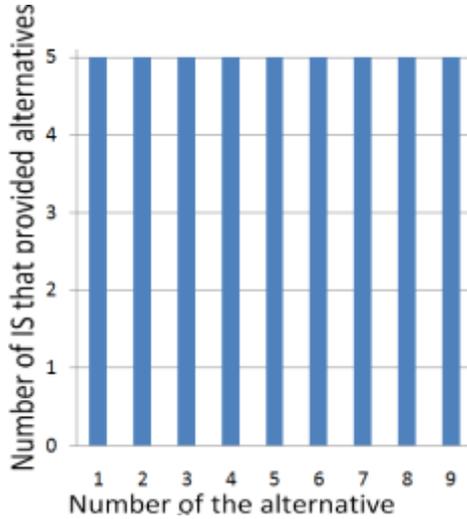

Fig. 5.11 Example of a unified set of alternatives, where each alternative is provided by all IS

In such a case, IS weights should be equal to $E_i$, because all IS are equally efficient in providing alternatives in response to information query. Their weights can be defined, again, only based on the previous history of information search sessions that is reflected by the expert estimate.

In order to devise a formal expression for $x_1$ and $x_2$, based on these intuitive considerations, let us introduce the indicator $\rho$, characterizing the alternative frequency function or density of representation of alternatives across all information sources.

$$\rho = \frac{\sum_{j=1}^{P} h_j}{nP}, \qquad (5.15)$$

where $h_j$ is the quantity of IS, that provided $j$-th alternative, while $n$ is the total quantity of IS.

In "extreme case 1" $\rho$ assumes the minimum value $\rho = \dfrac{1}{n}$.

In "extreme case 2" $\rho$ assumes maximum values $\rho = 1$.

In view of above-mentioned convexity requirements, we suggest calculating $x_2$ as:

$$x_2 = \rho \; ; \; x_1 = 1 - x_2 . \qquad (5.16)$$

In "extreme case 1" the normalized relative IS weight assumes the value of $w_i = \dfrac{E_i\left((1-\dfrac{1}{n})O_i + \dfrac{1}{n}V_i\right)}{\sum\limits_{j=1}^{n} E_j\left((1-\dfrac{1}{n})O_j + \dfrac{1}{n}V_j\right)}$, and in

"extreme case 2" it assumes the value of $w_i = E_i$.

***3) Statistical approach*** represents a modification of the previous approach. In order to take the quantity of information, provided by the IS ("objective" weight component in formula (5.13)) into account, we can normalize the numbers of alternatives across all IS (see formula (5.11)) and then calculate the importance of the respective component (i.e. value of $x_1$) as dispersion if this indicator (normalized number of alternatives).

$$x_1 = D(V); \quad x_2 = 1 - x_1 \qquad (5.17)$$

In this case, if all IS provide the same number of alternatives, the respective component of their relative weights can be neglected as it does not vary across different IS. The respective component will only come into play when the numbers of alternatives across IS are different.

## 5.6. Developing of information operation counteraction strategy

Probably, the term IO became popular after declassification of a set of documents of the US Department of Defense, in which IO were defined as actions aimed to influence the enemy information and information systems, and to protect one's own information and information systems. Later on, in the Roadmap of information operations [Roadmap, 2003] , the term was clarified and defined as integrated application of the key means of radio-electronic combat, operations in computer networks, psychological operations, military disguise and operations on ensuring security, in combination with associated capacities, aimed at impacting, destroying, elimination, or seizing of the decision-making process (both personal and automated) from the enemy with simultaneous protection of our means. The sense, input into the IO term, covers and elaborates information impact upon mass conscience (both enemy's and allied), impact upon information available to the enemy and necessary for decision-making, as well as enemy's information and analytical systems [Горбулін, 2009]. In modern realities an IO as an essential part of information warfare is considered as a certain type of combat, active counteraction in the information space, while information is considered a potential weapon for performing a strike.

Two basic IO types are singled out – offensive and defensive. However, in practice most IO are mixed ones, and most of the procedures they include are offensive and defensive at the same time. A peculiar feature of offensive IO (information attacks) is that objects of impact of such operations are clearly defined, and planning is based on rather precise information about these objects. Information attack, most often, calls for finding or creation of an information precedent (for defensive IO the precedent may be the enemy's information attack itself), boosting (promotion, propaganda) this precedent (counter-propaganda for defensive IO), as well as activities on counteracting information impacts. Thus, irrespectively of IO type, an IO can be divided into the following phases: estimation, planning, execution, and concluding phase. Following the purposes of our research, we are going to consider in greater detail a defensive IO, complying with the military doctrines of the majority of progressively developing states.

A typical defensive IO includes the following basic phases:

- Estimation (evaluation):
  - Analysis of potential vulnerabilities (targets);
  - Collection of information on possible operations;
  - Identification of potential "commissioners (customers)" of information impacts:
    - Identification of common interest spheres of the object and potential "customers";
    - Ranking of potential customers according to their interests;
- Planning:
  - Strategic planning of a defensive operation (explicit or implicit):
    - Identification of information influence criteria;
    - Modeling of information influences based on the object's connections, influence dynamics, "special" (critical) influence points
    - Prediction of further steps;
    - Calculation of implications;
  - Tactic planning of counter-operations;
- Execution – implementation of information impact:
  - Detection and "smoothing" of the information precedent;
  - Counter-propaganda;
  - Operative intelligence;
  - Evaluation of information environment;
  - Adjustment of information counteraction;
- Concluding phase:
  - Efficiency analysis;
  - Usage of positive results of information influence;
  - Counteracting negative results.

As we can see from the suggested detailed description of a defensive IO, the basic component of an IO is strategic planning. Obviously, there is no unified "standard" plan of IO implementation. We can only consider some model sequence of actions, based on generalization of some of the already implemented IO. We should note that the choice of an optimal set of such activities at a certain point in time depends, first and foremost, on the availability of resources for their completion at this given moment, as well as on the results of implementation of previously selected activities. Optimality in this context should be considered as efficiency of achievement of goals of some specific defensive IO.

The purpose of the current research is to improve the existing decision support (DS) mechanisms based on the peculiarities of strategic planning process in weakly structured subject domains. The current research has a more specific objective: development of a complex DS methodology, allowing us to improve the quality of strategic planning process for defensive IO. We suggest a methodology for building of a formal IO strategy involving a group of competent specialists from the given subject domain. Based on modern expert data-based DS methods, we suggest a technique for obtaining of thorough undistorted knowledge from specialists, and usage of this knowledge for building an adequate subject domain model. Application of the methodology will be demonstrated on an illustrative example.

## *Content and the key phases of strategy development process*

As we know, in the general sense, a strategy is a non-detailed plan of action, intended for a continuous period of time and targeted at achievement of a certain main goal. At the same time, the plan should be flexible, constructive, stable to the uncertainties of external situation, and allow for specification of the main goal through its decomposition.

In weakly structured domains, which include management, environmental protection, production, social services and others, there is a relevant problem of constructing long-term non-detailed plans of action. There is no doubt that during building of such strategic plans we should utilize all available knowledge from the given subject domain. As knowledge in every such domain is not entirely formalized and, to a large extent, stored within the minds of specialists, it would be unreasonable not to use the information, procured from experts. It would be even more unreasonable to reduce the estimates of planning options to solely quantitative (say, financial) indicators. In order to have realistic long-term plans at hand, they should be adapted to inevitable changes of current situation, and adjusted based on availability of resources required for their implementation at a specific moment. That is why strategic plans can be rational only on certain time intervals.

The purpose of the current research is to create a technology, which would include formal mechanisms for strategic plan development in weakly structured subject domains, involving groups of experts and knowledge engineers.

With the listed requirements to strategies (namely, the need for realistic and dynamically adjustable plans) taken into consideration, we suggest using the tools for limited resource distribution among specific suggested activities during strategic planning. Resources are distributed as of a given moment in time, depending on potential contribution of a certain activity into achievement of a strategic goal. In fact, the results of conducted work should provide an answer to the question: "which activities should be implemented in the current situation in order to ensure the most efficient achievement of the strategic goal?"

With these considerations in mind, the developed strategic planning technology envisions several phases.

1) Knowledge base (KB) building.

This phase is implemented in the form of a web-oriented software system, allowing a DM, knowledge engineers, and experts to work in remote mode to build a KB, without the need to gather at one place.

The phase includes several sub-phases:

a. Selection of expert groups to conduct the expert examination.

The task of selection of experts, in the general case, is placed upon a DM and knowledge engineers. Within the framework of a given examination different groups of specialists (representing specific sub-domains) are formed for resolution of particular issues.

b. Building of a hierarchy of goals (in the process of a dialogue with experts), describing the subject domain.

At this sub-phase a DM formulates a strategic goal, which is to be decomposed into local goals (factors), significantly influencing its achievement, in the process of expert examinations conducted by knowledge engineers. In the process of decomposition, experts reconcile their judgments about the content of the set of factors, influencing a particular goal, until they reach consensus in every issue under consideration. Knowledge engineers, authorized to organize expert examinations, form a separate expert group for decomposition of each local sub-goal.

The decision, regarding sufficient level of detail and termination of strategic goal decomposition process, is made by expert examination organizers, when the lowest goal hierarchy level includes only sub-goals (factors), representing specific activities (projects) ready for implementation.

The result of this strategic planning phase is a hierarchic structure, which thoroughly describes the subject domains, according to the given expert group.

c. Expert estimation of relative impacts of goals in the hierarchy.

Relative impact of each goal in the hierarchy graph is determined by the knowledge engineer, if credible information on the degree of this impact upon achievement of the given goal is available, or (otherwise) – by a group of experts through pair-wise comparisons of goal (factor) impacts.

The result of this sub-phase is the relative values of mutual impacts of goals, obtained in the process of aggregation of individual expert estimates, provided in the scales of different degrees of detail, and represented in the form of incomplete matrices of pair-wise comparisons (PCM). We suggest aggregating pair-wise comparisons using the combinatory method [Циганок, 2000], advantages of which over other methods in terms of efficiency are confirmed by the respective experimental research [Tsyganok, 2010a].

The combinatory aggregation method has several specific advantages over other existing approaches to processing of PCM:

• It uses the redundancy of information most thoroughly.

• It allows us to calculate the weights of alternatives when certain PCM elements are missing (not provided). That is, in order to calculate alternative weights, we do not necessarily need to have all pair-wise comparisons in the matrices. The only necessary condition is just the connectivity of the graph, representing the aggregate PCM.

• The method includes a single phase (in contrast to other known approaches applied for weight calculation in group estimation methods [Forman, 1998]). Aggregation of pair-wise comparisons in such group DS methods is a two-phase procedure: either (1) first, individual PCM are aggregated, and then, based on aggregate matrix, alternative weight vector is calculated, or (2) first, weight vector is calculated based on every PCM, and then all vectors are aggregated. In case (1) consistency of all individual PCM does not guarantee consistency of the aggregate PCM. In case (2) consistency of individual PCM does not guarantee compatibility of weight vectors calculated based on all PCM. If consistency level is insufficient for aggregation and weight calculation, then the two-phase nature of these procedures makes it impossible to organize feedback with experts to increase consistency. In the case when combinatory method is used there is no need for step-by-step achievement of desired consistency level, and that is why there is no conflict between two consecutive consistency improvement procedures. If we need to improve consistency of pair-wise comparisons, the respective elements of individual PCM are adjusted as required, providing the experts who built the respective PCM agree with the adjustments.

We should stress that aggregation of PCM is allowed only under sufficient consistency of expert judgments [Tsyganok, 2010b]. In order to evaluate pair-wise comparison consistency level, we suggest using the Double Entropy index [Olenko, 2016], allowing us to calculate the degree of consistency based on spectrums of expert estimates of all impact weights, and meeting all the requirements to consistency measures (which is its key advantage over other existing indices). In the case when consistency is insufficient, the method allows us to improve it through feedback with experts.

This is where the KB building phase ends, and the phase of optimal strategy building, based on knowledge input into the KB, starts.

2) Identifying the optimal strategy.

Obviously, the greater a project's weight is, the more significantly it influences the achievement of the strategic goal. That is why allocation of resources to this project will produce more significant and tangible results. At the same time, the amount of resources we allocate to a project shouldn't be smaller than the minimum amount we need to launch it. Consequently, we propose to define an optimal strategy as an optimal variant of resource distribution among the projects (i.e. the variant which ensures the most efficient achievement of the strategic goal).

The problem of finding an optimal distribution of resources among projects is a subject of a separate research. We should note that as projects can be characterized by different implementation terms, and, besides that, sub-goals can exert delayed impact upon the main goal, optimal resource distribution can be relevant only for a given moment of time. Thanks to implementation of complex target-oriented alternative estimation method [Тоценко, 2001; Циганок, 2013] within a particular strategic plan we can estimate and compare absolutely heterogeneous projects/activities, including the ones which produce results momentary and the ones, for which the effect becomes evident only in the long-term strategic prospect. Another important parameter, characterizing the projects, is the range of necessary resource volumes. For example, if a minimum funding volume for a project amounts to 1 million UAH, while the requested amount equals 2 million UAH, there is no need to allocate the amount, lying outside this range, for the project.

With the listed peculiarities of the technology in mind, the most rational way to solve the problem of resource distribution among projects at a given moment of time is the targeted enumeration of all possible resource distribution with a given precision (say, 10 thousand UAH), for example, using the Genetic algorithm [Holland, 1994].

Depending on the complexity of subject domain and the formulated goal to be achieved, the process of strategic plan development may be more simple or more complex. However, the suggested mathematical mechanisms and developed DS software tools allow us to develop rather complex and substantial, and, more important, realistic prospective plans, based on all available knowledge about a subject domain.

### *An example of strategy development*

Now let us consider a hypothetical example, illustrating the final stages of the process of building of an optimal strategic plan on counteracting information operations in the 5-year prospect, where the available funding amount is 200 million UAH.

Within the example we assume that a hierarchy with the main goal "Ensure sufficient level of counteracting information attacks" is already built, and the sub-phase c) of strategic plan development – estimation of relative impacts of projects upon some goal from the hierarchy graph – is under way.

Let us assume that the estimation is conducted by a group of three equally competent experts. Every expert has an opportunity to define preferences within each pair from among 4 projects, perform ordinal comparison („>"– more; „<"– less), select a verbal scale for comparisons, the number of grades in this scale, and a particular grade.

Table 5.2 contains the expert estimates of relative importance of activities, influencing the sub-goal "Implementation of information impact": $C_1$ – Detection and "smoothing" of the information precedent; $C_2$ – Counter-propaganda; $C_3$ – Operative intelligence; and $C_4$ – Evaluation of information environment. The asterisk symbol «*» in the matrices denotes elements, which are missing, because the experts did not provide the information (for one reason or another).

Table 5.2 An example: expert estimates of relative impacts of projects

|  | Expert 1 | | | | Expert 2 | | | | Expert 3 | | | |
|---|---|---|---|---|---|---|---|---|---|---|---|---|

**Ordinal preferences**

| | $C_1$ | $C_2$ | $C_3$ | $C_4$ |
|---|---|---|---|---|
| $C_1$ | 1 | > | > | > |
| $C_2$ | | 1 | < | > |
| $C_3$ | | | 1 | > |
| $C_4$ | | | | 1 |

| | $C_1$ | $C_2$ | $C_3$ | $C_4$ |
|---|---|---|---|---|
| $C_1$ | 1 | > | > | > |
| $C_2$ | | 1 | < | > |
| $C_3$ | | | 1 | * |
| $C_4$ | | | | 1 |

| | $C_1$ | $C_2$ | $C_3$ | $C_4$ |
|---|---|---|---|---|
| $C_1$ | 1 | * | > | > |
| $C_2$ | | 1 | < | < |
| $C_3$ | | | 1 | > |
| $C_4$ | | | | 1 |

**Number of grades in the scale**

| | $C_1$ | $C_2$ | $C_3$ | $C_4$ |
|---|---|---|---|---|
| $C_1$ | 1 | 5 | 9 | 9 |
| $C_2$ | | 1 | 5 | 9 |
| $C_3$ | | | 1 | 7 |
| $C_4$ | | | | 1 |

| | $C_1$ | $C_2$ | $C_3$ | $C_4$ |
|---|---|---|---|---|
| $C_1$ | 1 | 5 | 5 | 9 |
| $C_2$ | | 1 | 7 | 9 |
| $C_3$ | | | 1 | * |
| $C_4$ | | | | 1 |

| | $C_1$ | $C_2$ | $C_3$ | $C_4$ |
|---|---|---|---|---|
| $C_1$ | 1 | * | 8 | 9 |
| $C_2$ | | 1 | 9 | 2 |
| $C_3$ | | | 1 | 9 |
| $C_4$ | | | | 1 |

**Grade number**

| | $C_1$ | $C_2$ | $C_3$ | $C_4$ |
|---|---|---|---|---|
| $C_1$ | 1 | 2 | 3 | 5 |
| $C_2$ | | 1 | 2 | 2 |
| $C_3$ | | | 1 | 5 |
| $C_4$ | | | | 1 |

| | $C_1$ | $C_2$ | $C_3$ | $C_4$ |
|---|---|---|---|---|
| $C_1$ | 1 | 3 | 2 | 5 |
| $C_2$ | | 1 | 4 | 5 |
| $C_3$ | | | 1 | * |
| $C_4$ | | | | 1 |

| | $C_1$ | $C_2$ | $C_3$ | $C_4$ |
|---|---|---|---|---|
| $C_1$ | 1 | * | 4 | 8 |
| $C_2$ | | 1 | 2 | 2 |
| $C_3$ | | | 1 | 3 |
| $C_4$ | | | | 1 |

**Scale number**

| | $C_1$ | $C_2$ | $C_3$ | $C_4$ |
|---|---|---|---|---|
| $C_1$ | 1 | 1 | 3 | 1 |
| $C_2$ | | 1 | 2 | 1 |
| $C_3$ | | | 1 | 2 |
| $C_4$ | | | | 1 |

| | $C_1$ | $C_2$ | $C_3$ | $C_4$ |
|---|---|---|---|---|
| $C_1$ | 1 | 3 | 4 | 1 |
| $C_2$ | | 1 | 4 | 2 |
| $C_3$ | | | 1 | * |
| $C_4$ | | | | 1 |

| | C1 | C2 | C3 | C4 |
|---|---|---|---|---|
| C1 | 1 | * | 2 | 3 |
| C2 | | 1 | 3 | 5 |
| C3 | | | 1 | 1 |
| C4 | | | | 1 |

**Unified pair-wise comparison values**

| | $C_1$ | $C_2$ | $C_3$ | $C_4$ |
|---|---|---|---|---|
| $C_1$ | 1 | 2.500 | 1.732 | 5.000 |
| $C_2$ | | 1 | 0.739 | 2.000 |
| $C_3$ | | | 1 | 3.138 |
| $C_4$ | | | | 1 |

| | $C_1$ | $C_2$ | $C_3$ | $C_4$ |
|---|---|---|---|---|
| $C_1$ | 1 | 2.615 | 0.833 | 5.000 |
| $C_2$ | | 1 | 0.574 | 2.333 |
| $C_3$ | | | 1 | * |
| $C_4$ | | | | 1 |

| | $C_1$ | $C_2$ | $C_3$ | $C_4$ |
|---|---|---|---|---|
| $C_1$ | 1 | * | 2.011 | 6.839 |
| $C_2$ | | 1 | 0.760 | 0.534 |
| $C_3$ | | | 1 | 3.000 |
| $C_4$ | | | | 1 |

Based on unified pair-wise comparison values (lower row of matrices in the table 5.2), relative weights of project impacts are calculated (table 5.3).

Table 5.3 Calculated relative weights of project impacts

| Project marking | Normalized weight values |
|---|---|
| $C_1$ | 0.4455 |
| $C_2$ | 0.1743 |
| $C_3$ | 0.2919 |
| $C_4$ | 0.0883 |

For building of an optimal strategy for a 5-year prospect "Solon-3" DSS was used (see fig. 5.12).

Fig. 5.12 – Calculated distribution of resources among projects

From the fig. 5.12 we can see that for each project, eligible for funding expert estimates are input: minimum necessary resource volume for project implementation ($R\,min$), percentage of project implementation under minimum funding (% $min$), requested amount of resources ($R\,max$), and planned percentage of project implementation (% $max$ – usually, equals 100%). After the respective calculations, the amounts of allocated resources are shown in the "allocations" column.

The list of recommended actions of the DM (in the form of a set of projects with calculated funding volumes) will provide the basis for optimal strategic plan on ensuring the adequate level of counteracting information attacks within 5-year time slot under condition of sufficient funding.

## 5.7. A concept of an information and analytical system for information operation detection

Fig. 5.13 displays a conceptual Use Case diagram of an information and analytical system for information operation detection, which implements the above-listed methodologies through systematic integration of DEICPT, EES, DSS, and CMS.

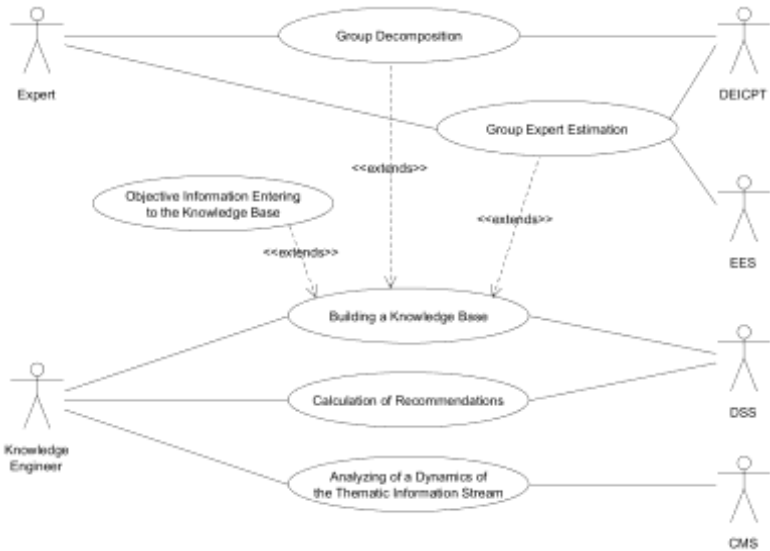

Fig. 5.13 – Use Case diagram of a concept of an information and analytical system for information operation detection

Actors:

- Expert — a specialist, who is invited or hired for financial incentive to provide a qualified conclusion, estimate, or judgment on specific matters.
- Knowledge engineer — a specialist, who builds the KB and (based on it) calculates recommendations for the DM using the DSS tools.
- DEICPT – a system for remote group work of experts in the global network.
- EES – a complex of software tools, which can flexibly adapt to the level of expert competence, and allow the user to obtain thorough and undistorted knowledge from the experts.
- DSS – a system, which calculates and returns the recommendations, based on knowledge (both objective and expert) input into its KB.
- CMS – a system for automated collection of information from websites in real-time mode, its structuring, grouping (clustering) according to semantic criteria, as well as topical selective distribution, accessing information databases in search modes, and analysis of topical information stream dynamics.

Application modes:

– Knowledge base building. The process is initiated by the knowledge engineer. As part of this process, a problem-oriented subject domain model is built, using the DSS tools, based on knowledge, obtained from experts, and on the results of global network content monitoring.

– Group decomposition. The process is initiated by the knowledge engineer. Experts participate in group decomposition of goals. According to the formulated main IO goal, the knowledge engineer forms a group of experts, i.e. specialists, who are competent in the issues under consideration. The process of group decomposition is organized as a step-by-step dialogue with the experts from the group, during which the main goal is gradually split into sub-components. At each decomposition stage every expert from the group is offered to formulate a set of goals (select goals from the list of already input ones, or input new ones), which directly influence the goal being decomposed. Then other goals are decomposed in a similar fashion. Decomposition process ends, when sub-goals represent specific IO components with specific formulations, providing the basis for respective queries to the CMS.

– Group expert estimate. The process is initiated by the knowledge engineer, once group decomposition is done. During this process, for each group of sub-goals the character (negative or positive, quantitative or qualitative) and the value of their impacts is determined.

– Input of objective information into the KB. The process is initiated by the knowledge engineer. During this process the information, which can be obtained through measurements of certain values, is input into the KB.

– Calculation of recommendations. During this process, relative efficiency of planted information precedents on each of IO components (i.e., "relative contribution" of each planted information precedent into achievement of the main IO goal) is calculated using the previously built KB. Besides that, an integral efficiency estimate of planted information precedents and the degree of goal achievement are calculated. Temporal dynamics is taken into account at this point. Based on retrospective analysis the target properties of the object are adjusted for the current period.

– The central component within the concept of an information and analytical system for information operation detection is the DSS. Since the key component of the DSS itself is the KB, let us address its structure in greater detail (see fig. 5.14). The basic elements of the KB are objects and connections between them. KB objects can be represented by goals and projects. In our case projects denote specific IO components.

– A KB object is denoted in the form of a specific formulation. It can be quantitative or qualitative, threshold-type or quasi-linear. Duration of implementation and required amounts of resources are specified for each project.

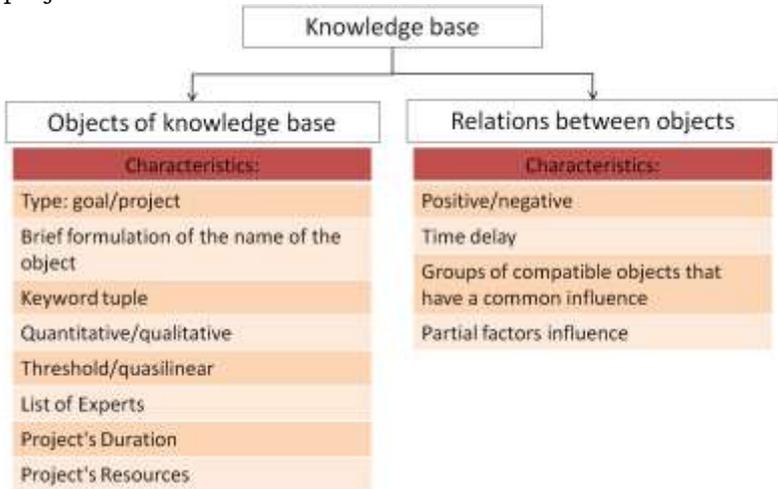

Fig. 5.14 – Structure of an expert DSS KB

Connection between KB objects can be positive or negative, have time delay, and denote whether the object belongs to some compatibility group. Besides that, it is characterized by a partial impact coefficient.

# Conclusion

For efficient analysis of present-day information processes through monitoring of information streams from global computer networks modern methods, based on non-linear analysis, should be used. Many of these methods are already successfully applied in natural sciences. Modern approaches allow us to use the methods, approbated, first and foremost, in natural sciences, even for analysis and modeling of societal and information systems. Analysis of information streams provides the foundation for such research areas as modeling, engineering, and forecasting.

The monograph suggests and substantiates the application of a methodology and an information technology for detection of information campaigns and operations based on analysis of statistical parameters of information message "life-cycle" characteristics' distribution.

The approaches, considered in the book, allow us to describe information processes and information impact processes; they are also applicable to description of general trends in the dynamics of information processes. At the same time, no progress in exploration of present-day information space is possible without the general understanding of the structure and dynamic properties of network information processes, which, in its turn, calls for detection and consideration of recurring patterns within them. We should note that often approaches based on precise methods and mathematical formalism, as well as on computer modeling, can produce predominantly qualitative results. This phenomenon can be explained by multi-parametric nature of real models. At the same time, in many cases, even these results can explain the reality better than traditional qualitative methods do.

Methods, algorithms, and analytic tools, considered in our work, are presented not only as a demonstrative basis for explanation of events and processes that actually take place, but also as essential components of their planning and forecasting.

At the same time, the listed models and methods are applicable to description of the general dynamic trends of information processes and information operation recognition, however, the problem of forecasting remains open. Evidently, more realistic models can be obtained if we consider an additional set of factors, most of which do not replicate themselves in time. Moreover, the structure of rules, providing the basis for functioning of the majority of available models, allows us to introduce the respective adjustments, for instance, artificially simulate random deviations. We should note that reproduction of results in time represents a serious challenge during modeling of information processes, and provides the basis for a scientific methodology. At present, only retrospective analysis of already implemented information operations remains the only relatively credible way of their verification.

International World Wide Web Conference, pp. 97-106, Chiba, Japan, 2005.

Международной конференции Моделирование-2016, Киев, 25-27 мая 2016 г. / ИПМЭ НАН Украины, 2016. - С. 198-201.